\documentclass{sig-alternate-2013}
\usepackage{cite}
\usepackage[ruled, vlined, linesnumbered]{algorithm2e}
\usepackage{enumitem}
\usepackage{times}
\usepackage{ifpdf}
\usepackage{xspace}
\usepackage{multirow}
\usepackage[table]{xcolor}
\usepackage{balance}
\usepackage{makecell}
\usepackage{arydshln}

\newcommand{\e}{\varepsilon}
\newcommand{\I}{\mathcal{I}}
\newcommand{\depth}{\text{\it depth}}
\newcommand{\Lap}{\text{\it Lap}}
\newcommand{\dom}{\text{\it dom}}
\newcommand{\hist}{\text{\it hist}}
\newcommand{\tree}{PrivTree\xspace}  
\newcommand{\para}[1]{\medskip \noindent {\bf #1}}
\newcommand{\edit}[1]{{\color{black} #1}}

\newtheorem{definition}{Definition}[section]

\newtheorem{theorem} {Theorem}[section]
\newtheorem{lemma} {Lemma}[section]

\newtheorem{corollary} {Corollary}
\newtheorem{claim} {Claim}

\def\done{\hspace*{\fill} {$\square$}}
\def\header{\vspace{2mm} \noindent}

\newfont{\mycrnotice}{ptmr8t at 7pt}
\newfont{\myconfname}{ptmri8t at 7pt}
\let\crnotice\mycrnotice%
\let\confname\myconfname%

\ifpdf
\fi

\begin{document}
\begin{sloppy}

\title{PrivTree: A Differentially Private Algorithm for Hierarchical Decompositions}

\author{
Jun Zhang$^{1}$ \hspace{15mm} Xiaokui Xiao$^1$ \hspace{15mm} Xing Xie$^2$\\
\and
\alignauthor
\affaddr{\hspace {-12mm}$^1$Nanyang Technological University $\qquad \qquad \qquad \qquad$ $^2$Microsoft Research} \\
\email{\hspace {-12mm} \{jzhang027, xkxiao\}@ntu.edu.sg \hspace {20mm} xingx@microsoft.com} \\
}
\maketitle

\begin{abstract}
Given a set $D$ of tuples defined on a domain $\Omega$, we study differentially private algorithms for constructing a histogram over $\Omega$ to approximate the tuple distribution in $D$. Existing solutions for the problem mostly adopt a {\em hierarchical decomposition} approach, which recursively splits $\Omega$ into sub-domains and computes a noisy tuple count for each sub-domain, until all noisy counts are below a certain threshold. This approach, however, requires that we (i) impose a limit $h$ on the recursion depth in the splitting of $\Omega$ and (ii) set the noise in each count to be proportional to $h$. The choice of $h$ is a serious dilemma: a small $h$ makes the resulting histogram too coarse-grained, while a large $h$ leads to excessive noise in the tuple counts used in deciding whether sub-domains should be split. Furthermore, $h$ cannot be directly tuned based on $D$; otherwise, the choice of $h$ itself reveals private information and violates differential privacy.

To remedy the deficiency of existing solutions, we present {\em \tree}, a histogram construction algorithm that adopts hierarchical decomposition but completely eliminates the dependency on a pre-defined $h$. The core of \tree is a novel mechanism that (i) exploits a new analysis on the Laplace distribution and (ii) enables us to use only {\em a constant amount of noise} in deciding whether a sub-domain should be split, without worrying about the recursion depth of splitting. We demonstrate the application of \tree in modelling spatial data, and show that it can be extended to handle sequence data (where the decision in sub-domain splitting is not based on tuple counts but a more sophisticated measure). Our experiments on a variety of real datasets show that \tree considerably outperforms the states of the art in terms of data utility.
\end{abstract}



\section{Introduction}\label{sec:intro}
Releasing sensitive data while preserving privacy is a problem that has attracted considerable attention in recent years. The state-of-the-art paradigm for addressing the problem is {\em differential privacy} \cite{D06}, which requires that the data released reveals little information about whether any particular individual is present or absent from the data. To fulfill such a requirement, a typical approach adopted by the existing solutions is to publish a noisy version of the data in place of the original one.

In this paper, we consider a fundamental problem that is frequently encountered in differentially private data publishing: Given a set $D$ of tuples defined over a domain $\Omega$, we aim to decompose $\Omega$ into a set $S$ of sub-domains and publish a noisy count of the tuples contained in each sub-domain, such that $S$ and the noisy counts approximate the tuple distribution in $D$ as accurately as possible. Applications of the problem include:
\begin{itemize}[topsep = 6pt, parsep = 6pt, itemsep = 0pt, leftmargin=18pt]
\item Private modelling of spatial data \cite{CPSSY12, QYL13} often requires generating a multi-dimensional histogram of the input data. 

\item For differentially private data mining (e.g., $k$-means \cite{SCLBJ15} and regression analysis \cite{Lei11}), one of the general approaches is to first coarsen the input data and inject noise into it, and then use the modified data to derive mining results.

\item Existing algorithms for sequence data publishing \cite{CFDS12} require identifying frequent patterns (e.g., prefixes) in a given set $D$ of sequences. This is equivalent to asking for a decomposition of the sequence domain $\Omega$ into a set of disjoint sub-domains, such that (i) each sub-domain includes all sequences in $D$ containing a particular pattern, and (ii) the number of sequences included in each sub-domain is larger than a given threshold.
\end{itemize}

To address the above decomposition problem, the prior art mostly adopts a {\em hierarchical} approach, which (i) recursively splits $\Omega$ into sub-domains and computes a noisy tuple count for each of them, and (ii) stops splitting a sub-domain when its noisy count is smaller than a threshold. This approach, albeit intuitive, requires a pre-defined limit $h$ on the maximum depth of recursion when splitting $\Omega$. The reason is that, to ensure differential privacy, the amount of noise injected in each tuple count has to be proportional to the maximum recursion depth, and hence, $h$ must be fixed in advance so that the algorithm can decide the correct noise amount to use.

Nevertheless, the choice of $h$ is a serious dilemma: for the algorithm to produce fine-grained sub-domains of $\Omega$, $h$ cannot be small; yet, increasing $h$ would lead to noisier tuple counts, and thus more errors in deciding whether a sub-domain should be split. As a consequence, no choice of $h$ could result in an accurate approximation of the input data. Furthermore, we cannot tune $h$ directly on the input dataset; otherwise, the choice of $h$ itself reveals private information and violates differential privacy. To mitigate these issues, existing work relies on heuristics to select an appropriate value of $h$, and to generate fine-grained decompositions even when $h$ is small. As we show in our experiments, however, those heuristics are rather ineffective when the input data follows a skewed distribution (which is often the case in practice).

\header
{\bf Contributions.} Motivated by the limitations of existing solutions, we present {\em \tree}, an algorithm for the decomposition problem that adopts the hierarchical approach but completely eliminates the dependency on a pre-defined $h$. In particular, \tree requires only {\em a constant amount of noise} in deciding whether a sub-domain should be split, which enables it to generate fine-grained decompositions without worrying about the recursion depth. Such a surprising improvement is obtained with a novel mechanism for differential privacy that exploits a non-trivial analysis on the Laplace noise \cite{DMNS06} to derive an extremely tight privacy bound. Its central insight is that, in the context of hierarchical decomposition, it is possible to publish a sequence $S$ of $0/1$ values using $O(1)$ noise, regardless of the {\em sensitivity} of $S$ \cite{DMNS06}. In contrast, the standard Laplace mechanism \cite{DMNS06} requires that noise amount must be proportional to $S$'s sensitivity.

To demonstrate the applications of \tree, we apply it to the private modeling of spatial data, and present a non-trivial extension to tackle sequence data, for which we adopt an advanced Markov model and utilize a sophisticated measure (instead of tuple counts) to decide whether a sub-domain should be split. We experimentally evaluate our algorithms on a variety of real data, and show that they considerably outperform the states of the art in terms of data utility. 

In addition, we present an in-depth analysis on the connection between \tree and the {\em support vector technique (SVT)} \cite{H11, DR13, LC14}, a technique widely adopted for data publishing under differential privacy. We show that there exists a variant of SVT \cite{LC14} that could have been used to implement \tree, if its privacy guarantees are as claimed in previous work \cite{LC14}. Nevertheless, we prove that the SVT variant \cite{LC14} does not satisfy differential privacy, which makes it inapplicable in our context.

In summary, we make the following contributions in this paper:
\begin{enumerate}[topsep = 4pt, parsep = 4pt, itemsep = 0pt, leftmargin=20pt]
\item We propose \tree, a differentially private algorithm for hierarchical decomposition that eliminates the dependency on a pre-defined threshold of the recursion depth.

\item We present applications of \tree in modeling spatial and sequence data. (Sections \ref{sec:alg} and \ref{sec:markov})

\item We analyze the connection between \tree and the SVT, and point out a misclaim about the latter in \cite{LC14}. (Section~\ref{sec:svt})

\item We conduct extensive experiments to demonstrate the superiority of \tree over the states of the art. (Section~\ref{sec:exp})
\end{enumerate}
\begin{figure*}[t]
\centering
\includegraphics[height=23mm]{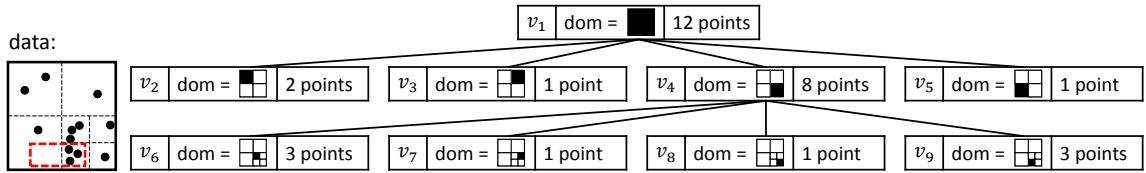}\\
\vspace{-2mm}
\caption{An illustration of a spatial decomposition tree.}
\label{fig:pre-spatial}
\vspace{-3mm}
\end{figure*}

\section{Preliminaries}\label{sec:pre}
In this section, we introduce the concepts behind {\em differential privacy} \cite{D06}, and define the problem of {\em spatial decomposition} \cite{S06book, D00book}, which we will address with our \tree algorithm in Section~\ref{sec:alg}.

\subsection{Differential Privacy}\label{sec:pre-dp}

Let $D$ be a sensitive dataset with $n$ tuples, and $\mathcal{A}$ be a data publishing algorithm that takes $D$ as input and releases a set of information $\mathcal{A}(D)$. Differential privacy requires that $\mathcal{A}(D)$ should be insensitive to the presence or absence of any particular tuple in $D$, so that an adversary cannot infer much private information from $\mathcal{A}(D)$. More formally, differential privacy is defined based on the concept of {\em neighboring datasets}, as shown in the following.

\begin{definition}[Neighboring Datasets \cite{D06}] \label{def:pre-neighbor}
Two datasets are neighboring if one of them can be obtained by inserting a tuple into the other.
\end{definition}

\begin{definition}[$\e$-Differential Privacy \cite{D06}] \label{def:pre-dp}
An algorithm $\mathcal{A}$ satisfies $\e$-differential privacy if, for any two neighboring datasets $D$ and $D'$ and for any possible output $O$ of $\mathcal{A}$,
\vspace{-1mm}
\begin{equation*} 
\ln\left(\frac{\Pr\left[\mathcal{A}(D) = O\right]}{\Pr\left[\mathcal{A}(D') = O \right]}\right) \le \varepsilon,
\vspace{-1mm}
\end{equation*}
where $\Pr[\cdot]$ denotes the probability of an event.
\end{definition}

There exist several mechanisms \cite{DMNS06, MT07, HT10} for achieving differential privacy, among which the most fundamental one is the {\em Laplace mechanism}. Specifically, the Laplace mechanism considers a function $f$ that takes $D$ as input and outputs a vector of real numbers, and it aims to release $f(D)$ with differential privacy. To achieve this objective, it adds i.i.d.\ noise into each value in $f(D)$, such that the noise $\eta$ follows a {\em Laplace distribution} with the following probability density function:
\vspace{-1mm}
\begin{equation} \label{eqn:pre-laplace}
\Pr[\eta = x] = \frac{1}{2\lambda} e^{-|x|/\lambda}.
\end{equation}
We denote the above distribution as $\Lap(\lambda)$, and refer to $\lambda$ as the {\em scale} (since the standard deviation of $\Lap(\lambda)$ is proportional to $\lambda$).
Dwork et al.\ \cite{DMNS06} prove that the Laplace mechanism achieves $(S(f)/\lambda)$-differential privacy,
where $S(f)$ is the {\em sensitivity} of $f$ defined as follows:
\begin{definition} [Sensitivity \cite{DMNS06}] \label{def:pre-sen}
Let $f$ be a function that maps a dataset $D$ into a vector of real numbers. The global sensitivity of $f$ is defined as
\begin{equation*} 
S(f) = \max_{D,D'} \left\|f(D) - f(D')\right\|_1,
\vspace{-1mm}
\end{equation*}
where $D$ and $D'$ are any two neighboring datasets, and $\|\cdot\|_1$ denotes the $L_1$ norm.
\end{definition}
Intuitively, $S(f)$ measures the maximum possible change in $f$'s output when we insert or remove one arbitrary tuple in $f$'s input.

An important property of differential private algorithms is that their composition also ensures differential privacy: 
\begin{lemma}[Composition Rule\cite{MM09}]\label{lmm:pre-composition}
Let $\mathcal{A}_1, \ldots, \mathcal{A}_k$ be $k$ algorithms, such that $\mathcal{A}_i$ satisfies $\e_i$-differential privacy ($i \in [1, k]$). Then, the sequential composition $(\mathcal{A}_1, \ldots, \mathcal{A}_t)$ satisfies $(\sum_{i=1}^k \e_i)$-differential privacy.
\end{lemma}
This lemma is particularly useful in proving that an algorithm ensures differential privacy: we can first decompose the algorithm into a few sequential components, and then analyze each component separately; after that, we can apply Lemma~\ref{lmm:pre-composition} to establish the overall privacy guarantee of the algorithm.

\subsection{Spatial Decompositions}\label{sec:pre-spatial}

Let $D$ be a set of data points in a multi-dimensional space $\Omega$. A {\em spatial decomposition} \cite{S06book, D00book} of $D$ consists of a tree-structured decomposition of $\Omega$ into its sub-domains, along with a partitioning of the data points among the leaves of the decomposition tree. For example, Figure~\ref{fig:pre-spatial} illustrates a spatial decomposition of a two-dimensional dataset $D$ that contains $12$ data points. The decomposition tree has $9$ nodes, namely, $v_1, v_2, \ldots, v_9$, each of which is associated with a sub-domain of $\Omega$ (denoted as ``$\dom$'' and visualized as a black rectangle in Figure~\ref{fig:pre-spatial}). We refer to each sub-domain as a {\em region}. The root of the tree, $v_1$, corresponds to a region that covers the entire $\Omega$; this region is recursively divided into four equal-size sub-regions in the lower levels of the tree, until each leaf node contains a sufficiently small number of data points.

The spatial decomposition in Figure~\ref{fig:pre-spatial} is referred to as a {\em quadtree} \cite{S06book, D00book}, and is widely adopted in spatial databases for efficient query processing. In particular, suppose that we are to use the quadtree to answer {\em range count queries}, i.e., queries that ask for the number of data points contained in a rectangle $q$. In that case, we can pre-compute, for each node $v$ in the quadtree, the number of data points contained in $v$'s region. Then, we can answer any range count query $q$ with a top-down traversal from the root node of the quadtree. Specifically, at the beginning of the traversal, we initialize the query answer as $ans = 0$. After that, for each node $v$ that we traverse, we examine $v$'s region $\dom(v)$, and differentiate four cases:
\begin{enumerate}[topsep = 4pt, parsep = 4pt, itemsep = 0pt, leftmargin=20pt]
\item If $\dom(v)$ is disjoint from $q$, we ignore $v$;

\item If $\dom(v)$ is fully contained in $q$, we increase $ans$ by the point count pre-computed for $v$;

\item If $\dom(v)$ partially intersects $q$ and $v$ is not a leaf node, then we visit every child of $v$ with a region not disjoint from $q$;

\item If $\dom(v)$ partially intersects $q$ and $v$ is a leaf node, then we inspect the data points in $\dom(v)$, and add to $ans$ the number of points contained in $q$.
\end{enumerate}
After the traversal terminates, we return $ans$ as the result. For instance, consider a range count query $q$ that corresponds to the dashed-line rectangle in Figure~\ref{fig:pre-spatial}. To answer $q$, we only need to examine four nodes, namely, $v_1, v_4, v_5, v_9$; the other nodes are all ignored since their regions are disjoint from $q$.

The efficiency of quadtrees results from its adaptiveness to the underlying distribution, i.e., it grows deep into the dense regions of $\Omega$ where there are a large number of data points (e.g., the region of $v_4$ in Figure~\ref{fig:pre-spatial}), and it ignores those regions that are sparse (e.g., the regions of $v_2, v_3, v_5$). Such adaptiveness has motivated existing work \cite{CPSSY12} to utilize quadtrees for generating private synopses of spatial data. Specifically, the technique in \cite{CPSSY12} first applies a differentially private algorithm to generate a quadtree, and then employs the Laplace mechanism to inject noise into the point count of each node. The quadtree and the noisy counts can then be used to answer any range-count query $q$ using the top-down traversal algorithm mentioned above, with two minor modifications. First, whenever we visit a node $v$ whose region is fully contained in $q$, we add the noisy count associated with $v$ (instead of the exact count) to the query answer $ans$. Second, if $v$ is a leaf node whose region $\dom(v)$ partially intersects $q$, then we multiply the noisy count of $v$ by $\frac{|q \, \cap \, \dom(v)|}{|\dom(v)|}$ before adding it to $ans$, where $|\cdot|$ denotes the area of a region. That is, given only the noisy count of $v$, we estimate the number of data points in $\dom(v)$ that are contained in $q$, by assuming that the points follow a uniform distribution. The rationale of this approach is that, given the adaptiveness of the quadtree, each leaf node $v$ should cover a region where the data distribution is not highly skewed; otherwise, $v$ should contain a dense sub-region, in which case the quadtree construction algorithm should have further split $v$ (instead of making $v$ a leaf node). This makes it relatively accurate to adopt a uniform assumption when estimating the contribution of $v$ to the answer of $q$. In Section~\ref{sec:alg}, we will present a more detailed analysis of the above quadtree approach, and then use it to motivate our \tree algorithm.

\begin{table} [t]
\centering
\vspace{-2mm}
\caption{Table of notations}\label{tbl:notation}
\vspace{2mm}
\renewcommand{\arraystretch}{1.1}
\begin{small}
 \begin{tabular} {|l|p{2.5in}|} \hline
   {\bf Notation} & {\bf Description} \\ \hline
   \hline
   $n, d$                   & the cardinality and dimensionality of the input dataset $D$ \\
   \hline
   $\Lap(\lambda)$          & a random variable following the Laplace distribution with $0$ mean and $\lambda$ scale  \\
   \hline
   $\dom(v)$                & the sub-domain of a node $v$\\
   \hline
   $\depth(v)$              & the hop distance from a node $v$ to the root of the tree\\
   \hline
   $\theta$                 & the threshold used to decide if a node should be split \\
   \hline
   $c(v), \hat{c}(v)$       & the point count of a node $v$, and its noisy version \\
   \hline
   \Gape[1pt][0pt]{$b(v), \hat{b}(v)$}       & the biased count of a node $v$, and its noisy version \\
   \hline
   $\rho(v)$                & the privacy risk of a node $v$ (see Equation~\eqref{eqn:alg-risk}) \\
   \hline
   \Gape[1.2pt][0pt]{$\rho^\top(v)$}           & an upper bound of $\rho$ (see Equation~\eqref{eqn:alg-ratio-bound}) \\
   \hline
   $\beta$                 & the fanout of the spatial decomposition tree \\
   \hline
   $\delta$                 & the decaying factor used by \tree \\
   \hline
   $\mathcal{I}$            & the set of distinct items in a given set $D$ of sequences \\
   \hline
\end{tabular}
\end{small}
\vspace{-0mm}
\end{table}

\section{Private Spatial Decompositions}\label{sec:alg}

This section presents our solution for constructing private spatial decompositions. We first revisit the private quadtree approach (in Section~\ref{sec:pre-spatial}) and discuss its limitations; after that, we elaborate our \tree algorithm, analyze its guarantees, and discuss its extensions. Table~\ref{tbl:notation} shows the notations that we frequently use.

\subsection{Private Quadtrees Revisited}\label{sec:alg:first}

Algorithm~\ref{alg:simple} presents a generic version of the private quadtree approach mentioned in Section~\ref{sec:pre-spatial}. The algorithm takes as input four parameters: (i) a set $D$ of spatial points defined over a multi-dimensional domain $\Omega$, (ii) the scale $\lambda$ of the Laplace noise to be used in the construction of the quadtree, (iii) the threshold $\theta$ used to decide whether a quadtree node should be split, and (iv) the threshold $h$ on the maximum height of the decomposition tree. The output of the algorithm is a quadtree $\mathcal{T}$ where each node $v$ comes with two pieces of information: the sub-domain of $\Omega$ corresponding to $v$ (denoted as $\dom(v)$), and a noisy version of the point count in $\dom(v)$ (denoted as $\hat{c}(v)$). We define the {\em depth} of $v$ as the hop distance between $v$ and the root of $\mathcal{T}$, and denote it as $\depth(v)$.

The algorithm starts by creating the root note $v_1$ of $\mathcal{T}$, after which it sets $\dom(v_1) = \Omega$ and marks $v_1$ as {\em unvisited} (Lines 1-2). The subsequent part of the algorithm consists of a number of iterations (Lines 3-9). In each iteration, we examine if there is an unvisited node $v$ in $\mathcal{T}$. If such $v$ exists, we mark $v$ as visited, and employ the Laplace mechanism to generate a noisy version $\hat{c}(v)$ of the number of points contained in $\dom(v)$ (Lines 4-6). After that, we split $v$ if the following two conditions simultaneously hold. First, $\hat{c}(v) > \theta$, i.e., $\dom(v)$ is likely to contain a sufficiently large number of points. Second, the height of the tree is smaller than $h$, which, as we discuss shortly, ensures that the noisy counts generated by the algorithm would not violate differential privacy. If both of the above conditions are met, then we generate $v$'s children and insert them into $\mathcal{T}$ as unvisited nodes (Lines 7-9); otherwise, $v$ becomes a leaf node of $\mathcal{T}$. When all of the nodes in $\mathcal{T}$ become visited, the algorithm terminates and returns $\mathcal{T}$.

\begin{algorithm}[t]
\caption{\label{alg:simple} {\bf SimpleTree} ($D$, $\lambda$, $\theta$, $h$)}
\BlankLine
    initialize a quadtree $\mathcal{T}$ with a root node $v_1$\;
    set $\dom(v_1) = \Omega$, and mark $v_1$ as unvisited\;
    \While{there exists an unvisited node $v$}
    {
        mark $v$ as visited\;
        compute the number $c(v)$ of points in $D$ that are contained in $\dom(v)$\;
        compute a noisy version of $c(v)$: $\;$ $\hat{c}(v) = c(v) + \Lap(\lambda)$\;
        \If{$\hat{c}(v) > \theta$ \textbf{and} $\depth(v) < h - 1$}
        {
            split $v$, and add its children to $\mathcal{T}$\;
            mark the children of $v$ as unvisited\;
        }
    }
    \Return $\mathcal{T}$\;
\end{algorithm}

\header
{\bf Privacy and Utility Analysis.} Algorithm~\ref{alg:simple} ensures $\e$-differential privacy if $\lambda \ge h/\e$. To understand this, suppose that we insert an arbitrary point $t$ into $D$. Then, $\mathcal{T}$ has only $h$ nodes whose exact point counts are affected by the insertion of $t$, i.e., the $h$ nodes whose sub-domains contain $t$. In addition, the point count of those nodes should change by one after $t$'s insertion. This indicates that the sensitivity (see Definition~\ref{def:pre-sen}) of all point counts in $\mathcal{T}$ equals $h$, and hence, adding i.i.d.\ Laplace noise of scale $\lambda \ge h/\e$ into the counts would achieve $\e$-differential privacy.


As we mention in Section~\ref{sec:intro}, however, requiring $\lambda \ge h/\e$ makes it rather difficult for Algorithm~\ref{alg:simple} to generate high-quality quadtrees. Specifically, if we set $h$ to a small value, the resulting quadtree $\mathcal{T}$ would not adapt well to the data distribution in $D$, due to the restriction on the tree height; meanwhile, increasing $h$ would also increase the amount of noise in each $\hat{c}(v)$, which makes Algorithm~\ref{alg:simple} more error-prone in deciding whether a node should be split, thus degrading the quality of $\mathcal{T}$. In other words, any choice of $h$ inevitably leads to inferior data utility. Furthermore, we cannot directly tune $h$ by (i) testing the performance of Algorithm~\ref{alg:simple} on $D$ under different settings of $h$, and then (ii) selecting the one that yields the best result. The reason is that such a tuning process violates differential privacy: when we change the input data from $D$ to a neighboring dataset $D'$, the tuning process may select a different $h$, in which case Algorithm~\ref{alg:simple} would use different noise scales for $D$ and $D'$, invalidating its privacy guarantee.


To alleviate the above issue, existing work \cite{CPSSY12,QYL13,QYL13hierarchy,SCLBJ15} resorts to heuristics to choose $h$ without violating differential privacy, and to enhance the performance of Algorithm~\ref{alg:simple}, e.g., by avoiding the generation of noisy counts for certain levels of the decomposition tree (so that $h$ can be reduced), and by exploiting correlations among the noisy counts to improve their accuracy \cite{HRMS10}. However, none of those heuristics is able to thoroughly address the limitations of Algorithm~\ref{alg:simple}. As we shown in our experiments in Section~\ref{sec:exp}, existing approaches tend to provide inferior data utility, especially when the input data follows a skewed distribution.

\color{black}

\subsection{Rationale Behind Our Solution}\label{sec:alg:rationale}

To remedy the deficiency of Algorithm~\ref{alg:simple}, we aim to eliminate the requirement that $\lambda \ge h/\e$, and make $\lambda$ a constant instead. This would not only resolve the dilemma in choosing $h$, but also unleash the potential of quadtrees as the tree height is no longer restricted. Towards this end, we first make a simple observation: after we finish constructing the quadtree $\mathcal{T}$ in Algorithm~\ref{alg:simple}, we could remove all noisy counts associated with the intermediate nodes, and release only the noisy counts for the leaf nodes as well as the sub-domains of all nodes. The released tree, denoted as $\mathcal{T}'$, could still be used for query processing, since we can re-generate an alternative count for each intermediate node $v$ in $\mathcal{T'}$ by summing up the published noisy counts of the leaf nodes under $v$. Intuitively, $\mathcal{T'}$ reveals less information than $\mathcal{T}$ does, and hence, we might use less noise in $\mathcal{T}'$ to achieve $\e$-differential privacy.

However, the above intuition does not hold in general, as $\mathcal{T'}$ and $\mathcal{T}$ require the same amount of noise to enforce the same privacy guarantee. To explain, consider two neighboring datasets $D$ and $D'$, such that $D$ is obtained by inserting a point $t$ into $D'$. Let $v_1, v_2, \ldots, v_{h}$ be the $h$ nodes in $\mathcal{T'}$ whose sub-domains contain $t$. Then, these $h$ nodes should form a path from the root of $\mathcal{T}$ to a leaf node. Furthermore, for any $v_i$, the point count of $v_i$ is decreased by one when we change the input dataset from $D$ to $D'$. We use $c(v_i)$ to denote $v_i$'s exact point count on $D$.

Without loss of generality, assume that $v_h$ is a leaf node (i.e., $v_1, \ldots, v_{h-1}$ are all intermediate nodes). Let $\Pr[D\rightarrow \mathcal{T'}]$ (resp.\ $\Pr[D'\rightarrow \mathcal{T'}]$) denote the probability that we obtain $\mathcal{T'}$ from $D$ (resp.\ $D'$) given fixed $\lambda$, $\theta$, and $h$. Then,
\vspace{-1mm}
\small
\begin{align*} 
\ln\left( \! \frac{\Pr[D\rightarrow \mathcal{T'}]}{\Pr[D'\rightarrow \mathcal{T'}]} \! \right)  = & \sum_{i = 1}^{h-1} \ln\left( \! \frac{\Pr[c(v_i) + \Lap(\lambda) > \theta]}{\Pr[c(v_i) - 1 + \Lap(\lambda) > \theta]} \! \right) \nonumber \\
  &  \! {} + \ln\left( \! \frac{\Pr[c(v_{h}) + \Lap(\lambda) = \hat{c}(v_h)]}{\Pr[c(v_h) - 1 + \Lap(\lambda) = \hat{c}(v_h)]} \! \right).
\vspace{-1mm}
\end{align*}
\normalsize
By Equation~\eqref{eqn:pre-laplace}, for any $c(v_{h}) < \hat{c}(v_{h})$, 
\vspace{-1mm}
\small
\begin{equation} \label{eqn:alg-leaf-ratio}
\ln\left(\frac{\Pr[c(v_{h}) + \Lap(\lambda) = \hat{c}(v_h)]}{\Pr[c(v_h) - 1 + \Lap(\lambda) = \hat{c}(v_h)]}\right) = \frac{1}{\lambda}.
\vspace{-1mm}
\end{equation}
\normalsize
In addition, for any $v_i$ ($i \in [1, h-1]$) with $c(v_i) \le \theta$,
\vspace{-1mm}
\small
\begin{equation} \label{eqn:alg-worst-ratio}
\ln\left(\frac{\Pr[c(v_i) + \Lap(\lambda) > \theta]}{\Pr[c(v_i) - 1 + \Lap(\lambda) > \theta]}\right) = \frac{1}{\lambda}.
\vspace{-1mm}
\end{equation}
\normalsize
Therefore, when $c(v_i) \le \theta$ holds for every $v_i$ ($i \in [1, h-1]$),
\vspace{-1mm}
\small
\begin{equation} \label{eqn:alg-worst-sum-ratio}
\ln\left(\frac{\Pr[D\rightarrow \mathcal{T'}]}{\Pr[D'\rightarrow \mathcal{T'}]}\right) = \frac{h}{\lambda}.
\vspace{-1mm}
\end{equation}
\normalsize
By Definition~\ref{def:pre-dp}, this indicates that $\lambda$ must be at least $h/\e$ to ensure that $\mathcal{T'}$ achieves $\e$-differential privacy.

In summary, $\mathcal{T'}$ is no better than $\mathcal{T}$ in terms of the amount of noise required, because of the negative result in Equations \eqref{eqn:alg-leaf-ratio} and \eqref{eqn:alg-worst-ratio}. In other words, {\em in the worst case}, releasing the boolean result of $c(v_i) + \Lap(\lambda) > \theta$ incurs the same privacy cost as releasing $c(v_i) + \Lap(\lambda)$ directly. That said, if $c(v_i) > \theta$ for some $v_i$, then Equation~\eqref{eqn:alg-worst-ratio} does not hold, in which case $\mathcal{T'}$ could entail a smaller privacy cost than $\mathcal{T}$ does. To illustrate this, we denote the l.h.s.\ of Equation~\eqref{eqn:alg-worst-ratio} as a function $\rho$ of $c(v_i)$, i.e.,
\begin{equation} \label{eqn:alg-risk}
\rho(x) = \ln\left(\frac{\Pr[x + \Lap(\lambda) > \theta]}{\Pr[x - 1 + \Lap(\lambda) > \theta]}\right),
\end{equation}
and we plot $\rho$ in Figure~\ref{fig:bound}. (Note that the y-axis of Figure~\ref{fig:bound} is in a logarithmic scale.) Observe that, when $x = c(v) \ge \theta + 1$, $\rho(x)$ decreases exponentially with the increase of $x$. This indicates that $\ln\left(\frac{\Pr[D\rightarrow \mathcal{T'}]}{\Pr[D'\rightarrow \mathcal{T'}]}\right)$ could be much smaller than $h/\lambda$, if $c(v_i) \ge \theta + 1$ holds for all $i \in [1, h-1]$. For example, if $c_{h-1} \ge \theta + 1$ and $c(v_i) - c(v_{i+1})$ is at least a constant for all $i \in [1, h-2]$, then
\vspace{-1mm}
\begin{equation} \label{eqn:alg-good-ratio}
\sum_{i=1}^{h-1} \rho\Big(c(v_i)\Big) = \Theta\left(\frac{1}{\lambda}\right),
\vspace{-0mm}
\end{equation}
due to the exponential decrease of $\rho(v_i)$. In that case, we have $\ln\left(\frac{\Pr[D\rightarrow \mathcal{T'}]}{\Pr[D'\rightarrow \mathcal{T'}]}\right) = \Theta\left(1/\lambda\right)$ instead of $\ln\left(\frac{\Pr[D\rightarrow \mathcal{T'}]}{\Pr[D'\rightarrow \mathcal{T'}]}\right) = h/\lambda$, which would enable us to set $\lambda$ as a constant independent of $h$.

\begin{figure}[t]
\centering
\includegraphics[height=25mm]{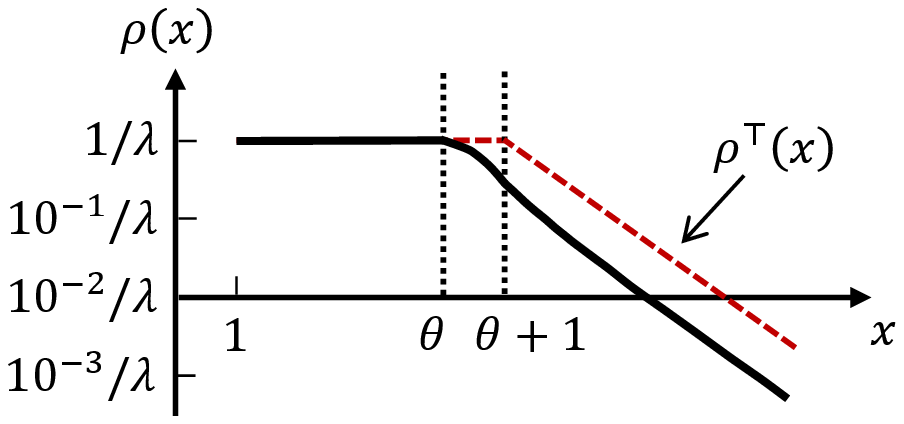} \\
\vspace{-2mm}
\caption{An illustration of $\boldsymbol{\rho(x)}$ and $\boldsymbol{\rho^\top(x)}$.}
\label{fig:bound}
\vspace{-1mm}
\end{figure}

The above analysis leads to an interesting question: can we ensure that Equation~\eqref{eqn:alg-good-ratio} holds for any input dataset? In Section~\ref{sec:alg:privtree}, we will give an affirmative answer to this question. The basic idea of our method is to add a bias term to each $c(v_i)$, so that $c(v_i) - c(v_{i+1})$ ($i \in [1, h-2]$) is larger than a constant of choice. In addition, the bias term is independent of the input data, which guarantees that its usage does not leak any private information. The derivation of the bias term requires a careful analysis of $\rho(x)$. To simplify our analysis, we devise a simple upper bound of $\rho(x)$: 
\begin{lemma} \label{lmm:alg-bound}
Let $\rho^\top$ be a function such that
\begin{equation} \label{eqn:alg-ratio-bound}
\rho^\top(x) =
\begin{cases}
1/\lambda, & \text{if $x < \theta + 1$}\\
\frac{1}{\lambda} \exp\left(\frac{\theta + 1 - x}{\lambda}\right), & \text{otherwise}
\end{cases}
\end{equation}
Then, $\rho(x) \le \rho^\top(x)$ for any $x$.
\end{lemma}
Figure~\ref{fig:bound} shows $\rho^\top$ with a dashed line. Observe that it closely captures the exponential decrease of $\rho$ when $c(v) \ge \theta + 1$.

\subsection{The {PrivTree} Algorithm}\label{sec:alg:privtree}

Algorithm~\ref{alg:privtree} presents our \tree technique for private spatial decomposition. As with Algorithm~\ref{alg:simple}, \tree asks for a spatial dataset $D$, the scale $\lambda$ of Laplace noise to be used, and a threshold $\theta$ for deciding whether a node should be split. However, it does not request a threshold $h$ on the maximum tree height; instead, it requires a positive number $\delta$, the usage of which will be clarified shortly. The output of \tree is a quadtree $\mathcal{T}$, with the point count associated with each node removed. That is, $\mathcal{T}$ reveals the sub-domain of each node $v$, but conceals all information about $c(v)$. In Section~\ref{sec:alg:parameter}, we will explain how we obtain the point count of each node, as well as our choices of $\theta$ and $\delta$.

In a nutshell, \tree is similar to Algorithm~\ref{alg:simple} in that it also (i) generates $\mathcal{T}$ by recursively splitting a root node $v_1$ whose region $\dom(v_1)$ covers the whole data space $\Omega$, and (ii) decides whether a node $v$ should be split based on a noisy point count of $v$. However, the method for obtaining noisy counts marks the crucial difference between the two algorithms. Specifically, given a node $v$, \tree does not generate its noisy count by directly adding Laplace noise to $c(v)$. Instead, \tree first computes a biased count ${b(v) = c(v) - \depth(v)\cdot \delta}$, and checks if it is smaller than $\theta - \delta$; if it is, then \tree increases it to $\theta - \delta$. In other words,
\begin{equation} \label{eqn:alg-bv}
b(v) = \max\Big\{\theta - \delta, \;\; c(v) - \depth(v)\cdot \delta \Big\}.
\end{equation}
After that, \tree produces a noisy count $\hat{b}(v) = b(v) + \Lap(\lambda)$, and splits $v$ if $\hat{b}(v)$ is larger than the given threshold $\theta$. Notice that \tree does not restrict the height of $\mathcal{T}$, as the decision to split any node $v$ solely depends on $\hat{b}(v)$.

\begin{algorithm}[t]
\caption{\label{alg:privtree} {\bf PrivTree} ($D$, $\lambda$, $\theta$, $\delta$)}
\BlankLine
    initialize a quadtree $\mathcal{T}$ with a root node $v_1$\;
    set $\dom(v_1) = \Omega$, and mark $v_1$ as unvisited\;
    \While{there exists an unvisited node $v$}
    {
        mark $v$ as visited\;
        compute a biased point count for $v$ with decaying factor $\delta$: $b(v) = c(v) - \depth(v)\cdot \delta$\;
        adjust $b(v)$ if it is excessively small: $b(v) = \max\left\{b(v), \theta - \delta\right\}$\;
        compute a noisy version of $b(v)$: $\;$ $\hat{b}(v) = b(v) + \Lap(\lambda)$\;
        \If{$\hat{b}(v) > \theta$}
        {
            split $v$, and add its children to $\mathcal{T}$\;
            mark the children of $v$ as unvisited\;
        }
    }
    \Return $\mathcal{T}$ with all point counts removed\;
\end{algorithm}

\header
{\bf Privacy Analysis.} Consider any quadtree $\mathcal{T}$ output by \tree, and any two neighboring datasets $D$ and $D'$, such that $D$ is obtained by inserting a point $t$ into $D'$. In what follows, we show that setting $\lambda = \Theta(1/\e)$ is sufficient for $\e$-differential privacy, i.e.,
\vspace{-1mm}
\begin{small}
\begin{equation} \label{eqn:alg-obj}
-\e \le \ln\left( \! \frac{\Pr[D\rightarrow \mathcal{T}]}{\Pr[D'\rightarrow \mathcal{T}]} \! \right)  \le \e.
\end{equation}
\end{small}
\vspace{0mm}

The proof for the first inequality in Equation~\eqref{eqn:alg-obj} is relatively straightforward. For any node $v$ in $\mathcal{T}$, let $c(v)$ be $v$'s point count on $D$, and $b(v)$ be the biased version of $c(v)$ generated from Equation \eqref{eqn:alg-bv}. Let $c'(v)$ and $b'(v)$ be the counterparts of $c(v)$ and $b(v)$, respectively, given $D'$ as the input. Then, we have $c(v) = c'(v)$ for all nodes $v$ in $\mathcal{T}$, except for the nodes whose sub-domains contain $t$. Note that those nodes should form a path from the root of $\mathcal{T}$ to a leaf. Let $k$ be the length of the path, and $v_i$ be $i$-th node in the path, with $v_1$ denoting the root of $\mathcal{T}$. We have $c(v_i) = c'(v_i) + 1$ and
\begin{equation} \label{eqn:alg-bv-compare}
b(v_i) =
\begin{cases}
b'(v_i) + 1, & \text{if $b(v_i) \ge \theta  - \delta + 1$}\\
b'(v_i), & \text{otherwise}
\end{cases}
\end{equation}
Then, by Equation~\eqref{eqn:pre-laplace},

\vspace{-4mm}
\begin{small}
\begin{align*} 
\ln\left( \! \frac{\Pr[D\rightarrow \mathcal{T}]}{\Pr[D'\rightarrow \mathcal{T}]} \! \right) \; = \; & \sum_{i = 1}^{k-1} \ln\left( \! \frac{\Pr[b(v_i) + \Lap(\lambda) > \theta]}{\Pr[b'(v_i) + \Lap(\lambda) > \theta]} \! \right) \\
 & {} + \ln\left( \! \frac{\Pr[b(v_k) + \Lap(\lambda) \le \theta]}{\Pr[b'(v_k) + \Lap(\lambda) \le \theta]} \! \right) \\
\ge \; &  0 -\frac{1}{\lambda} \;\; = -\frac{1}{\lambda}.
\end{align*}
\end{small}
\vspace{-2mm}

\noindent
This indicates that $\lambda \ge 1/\e$ ensures the first inequality in Equation~\eqref{eqn:alg-obj}.

Next, we prove the second inequality in Equation~\eqref{eqn:alg-obj} by analyzing $\ln\left( \! \frac{\Pr[b(v_i) + \Lap(\lambda) > \theta]}{\Pr[b'(v_i) + \Lap(\lambda) > \theta]} \! \right)$, which we refer to as the {\em privacy cost} of $v_i$. The high-level idea of our proof is as follows. First, due to the way that we generate biased counts, each node $v_i$'s bias count $b(v_i)$ is at least a constant $\delta$ smaller than that of its parent $v_{i-1}$, as long as $b(v_i) \ge \theta + 1$. Based on this observation and Lemma~\ref{lmm:alg-bound}, we show that all nodes $v_i$ with $b(v_i) \ge \theta + 1$ incur a total privacy cost of $\Theta(1/\e)$. After that, we prove that the total privacy cost of the remaining nodes is also $\Theta(1/\e)$.

By the definition of $v_1, \ldots, v_k$, we have $c(v_i) \ge c(v_{i+1})$ and $\depth(v_i) = \depth(v_{i+1}) - 1$ for any $i \in [1, k-1]$. This indicates that $b(v_i) \ge b(v_{i+1}) \ge \theta - \delta$, due to Equation~\eqref{eqn:alg-bv}. Without loss of generality, assume that there exists $m \in [1, k-1]$, such that $b(v_m) \ge \theta - \delta + 1$ and $b(v_{m+1}) = \theta - \delta$. Then,
\begin{equation} \label{eqn:alg-bv-sequence}
\begin{cases}
b(v_{i-1}) \, \ge \, b(v_i) + \delta \, \ge \, \theta + 1, & \text{if $i \in [2, m]$}\\
b(v_i) \, =  \, \theta - \delta, & \text{otherwise}
\end{cases}
\end{equation}
Combining Equations \eqref{eqn:alg-bv-compare} and \eqref{eqn:alg-bv-sequence}, we have $b'(v_i) = b(v_i)$ when $i > m$, and $b'(v_i) = b(v_i) - 1$ otherwise. Therefore,

\vspace{-4mm}
\begin{small}
\begin{align*} 
\ln\left( \! \frac{\Pr[D\rightarrow \mathcal{T}]}{\Pr[D'\rightarrow \mathcal{T}]} \! \right) \; = \;  & \sum\nolimits_{i = 1}^{k-1} \ln\left( \! \frac{\Pr[b(v_i) + \Lap(\lambda) > \theta]}{\Pr[b'(v_i) + \Lap(\lambda) > \theta]} \! \right) \\
 & {} + \ln\left( \! \frac{\Pr[b(v_k) + \Lap(\lambda) \le \theta]}{\Pr[b'(v_k) + \Lap(\lambda) \le \theta]} \! \right) \\
\le \; &  \sum\nolimits_{i = 1}^{k-1} \ln\left( \! \frac{\Pr[b(v_i) + \Lap(\lambda) > \theta]}{\Pr[b'(v_i) + \Lap(\lambda) > \theta]} \! \right) \\
= \; & \sum\nolimits_{i = 1}^{m} \ln\left( \! \frac{\Pr[b(v_i) + \Lap(\lambda) > \theta]}{\Pr[b(v_i) - 1 + \Lap(\lambda) > \theta]} \! \right) \\
= \; & \sum\nolimits_{i = 1}^{m} \rho\Big(b(v_i)\Big),
\end{align*}
\end{small}
\vspace{-2mm}

\noindent
where $\rho(\cdot)$ is as defined in Equation~\eqref{eqn:alg-risk}.
By Lemma~\ref{lmm:alg-bound} and Equation~\eqref{eqn:alg-bv-sequence},

\vspace{-4mm}
\begin{small}
\begin{align*} 
\sum\nolimits_{i = 1}^{m} \rho\Big(b(v_i)\Big) \, \le \; & \sum\nolimits_{i = 1}^{m} \rho^\top\Big(b(v_i)\Big) \\
 = \; & \rho^\top\Big(b(v_m)\Big) + \sum\nolimits_{i = 1}^{m-1} \frac{1}{\lambda}\exp\left(\frac{\theta + 1 - b(v_i)}{\lambda}\right) \\
\le \; & \frac{1}{\lambda} + \frac{1}{\lambda}\cdot \frac{1}{1 - \exp(-\delta/\lambda)} \\
 = \; & \frac{1}{\lambda} \cdot \frac{2e^{\delta/\lambda} - 1}{e^{\delta/\lambda} - 1}.
\end{align*}
\end{small}
\vspace{-2mm}

\noindent
Therefore, if we set $\delta = \gamma \cdot \lambda$, where $\gamma$ is a constant, then
\begin{small}
\begin{align*} 
\ln\left( \! \frac{\Pr[D\rightarrow \mathcal{T}]}{\Pr[D'\rightarrow \mathcal{T}]} \! \right) \; = \; \sum_{i = 1}^{m} \rho\Big(b(v_i)\Big) \; \le \; \frac{1}{\lambda} \cdot \frac{2e^{\gamma} - 1}{e^{\gamma} - 1} \; = \; \Theta\left(\frac{1}{\lambda}\right).
\end{align*}
\end{small}
Summing up the above analysis, we have the following theorem:
\begin{theorem}\label{thm:alg:privacy}
\tree satisfies $\e$-differential privacy if ${\lambda \ge \frac{2e^{\gamma} - 1}{e^{\gamma} - 1} \cdot \frac{1}{\e}}$ and $\delta = \gamma \cdot \lambda$ for some $\gamma > 0$.
\end{theorem}


\subsection{Noisy Counts and Parameterization}\label{sec:alg:parameter}

\vspace{1mm}

\noindent
{\bf Generation of Noisy Counts.} Recall that \tree outputs a quadtree with the point count for each node removed. However, if a quadtree with noisy counts is needed, we can easily obtain it by adding a postprocessing step to \tree. In particular, given a dataset $D$, we first invoke \tree to produce a $\frac{\e}{2}$-differentially private quadtree $\mathcal{T}$. After that, for each leaf node $v$ of $\mathcal{T}$, we publish a noisy version of $v$'s point count using Laplace noise of scale $2/\e$. It can be verified that this postprocessing step satisfies $\frac{\e}{2}$-differential privacy. Then, by Lemma~\ref{lmm:pre-composition}, the generation of $\mathcal{T}$ and the noisy counts as a whole achieves $\e$-differential privacy. Finally, we compute a noisy count for each intermediate node $v$ in $\mathcal{T}$, by taking the sum of the noisy counts of all leaf nodes under $v$.

\header
{\bf Choice of $\boldsymbol \delta$.} As shown in Theorem~\ref{thm:alg:privacy}, when $\delta = \gamma \cdot \lambda$, \tree needs to use a noise scale $\lambda \ge \frac{2e^{\gamma} - 1}{e^{\gamma} - 1} \cdot \frac{1}{\e}$ to achieve $\e$-differential privacy. Intuitively, the choice of $\delta$ is a balancing act between the amount of bias and the amount of noise in each biased noisy count $\hat{b}(v)$ used by \tree. In particular, if $\delta$ is small with respect to $\lambda$, then the bias term in each $\hat{b}(v)$ is small, but the noise amount in $\hat{b}(v)$ would need to be large, since $\frac{2e^{\gamma} - 1}{e^{\gamma} - 1} \cdot \frac{1}{\e}$ increases when $\gamma = \delta/\lambda$ decreases. In contrast, if $\delta$ is large with respect to $\lambda$, then each $\hat{b}(v)$ would have small noise but a large bias.

That said, we observe that there is a more important factor in choosing $\delta$. To explain, consider a node $v$ with a biased count $b(v) = \theta - \delta$. Ideally, we would like \tree to avoid splitting such a node $v$, as its point count is likely to be small. Nevertheless, if $b(v) + \Lap(\lambda) > \theta$, then \tree would split $v$ and insert its children into the quadtree. In turn, each of $v$'s children also has a certain probability to be split, and so on. If $\delta$ is excessively small, then each offspring of $v$ has a relatively large splitting probability, in which case the splitting process may not {\em converge}, i.e., \tree may keep generating offsprings of $v$ and does not terminate.

To address the above issue, we set $\delta$ in such a way to ensure that if $b(v) = \theta - \delta$, then in expectation, only $2$ nodes would be generated from the subtree under $v$ (including $v$ itself). Specifically, we set $\delta = \lambda \cdot \ln \beta$, where $\beta$ denote the {\em fanout} of $\mathcal{T}$, i.e., the number of children that each intermediate node in $\mathcal{T}$ has. (For example, $\beta = 4$ if $\mathcal{T}$ is a two-dimensional quadtree.) By Equation~\eqref{eqn:pre-laplace}, this setting of $\delta$ guarantees that any node $v$ with $b(v) = \theta - \delta$ has $\frac{1}{2\beta}$ probability to be split. Formally, we have the following lemma:
\begin{lemma} \label{lmm:alg-converge}
Let $\mathcal{T}$ be the output of \tree given $\delta = \lambda \cdot \ln \beta$, and $\mathcal{T^*}$ be the output of \tree when it sets $\hat{b}(v) = c(v)$ for each node $v$ (i.e., no noise or bias is introduced in the split decisions). Then, $\mathbb{E}[|\mathcal{T}|] \le 2 \cdot |\mathcal{T^*}|$ whenever $|T^*| > 1$, where $|\cdot|$ denotes the number of nodes in a tree.
\end{lemma}
The above setting of $\delta$ leads to the following corollary:
\begin{corollary}\label{coro:alg:privacy}
\tree satisfies $\e$-differential privacy if ${\lambda \ge \frac{2\beta - 1}{\beta - 1} \cdot \frac{1}{\e}}$ and $\delta = \lambda \cdot \ln \beta$, where $\beta$ is the fanout of $\mathcal{T}$.
\end{corollary}

\header
{\bf Choice of $\boldsymbol \theta$.} Intuitively, the threshold $\theta$ serves the purpose of ensuring that each leaf node $v$ of $\mathcal{T}$ contains a sufficiently large point count $c(v)$, so that if we choose to output a noisy version of $c(v)$, it would not overwhelmed by the Laplace noise injected. The choice of $\theta$, however, is complicated by the fact that \tree adds a negative bias to the point count of a node when it decides whether or not to split the node. In particular, due to the negative bias, even $\theta = 0$ could ensure that a node $v$ with $b(v) > \theta$ has a sufficient large point count. Therefore, we use $\theta = 0$ in our implementation of \tree, and we observe that it leads to reasonably good results in our experiments.

\subsection{Extensions}\label{sec:alg:extend}

Although we have presented \tree in the context of spatial decomposition, we note that it could be extended in several different aspects for other applications. First, the decomposition tree used by \tree does not have to be a quadtree, but can be any other tree structure instead. For example, suppose that we are given a multi-dimensional dataset $D$ containing both numeric and categorical attributes, and that each categorical attribute has a taxonomy. Then, we can still apply \tree on $D$ to generate a private synopsis of $D$, by splitting each numeric dimension of $D$ according to a binary tree and each categorical dimension based on its taxonomy.

Second, when \tree decides whether or not a node $v$ should be split, the decision does not have to be based on the count of tuples contained in $\dom(v)$, but can also be based on any other score function $\mu(v)$ that is {\em monotonic}, i.e., $\mu(v) \le \mu(u)$ whenever $v$ is a child node of another node $u$. The rationale is that, as long as $\mu$ is monotonic, we can add a bias to the score of each node $v$ to ensure that it is at least a constant smaller that the score of $v$'s parent. Then, we can apply Lemma~\ref{lmm:alg-bound} to show that \tree guarantees differential privacy, given that $\lambda$ is properly set based on the sensitivity of the score function $\mu$. In Section~\ref{sec:markov}, we will apply this idea to extend \tree for private modeling of sequence data.

Finally, although the privacy analysis of \tree (in Section~\ref{sec:alg:privtree}) assumes that the presence or absence of a tuple $t$ only affects one leaf node and its ancestors, it can be extended to the case when multiples leaf nodes and their ancestors are impacted. In particular, if at most $x$ leaf nodes can be affected, then we can apply \tree with the noise scale $\lambda$ enlarged $x$ times. The intuition is that each affected leaf node, along its ancestors, incurs one unit of privacy cost, which in turn requires one unit of noise to mitigate; as such, when there are $x$ affected leaf nodes, we need $x$ units of noise for sufficient privacy protection.

\section{Private Markov Models}\label{sec:markov}

This section presents an extension of \tree for constructing Markov models on sequence data. We first introduce the basic concepts of sequences and Markov models in Section~\ref{sec:markov-concept}. After that, we elaborate our \tree extension in Section~\ref{sec:markov-tree}, and compare it with existing solutions in Section~\ref{sec:markov-compare}.

\begin{figure*}[!t]
\centering
\hspace{-0mm}\includegraphics[height=21mm]{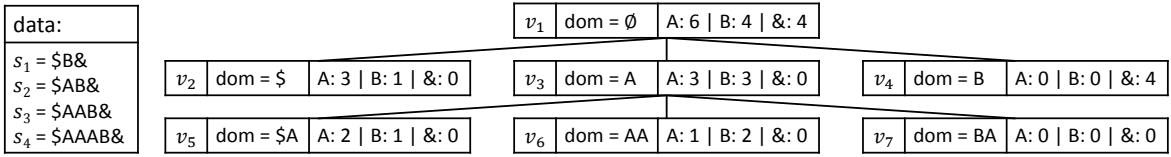}\\
\vspace{-1mm}\caption{An illustration of a prediction suffix tree (PST).}
\label{fig:markov}
\vspace{-1mm}
\end{figure*}

\subsection{Sequence Data and Markov Models} \label{sec:markov-concept}

Given a finite alphabet $\I$, a {\em sequence} $s$ of length $l$ over $\I$ is an ordered list $x_1x_2\cdots x_l$, where each $x_i$ ($i \in [1, l]$) is a symbol in $\I$. For convenience, we abuse notation and write $s = \$x_1x_2\cdots x_l\&$, where $\$$ and $\&$ are two special symbols that mark the beginning and the end of a sequence, respectively. Sequences are frequently used to represent user behavioral data, such as trajectories, web navigation traces, and product purchasing histories.


Markov models are a type of stochastic models commonly used to characterize sequence data. They assume the {\em Markov property} \cite{RST96}, i.e., a symbol $x$ in a sequence $s$ is decided by a few symbols that immediately proceeds $x$ in $s$, but not any others. A Markov model over $D$ is often represented as {\em prediction suffix tree (PST)} \cite{RST96, BEY04}, where each node $v$ is associated with a {\em predictor string} $\dom(v)$, as well as a {\em prediction histogram} $\hist(v)$. In particular, $\dom(v)$ consists of symbols in $\I \cup \{\$\}$, while $\hist(v)$ contains a count for each symbol $x$ in $\I \cup \{\&\}$. The count, denoted as $\hist(v)[x]$, is computed as follows. We first inspect all occurrences of $\dom(v)$ in the sequences in $D$, and count the number $y$ of occurrences when $\dom(v)$ is immediately followed by the symbol $x$; after that, we set $\hist(v)[x] = y$. In other words, $\hist(v)[x]$ indicates how often an appearance of $\dom(v)$ would immediately lead to an appearance of $x$ in a sequence.

For example, Figure~\ref{fig:markov} illustrates a PST constructed over a set $D$ containing four sequences $s_1, s_2, s_3, s_4$, over an alphabet $\I = \{A, B\}$. The node $v_6$ has a predictor string $\dom(v_6) = AA$, and prediction histogram $\hist(v_6)$ that contains a count for each element in $\{A, B, \&\}$. The counts in the histogram sum up to $3$, since the string $AA$ appears $3$ times in $D$, i.e., once in $s_3$ and twice in $s_4$. In addition, $\hist(v_6)[A] = 1$ because, among the $3$ occurrences of $AA$, only one is immediately followed by $A$ (i.e., the first occurrence of $AA$ in $s_4$). The root node $v_1$ has an empty predictor string $\dom(v_1) = \emptyset$, and its prediction histogram counts the occurrences of each individual symbol in $\I \cup \{\&\}$, e.g., $\hist(v_1)[A] = 6$ since the symbol $A$ appears $6$ times in total in $D$.

The nodes in a PST are organized in such a way that each node $v$ has $|\I|+1$ children. Furthermore, for each child $v'$ of $v$, $\dom(v')$ is obtained by adding a symbol in $\I \cup \{\$\}$ in the beginning of $\dom(v)$. That is, $\dom(v)$ is a suffix of $\dom(v')$. For example, in the PST in Figure~\ref{fig:markov}, $v_3$ is a parent of $v_5$; accordingly, $\dom(v_5) = \$A$ is obtained by adding the symbol $\$$ to the beginning of $\dom(v_3) = A$. The intuition here is that (i) each node $v$ in a PST provides a way to predict the ``next symbol'' in a sequence based on a ``predicate'' $\dom(v)$, and (ii) when we split $v$, each child node would have a longer ``predicate'' that provides a more specific predication.

A PST $\mathcal{T}$ can be used to support a wide range of queries, such as estimating the number of times that a query string $s_q$ appears in the sequences in $D$. Specifically, given $s_q = x_1x_2\ldots x_l$, we first inspect the root node $v_1$'s prediction histogram $\hist(v_1)$, and then initialize a temporary answer $ans = \hist(v_1)[x_1]$. After that, we examine $x_i$ ($i \in [2, l]$) in ascending order of $i$. For each $x_i$, we consider the length-$(i-1)$ prefix of $s_q$, i.e., $s^*_i = x_1x_2 \ldots x_{i-1}$. We identify the node $v$ in $\mathcal{T}$ whose predictor string is the longest suffix of $s^*_i$. Then, we compute the sum of the counts in $v$'s prediction histogram $\hist(v)$, referred to as the {\em magnitude} of the histogram and denoted as $\|\hist(v)\|_1$. After that, we set\\
\vspace{-1mm}
\begin{equation}
ans = ans \cdot \frac{\hist(v)[x_i]}{\|\hist(v)\|_1},
\end{equation}
i.e., we multiple $ans$ by the probability that the ``next symbol'' equals $x_i$, as predicted by $\hist(v)$. When all $x_i$ ($i \in [1, l]$) are examined, we return $ans$ as the query answer.

For example, consider a query sequence $s_q = AB$ on the PST in Figure~\ref{fig:markov}. We first visit the root node $v_1$, and initialize $ans = \hist(v_1)[A] = 6$. After that, we consider the length-$1$ prefix of $s_q$, i.e., $s^*_2 = A$. We identify $v_3$ as the node whose predictor string is the longest suffix of $s^*_2$, and we set $ans = ans \cdot \frac{\hist(v_3)[B]}{\|\hist(v_3)\|_1} = 3$. Finally, we return $ans = 3$ as the answer.

In addition to the aforementioned query type, we can also utilize a PST $\mathcal{T}$ to generate a {\it synthetic sequence dataset}, by sampling sequences from $\mathcal{T}$ one by one. Specifically, to generate a sequence, we start from an initial sequence $s_0 = \$$ and insert symbols into $s_0$ iteratively. In the $i$-th iteration ($i \ge 1$), we inspect the sequence $s_{i-1}$, and identify the node $v$ in $\mathcal{T}$ whose predictor string is the longest suffix of $s_{i-1}$. Then, we sample a symbol $x_i$ from the symbol distribution represented by $\hist(v)$, i.e., $\Pr[x_i = x] = \frac{\hist(v)[x]}{\|\hist(v)\|_1}$. After that, we insert $x_i$ to the end of $s_{i-1}$, and denote the resulting sequence as $s_i$. If $x_i$ happens to be $\&$, then we return $s_i$ as the result.

\subsection{Extension of \tree} \label{sec:markov-tree}

To construct a PST $\mathcal{T}$ on a sequence dataset $D$, we can start from a root node $v_1$ with a predictor string $\dom(v_1) = \emptyset$, and then recursively split $v_1$. This motivates us to adopt \tree for the generation of differentially private PSTs. However, we can no longer use a node $v$'s noisy count $c(v)$ to decide whether $v$ should be split, since $c(v)$ is undefined on a PST. Instead, as discussed in Section~\ref{sec:alg:extend}, we can redefine $c(v)$ as a score function that measures the suitability of $v$ for splitting. In the non-private setting, existing work \cite{RST96} typically avoids splitting a node $v$ if any of the following conditions is satisfied:
\begin{enumerate}[topsep = 6pt, parsep = 6pt, itemsep = 0pt, leftmargin=26pt]
\item[C1.] {\em $\dom(v)$ starts with $\$$.} In this case, no more symbol can be added to the beginning of $\dom(v)$; thus, $v$ cannot be split.

\item[C2.] {\em The magnitude of $\hist(v)$ is small.} The rationale is that, when $\|\hist(v)\|_1$ is small, further splitting $v$ results in child nodes $v'$ whose prediction histograms $\hist(v')$ have even smaller magnitudes. In that case, the symbol distribution captured by $\hist(v')$ is obtained from a tiny sample set of sequences, which leads to poor prediction accuracy.

\item[C3.] {\em The entropy\footnote{Here we treat $\hist(v)$ as a probability distribution.} of $\hist(v)$ is small.} This is because when $\hist(v)$ has a small entropy, there is little uncertainty in the symbol prediction given by $\hist(v)$; as such, there is little benefit in splitting $v$. (See $v_4$ in Figure~\ref{fig:markov} for an example.)
\end{enumerate}

Suppose that we are to adopt the above conditions into \tree. Condition C1 can be straightforwardly applied, since it only depends on $\dom(v)$ and does not rely on $D$, i.e., it does not leak private information. In contrast, conditions C2 and C3 cannot be directly adopted since the counts in $\hist(v)$ depend on $D$. To address this issue, we aim to design a score function $c(\cdot)$ for \tree with the following two properties:
\begin{enumerate}[topsep = 6pt, parsep = 6pt, itemsep = 0pt, leftmargin=26pt]
\item[P1.] $c(\cdot)$ is monotonic, i.e., $c(v) \le c(u)$ for any node $v$ and its parent $u$. This, as discussed in Section~\ref{sec:alg:extend}, is required to ensure that \tree satisfies differential privacy.

\item[P2.] If a node $v$'s prediction histogram has a small magnitude or a small entropy, then $c(v)$ tends to be small. This is motivated by conditions C2 and C3 mentioned above.
\end{enumerate}

Our construction of $c(\cdot)$ is based on the following observation: if a prediction histogram has a small entropy, it often has one symbol count that dominates the others, because a small entropy implies that the distribution of symbols in the histogram is skewed. ($v_4$ in Figure~\ref{fig:markov} shows an example.) Motivated by this, we define $c(v)$ as
\begin{equation} \label{eqn:markov-newc}
c(v) = \|\hist(v)\|_1 - \max_{x\in\I\cup\left\{\&\right\}}\hist(v)[x],
\end{equation}
i.e., $c(v)$ equals the magnitude of $\hist(v)$ minus the largest count in $\hist(v)$. The intuition is that if the magnitude of $\hist(v)$ is small, then $c(v)$ must be small, regardless of the largest count in $\hist(v)$; on the other hand, if the entropy of $\hist(v)$ is small, then the largest count in $\hist(v)$ tends to be close to the magnitude of $\hist(v)$ (since the count often dominates all other counts in $\hist(v)$), which results in a small $c(v)$ as well. Thus, $c(v)$ fulfills property P2. The following lemma show that $c(v)$ also satisfies property P1.
\begin{lemma}\label{lmm:newc:monotone}
$c(\cdot)$ is a monotonic function.
\end{lemma}

In summary, we can construct a private PST on a sequence dataset $D$ using \tree (i.e., Algorithm~\ref{alg:privtree}), with three minor changes. First, in Line 1 of Algorithm~\ref{alg:privtree}, $\mathcal{T}$ is a PST with a fanout $|\I| + 1$ (instead of a quadtree), and $v_1$ is a PST node with a predictor string $\dom(v_1) = \emptyset$ and a prediction histogram $\hist(v_1)$. Second, in Line 5, $c(v)$ is as defined in Equation~\eqref{eqn:markov-newc}. Third, in Line 11, we return $\mathcal{T}$ after removing the biased score $\hat{b}(v)$ and the prediction histogram $\hist(v)$ of each node $v$.

After we obtained the PST $\mathcal{T}$ (without prediction histograms) from the modified \tree, we can postprocess $\mathcal{T}$ to recover the prediction histograms. Specifically, for each leaf node $v$ in $\mathcal{T}$, we derive the prediction histogram $\hist(v)$ from $D$, and then compute a noisy version of $\hist(v)$, denoted as $\widehat{\hist}(v)$, by adding Laplace noise into each histogram count. After that, for any non-leaf node $v'$, we construct a noisy prediction $\widehat{\hist}(v')$, such that for any symbol $x \in \I \cup \{\&\}$,
\begin{equation*}
\widehat{\hist}(v')[x] = \sum_{\text{\small $v$ is a leaf node under $v'$}} \widehat{\hist}(v)[x].
\end{equation*}
Finally, if any noisy histogram in $\mathcal{T}$ has a negative count, we reset the count to zero. This is to ensure that each histogram represents a distribution of symbols.

\header
{\bf Privacy Analysis and Parameterization.} To analyze the privacy guarantee of the modified \tree, we first introduce an assumption that is also adopted in prior work \cite{CAC12} on sequence data publication under differential privacy: we assume that the length of each sequence in $D$, when taking into account $\&$ but not $\$$, is at most $l^\top$, where $l^\top$ is a known constant. To explain why this assumption is needed, consider that we insert an {\em infinite} sequence $s$ into $D$ to obtain a neighboring dataset $D'$. In that case, the insertion of $s$ incurs unbounded changes in the histogram counts of the PST, which makes it impossible to achieve differential privacy. In general, if $l^\top$ is unknown, we may choose an appropriate $l^\top$ and {\em truncate} any sequences $s$ that is excessively long\footnotemark. Specifically, if $s = \$x_1x_2\ldots x_{l^\top} \&$, then we truncate it to $s = \$x_1x_2\ldots x_{l^\top}$, i.e., $s$ becomes an open-ended sequence. Note that the removal of $\&$ from $s$ does not affect the construction of the PST.

\footnotetext{Such $l^\top$ can be chosen by first identifying the $90\%$ or $95\%$ quantile of the sequence lengths in $D$, and then computing a differentially private version of the quantile \cite{ZNC12}.}

Given the above assumption,
we prove the privacy guarantee of the modified \tree as follows.
\begin{theorem} \label{thrm:markov:privtree}
Let $\beta = |\I| + 1$. The modified \tree ensures $\e$-differential privacy when ${\lambda \ge \frac{2\beta - 1}{\beta - 1} \cdot \frac{l^\top}{\e}}$ and $\delta = \lambda \cdot \ln \beta$.
\end{theorem}
In addition, we prove that the postprocessing of \tree's output (i.e., adding Laplace noise to the histogram counts of the leaf nodes) also achieves $\e$-differential privacy.
\begin{theorem} \label{thrm:markov:post}
Postprocessing \tree's output with Laplace noise of scale $\lambda$ achieves $\e$-differential privacy, when $\lambda \ge \frac{l^\top}{\e}$.
\end{theorem}


Finally, we clarify how we set $\theta$ and divide the privacy budget $\e$ between \tree and its postprocessing step. First, we set $\theta = 0$, following our analysis in Section~\ref{sec:alg:parameter}. Second, we set the noise scale in \tree and its postprocessing step, such that \tree achieves $\frac{\e}{\beta}$-differential privacy and the postprocessing procedure ensures $\frac{\e \cdot(\beta - 1)}{\beta}$-differential privacy. To explain, recall that in \tree, we inject Laplace noise into each node $v$'s score $c(v)$, which equals the sum of $\beta - 1$ counts in $v$'s prediction histogram $\hist(v)$ (i.e., all counts except the largest one). Meanwhile, in the postprocessing step, we add Laplace noise to each count $y$ in the prediction histograms of $\mathcal{T}$'s leaf nodes. Intuitively, $c(v)$ is roughly $\beta - 1$ times more resilient to noise than $y$. Therefore, we set the privacy budget for the postprocessing step to be $\beta - 1$ times the budget for \tree, so as to balance the relative accuracy of $c(v)$ and $y$ after noise injection.




\subsection{Comparison with Previous Work} \label{sec:markov-compare}

There exist two differentially private methods \cite{CFDS12,CAC12} for modeling sequence data, and they both utilize hierarchical decompositions for model construction. However, they considerably differ from \tree in three aspects. First, they model sequences based on their prefixes \cite{CFDS12} or $n$-grams \cite{CAC12}, while \tree is based on a PST representation of the {\em variable length Markov chain model} \cite{RST96}. Second, their algorithms for hierarchical decompositions are similar in spirit to Algorithm~\ref{alg:simple}, due to which they also require a pre-defined threshold $h$ on the maximum height of the decomposition tree. Consequently, they suffer from similar deficiencies to those of Algorithm~\ref{alg:simple}, i.e., they cannot generate accurate models because of the dependency on $h$. Third, when constructing a decomposition tree, the methods in \cite{CFDS12,CAC12} decide whether a node be split based only on a count associated with the node, whereas \tree adopts a more advanced strategy that takes into account three conditions commonly considered in the non-private setting. The above differences make \tree an effective approach for modeling sequence data, as we demonstrate in our experiments in Section~\ref{sec:exp}.



\section{Connections to SVT}\label{sec:svt}

In this section, we investigate the connection between \tree and the {\em sparse vector techniques (SVTs)} \cite{H11, DR13, LC14}, which are a type of differentially private algorithms widely adopted in the literature \cite{LC14,CXZX15,LXJL15}. They take as input a sequence of queries and a threshold $\theta$, and output either a set of queries whose results are likely to be larger than $\theta$ \cite{DR13,LC14}, or a noisy version of the answers for such a query set \cite{H11}. Intuitively, SVTs are similar in spirit to \tree since they both aim to identify some elements in a set (e.g., a node set or a query set) with ``scores'' above a given threshold. Motivated by this, in the following, we examine whether SVTs can be adopted for the hierarchical decomposition problem. Among the three existing variants of SVTs \cite{H11, DR13, LC14} that satisfy $\e$-differential privacy\footnotemark, we will focus on a variant dubbed {\em the binary SVT}, as it is most relevant to our problem. Interested readers are referred to Appendix~\ref{appen:svt} for discussions on the other two variants.

\footnotetext{There exist other variants of SVT that satisfy a relaxed version of $\e$-differential privacy \cite{HR10}. We do not consider those variants.}

\begin{algorithm}[t]
\caption{\label{alg:svt-binary} {\bf BinarySVT} ($D$, $Q = \left\{q_1, q_2, \ldots\right\}$, $\theta$, $\lambda$)}
\BlankLine
    compute a noisy version of $\theta$: $\;$ $\hat{\theta} = \theta + \Lap\left(\lambda\right)$\;
    \For{$i  = 1, 2, \ldots$}
    {
        compute a noisy version of $q_i(D)$: $\;$ $\hat{q_i}(D) = q_i(D) + \Lap\left(\lambda\right)$\;
        \If{$\hat{q_i}(D) > \hat{\theta}$}
        {
            {\bf output} $o_i = 1$ and {\bf continue}\;
        }
        \Else
        {
            {\bf output} $o_i = 0$ and {\bf continue}\;
        }
    }
    \Return\;
\end{algorithm}

Algorithm~\ref{alg:svt-binary} presents a generic version of the binary SVT. Its input includes (i) a dataset $D$, (ii) a sequence $Q = \{q_1,q_2, \ldots\}$ of counting queries such that each $q_i$ has sensitivity $1$, (iii) a threshold $\theta$, and (iv) a noise scale $\lambda$. Its output is a sequence of binary variables $\{o_1,o_2,\ldots \}$, such that $o_i=1$ indicates that the result of $q_i$ is larger than $\theta$, and $o_i=0$ indicates otherwise. The algorithm is fairly simple. It first computes a noisy threshold $\hat{\theta} = \theta + \Lap(\lambda)$, and then, for each query $q_i$ in the sequence, it generates a noisy query answer $\hat{q_i}(D)$ using Laplace noise of scale $\lambda$ (Lines 1-3). If $\hat{q_i}(D) > \theta$, then algorithm outputs $o_i = 1$; otherwise, the algorithm outputs $o_i = 0$ (Lines 4-7). Previous work \cite{LC14} makes the following claim about the privacy assurance of the algorithm:
\begin{claim} \label{clm:svt-binary}
Algorithm~\ref{alg:svt-binary} ensures $\e$-differential privacy if $\lambda \ge \frac{2}{\e}$.
\end{claim}
In other words, the noise scale required by the algorithm is $\Theta(\frac{1}{\e})$ and independent of the number of queries.

If Claim~\ref{clm:svt-binary} holds, Algorithm~\ref{alg:svt-binary} could yield highly competitive solutions for the problems that we consider. For example, consider the spatial decomposition problem studied in Section~\ref{sec:alg}. Given a threshold $\theta$ and a set $D$ of spatial points in a multi-dimensional space $\Omega$, we first initialize (i) a quadtree $\mathcal{T}$ containing only a root node $v_1$ with $\dom(v_1) = \Omega$, and (ii) a query sequence $Q = \{ c(v_1) \}$, i.e., $Q$ contains only one query that asks the number of points in $\dom(v_1)$ (we will dynamically append queries to $Q$ during the construction of the quadtree). After the initialization, we invoke the binary SVT to inspect each query in $Q$ one by one; if the binary SVT outputs $1$ for a query $c(v)$, then we split the node $v$ in $\mathcal{T}$, and append a query $c(v')$ to the end of $Q$ for each child node $v'$ of $v$. When all queries in $Q$ are inspected, we return the quadtree $\mathcal{T}$ obtained. By Claim~\ref{clm:svt-binary}, $\mathcal{T}$ ensures $\e$-differential privacy, as long as the binary SVT uses Laplace noise of scale $\lambda \ge \frac{2}{\e}$ when generating the noisy versions of $c(v_i)$. In contrast, \tree requires injecting Laplace noise of scale $\lambda \ge \frac{2\beta - 1}{\beta - 1}\cdot \frac{1}{\e} > \frac{2}{\e}$, which indicates that the solution based on the binary SVT is more favorable.

Unfortunately, we show that Claim~\ref{clm:svt-binary} does not hold: in the worst case, Algorithm~\ref{alg:svt-binary} requires $\lambda = \Omega(\frac{k}{\e})$ to achieve $\e$-differential privacy, where $k$ denotes the number of queries.
\begin{lemma} \label{lmm:svt-binary}
There exists a sequence $Q$ of $k$ count queries for which Algorithm~\ref{alg:svt-binary} violates $\e$-differential privacy if $\lambda \le \frac{k}{4\e}$.
\end{lemma}
Lemma~\ref{lmm:svt-binary} invalidates all solutions based on the binary SVT, including those in previous work \cite{LC14,CXZX15,LXJL15}. In concurrent work \cite{CM15}, Chen and Machanavajjhala present a similar analysis on the binary SVT, and also come to the conclusion that it is not differentially private. In Appendix~\ref{appen:svt}, we discuss the other two variants of SVT \cite{H11, DR13}, and show that one of them \cite{H11} also violates differential privacy, while the other \cite{DR13} does not yield a competitive solution for our problem (even after we improve it with an optimization that yields better data utility).

\section{Experiments}\label{sec:exp}

This section evaluates \tree against the states of the art on differentially private modelling of spatial and sequence data.

\subsection{Experiments on Spatial Data} \label{sec:exp-spatial}

\begin{table*}[t]
\centering
\caption{Characteristics of spatial datasets.}
\vspace{1mm}
\begin{small}
 \begin{tabular}{|l|r|r|l|} 
 \hline
 {\bf Name} & {\bf Dimensionality $\boldsymbol{d}$} & {\bf Cardinality $\boldsymbol{n}$}  & {\bf Description}\\
 \hline \hline
 {\sf road} & $2$ & $1,634,165$ & Coordinates of road intersections in the states of Washington and New Mexico \\
 \hline
 {\sf Gowalla} & $2$ & $107,091$ & Check-in locations shared by users of a location-based social networking website  \\
 \hline
  {\sf NYC} & $4$ & $98,013$ & Pickup and drop-off locations of NYC taxis \\
 \hline
 {\sf Beijing} & $4$ & $30,000$ & Pickup and drop-off locations of Beijing taxis\\
 \hline
\end{tabular}
\end{small}
\label{tbl:datasets:spatial}
\vspace{0mm}
\end{table*}

\begin{figure*}[t]
\centering
\begin{small}
\begin{tabular}{cccc}
\hspace{3mm}\frame{\includegraphics[height=21.2mm]{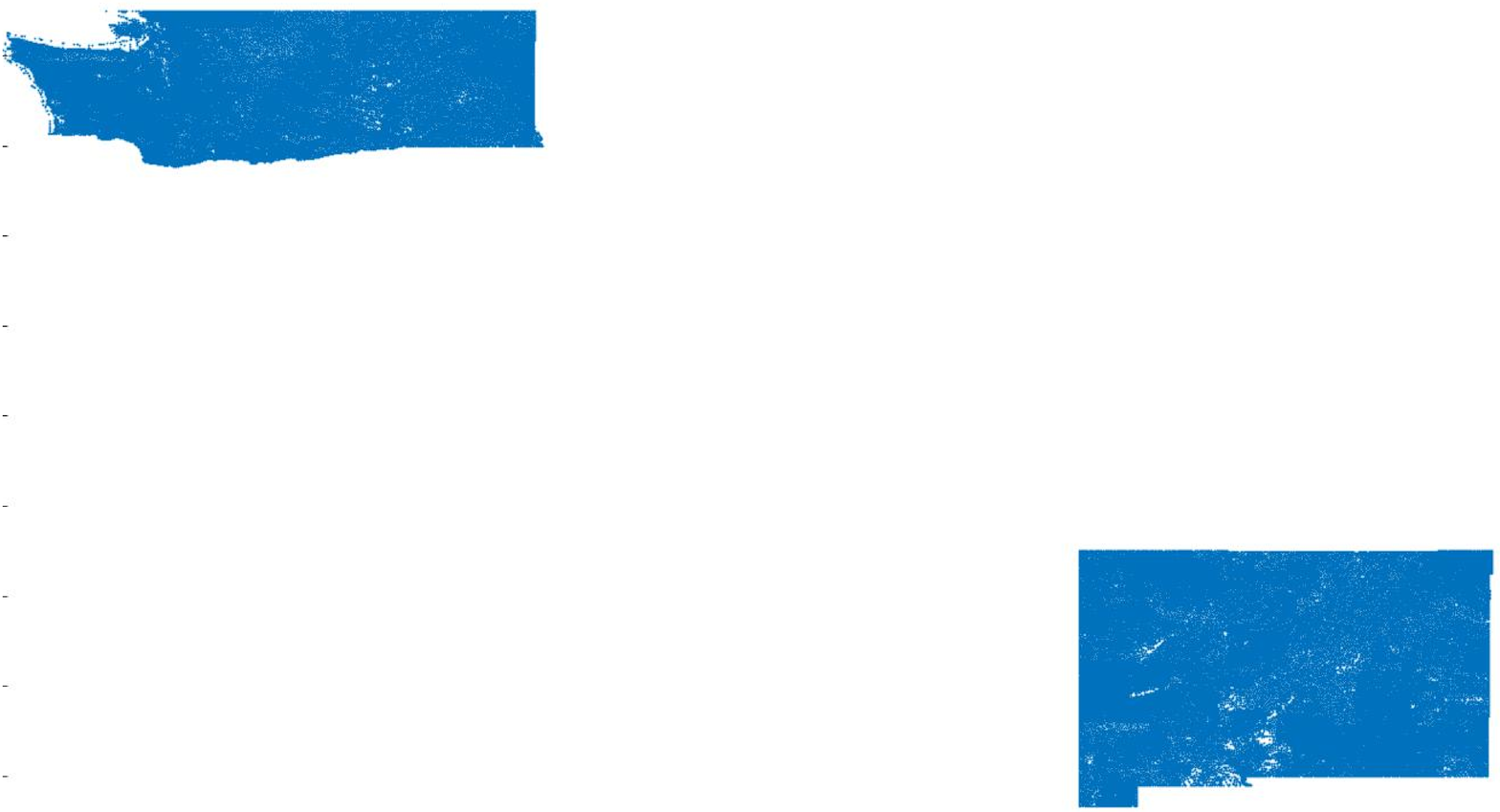}}
&
\hspace{-0mm}\frame{\includegraphics[height=21.2mm]{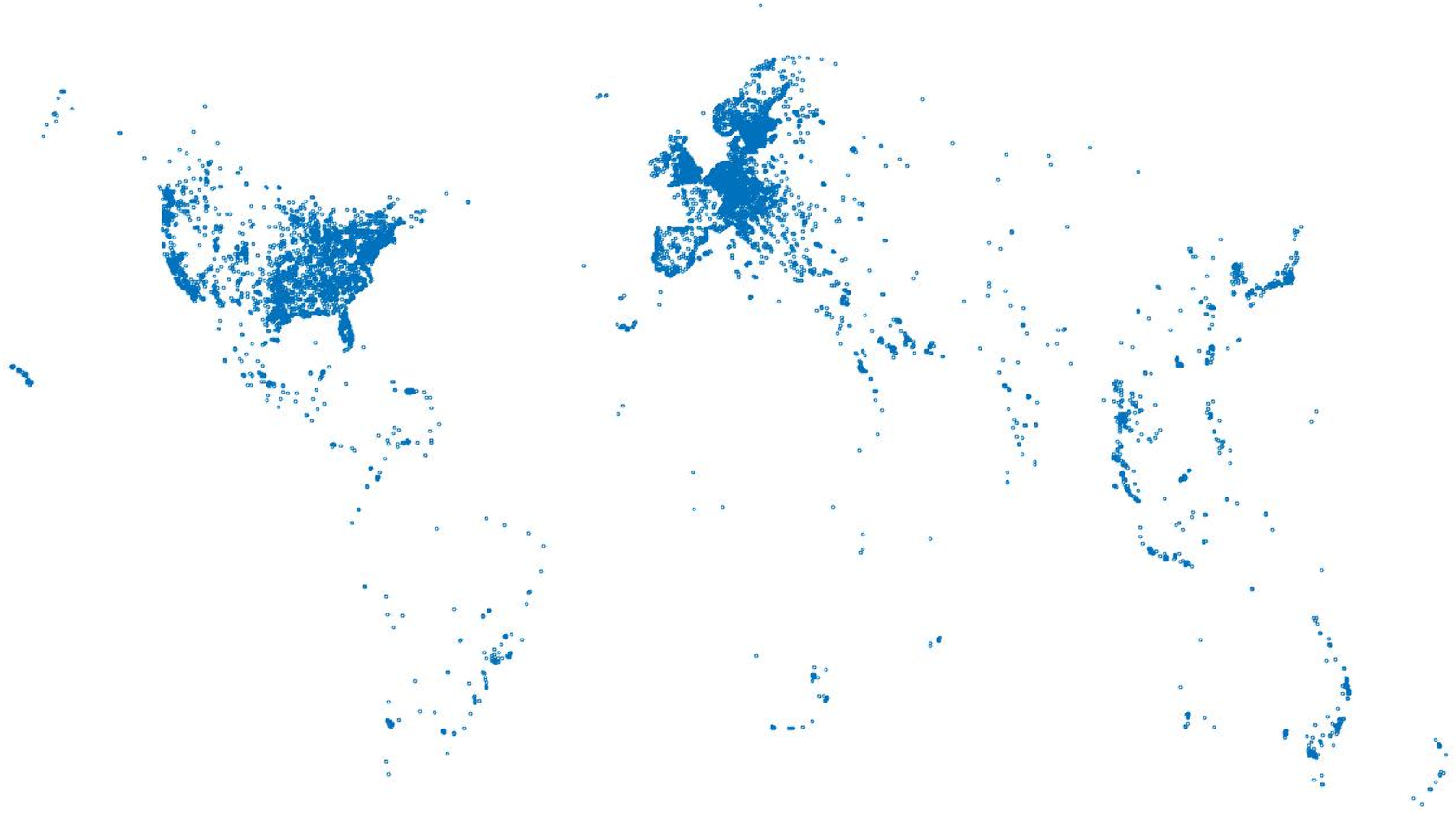}}
&
\hspace{-0mm}\frame{\includegraphics[height=21.2mm]{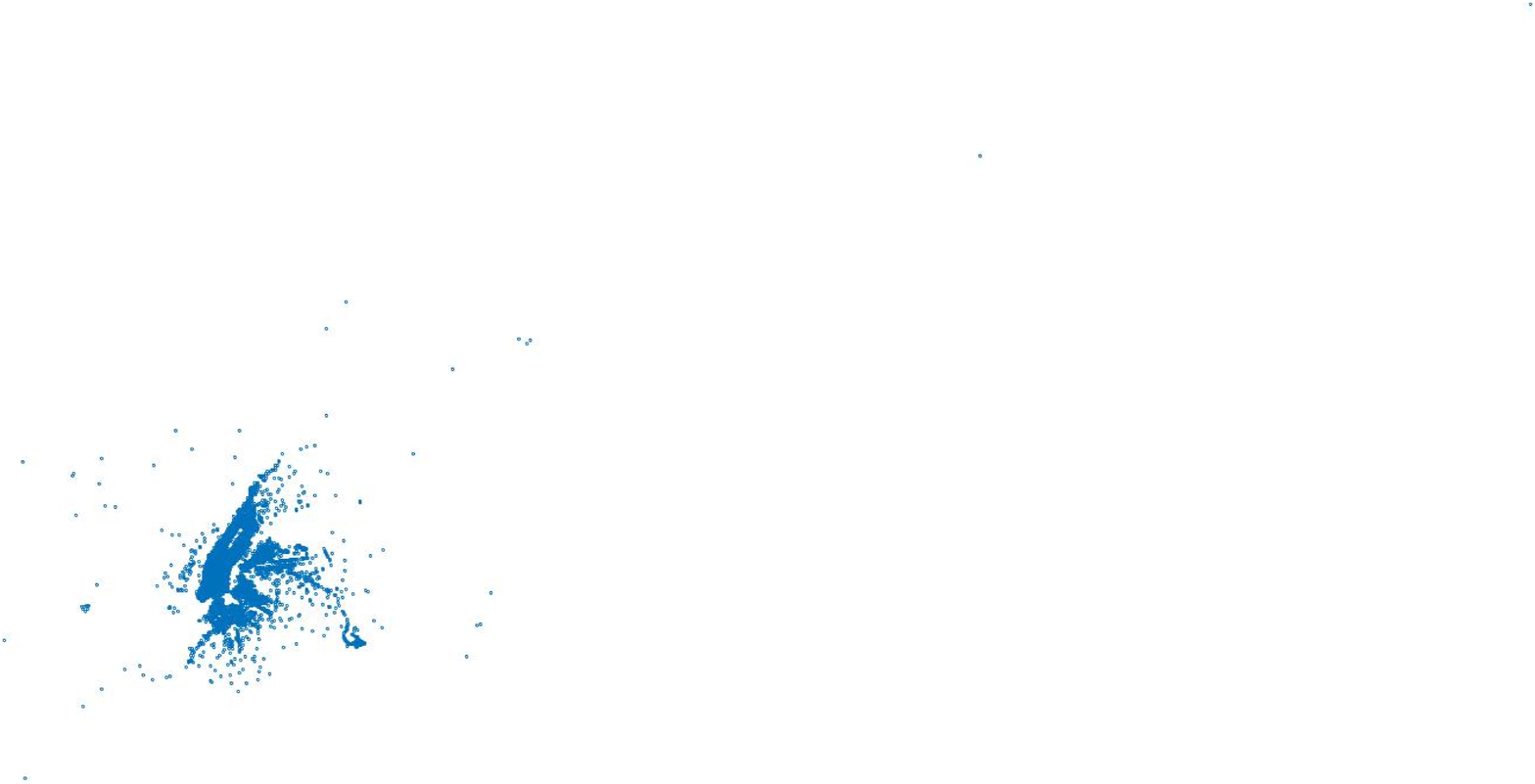}}
&
\hspace{-0mm}\frame{\includegraphics[height=21.2mm]{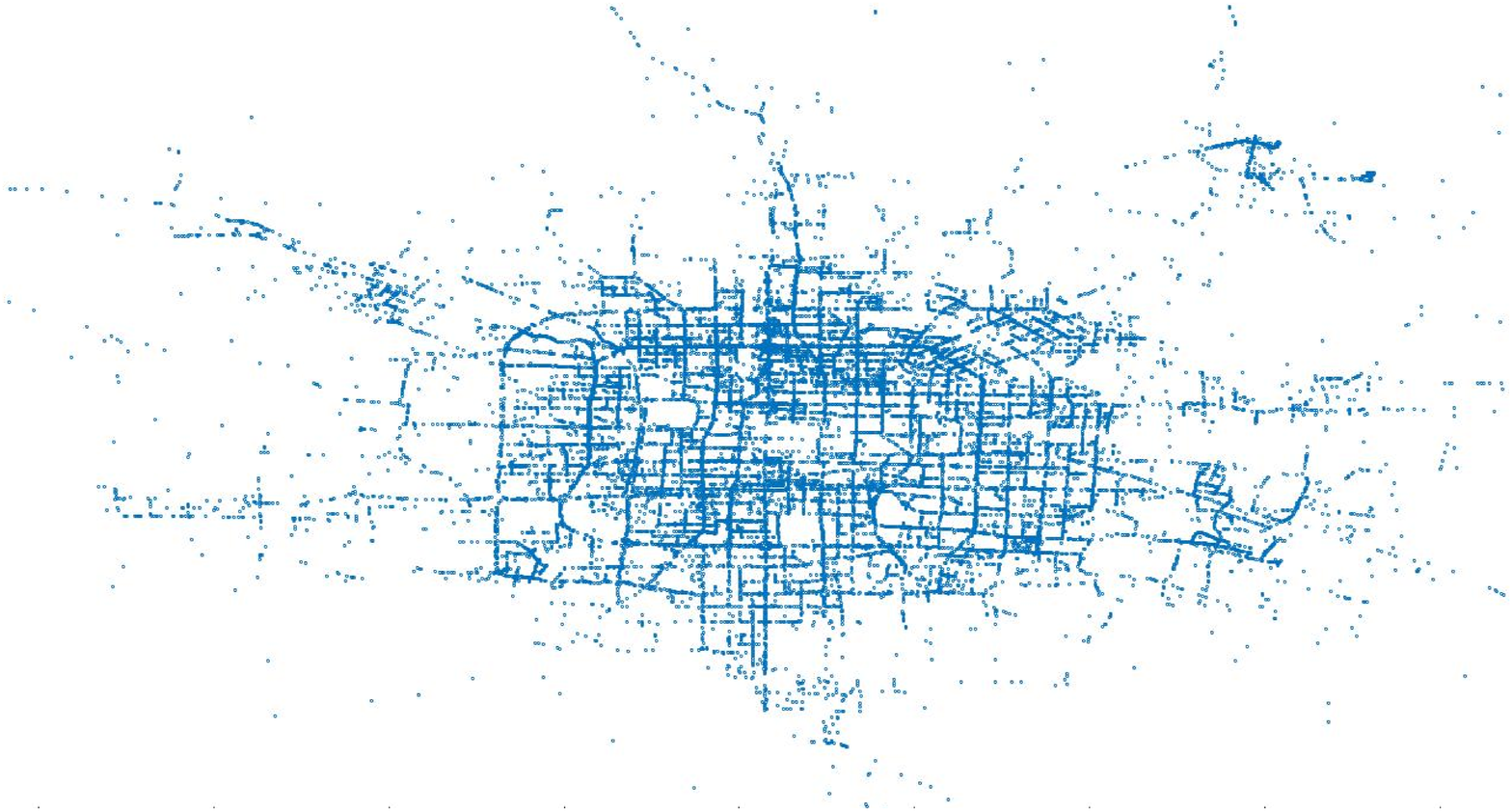}}\\
\hspace{-0mm}(a) {\sf road} & \hspace{-0mm}(b) {\sf Gowalla} & \hspace{-0mm}(c) {\sf NYC - pickup} & \hspace{-0mm}(d) {\sf Beijing - pickup}
\end{tabular}
\end{small}
\vspace{-3mm}
\caption{Visualization of datasets}
\label{fig:visual}
\vspace{-1mm}
\end{figure*}

%
%
%
%

\begin{figure*}[!t]
\centering
\edit{
\begin{small}
\begin{tabular}{c@{\hspace{-0.5mm}}:c@{\hspace{-0.5mm}}:c@{\hspace{-0.5mm}}:c}
\multicolumn{4}{c}{ \vspace{-0mm}\hspace{-5mm} \includegraphics[height=3.8mm]{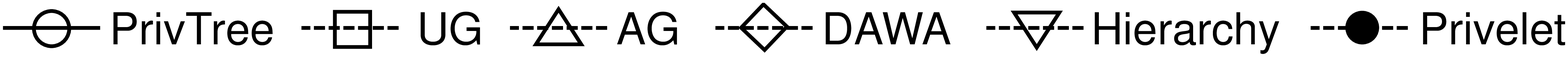}} \\

\hspace{-3mm}\includegraphics[height=31.5mm]{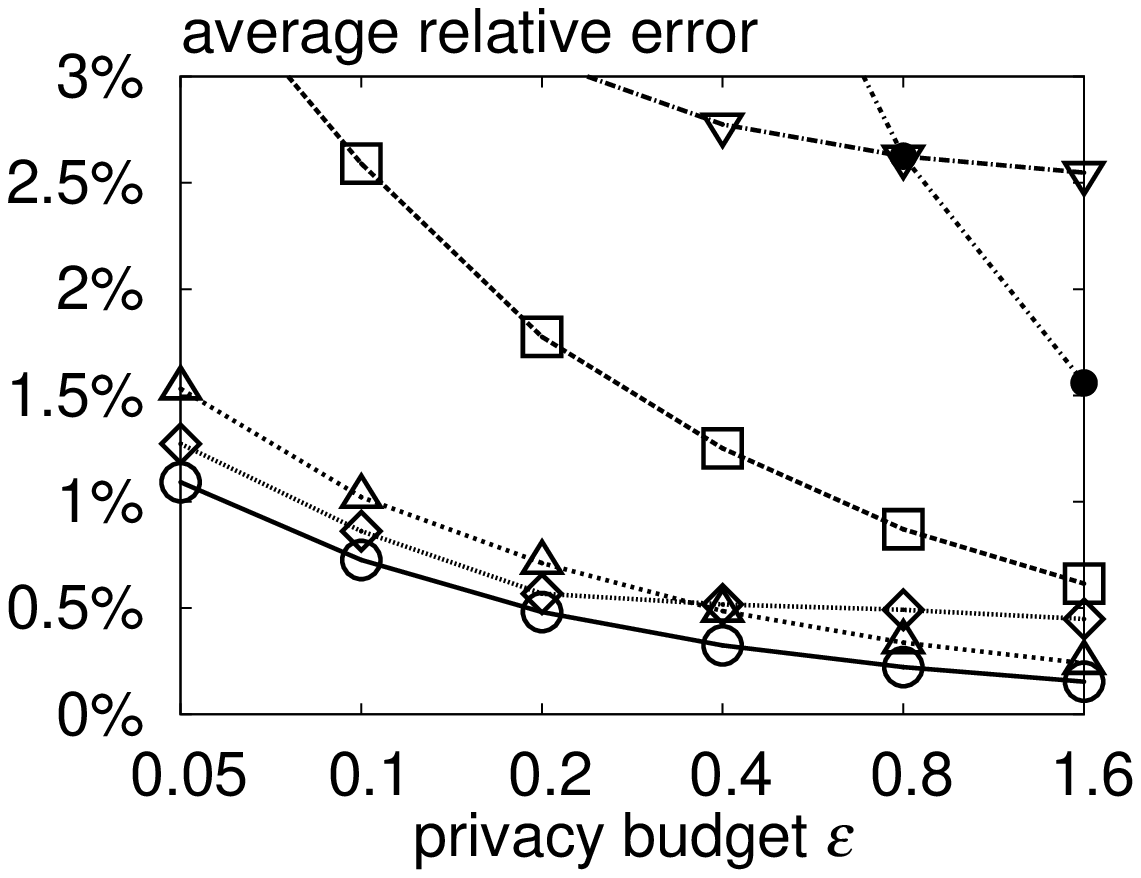}
&
\hspace{-1mm}\includegraphics[height=31.5mm]{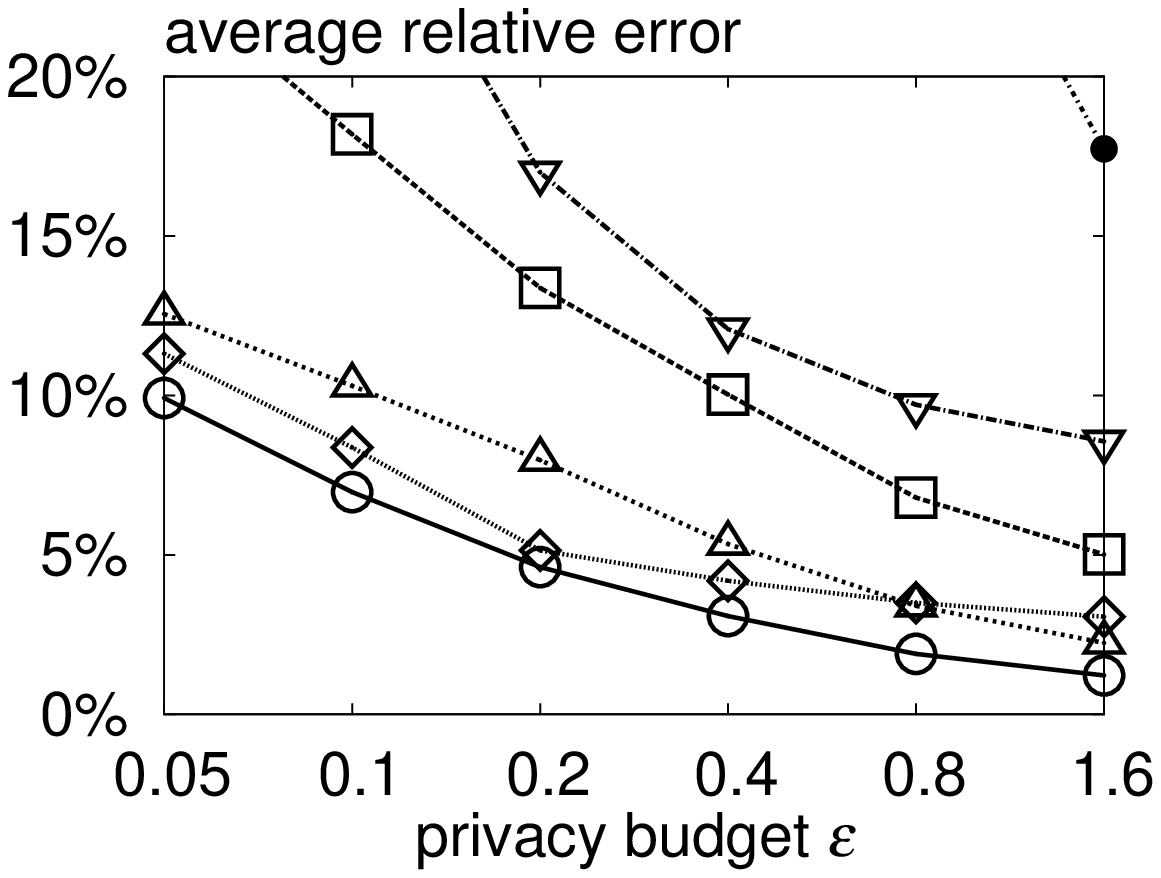}
&
\hspace{-1mm}\includegraphics[height=31.5mm]{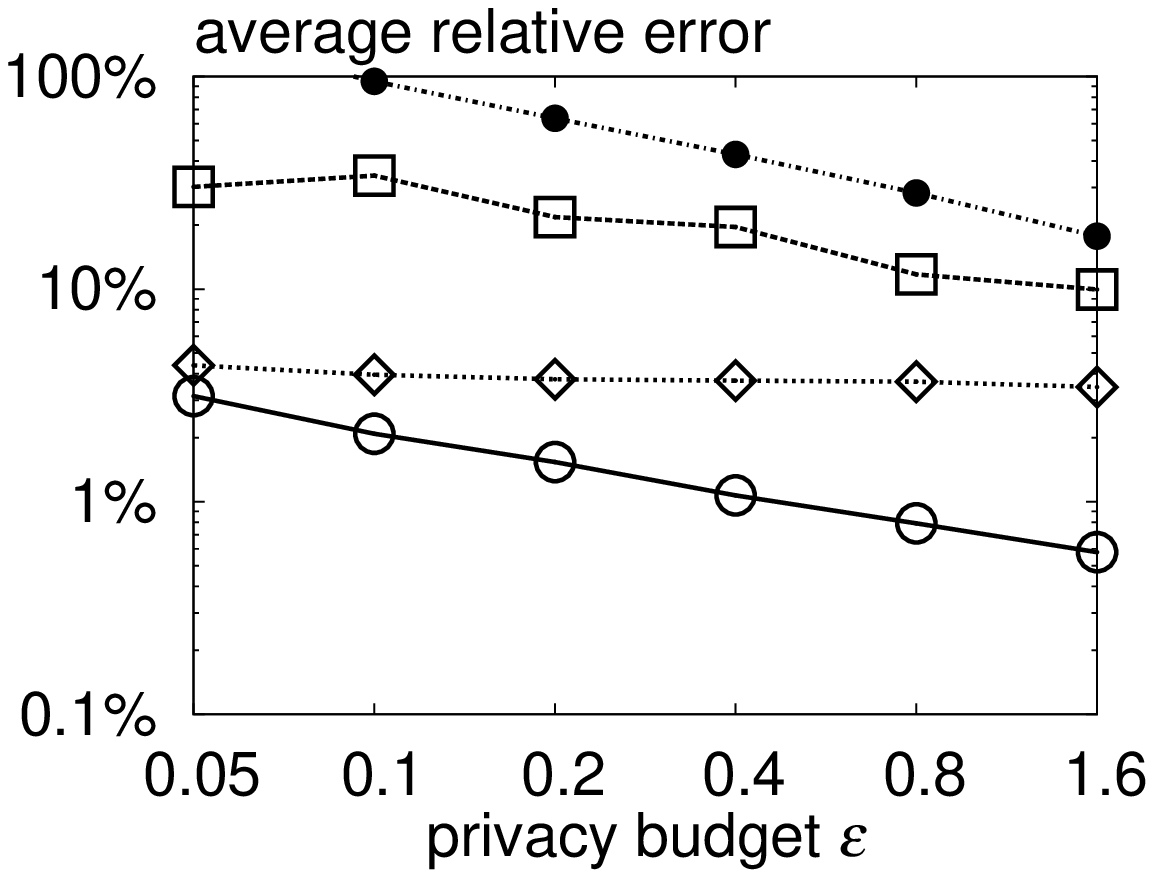}
&
\hspace{-1mm}\includegraphics[height=31.5mm]{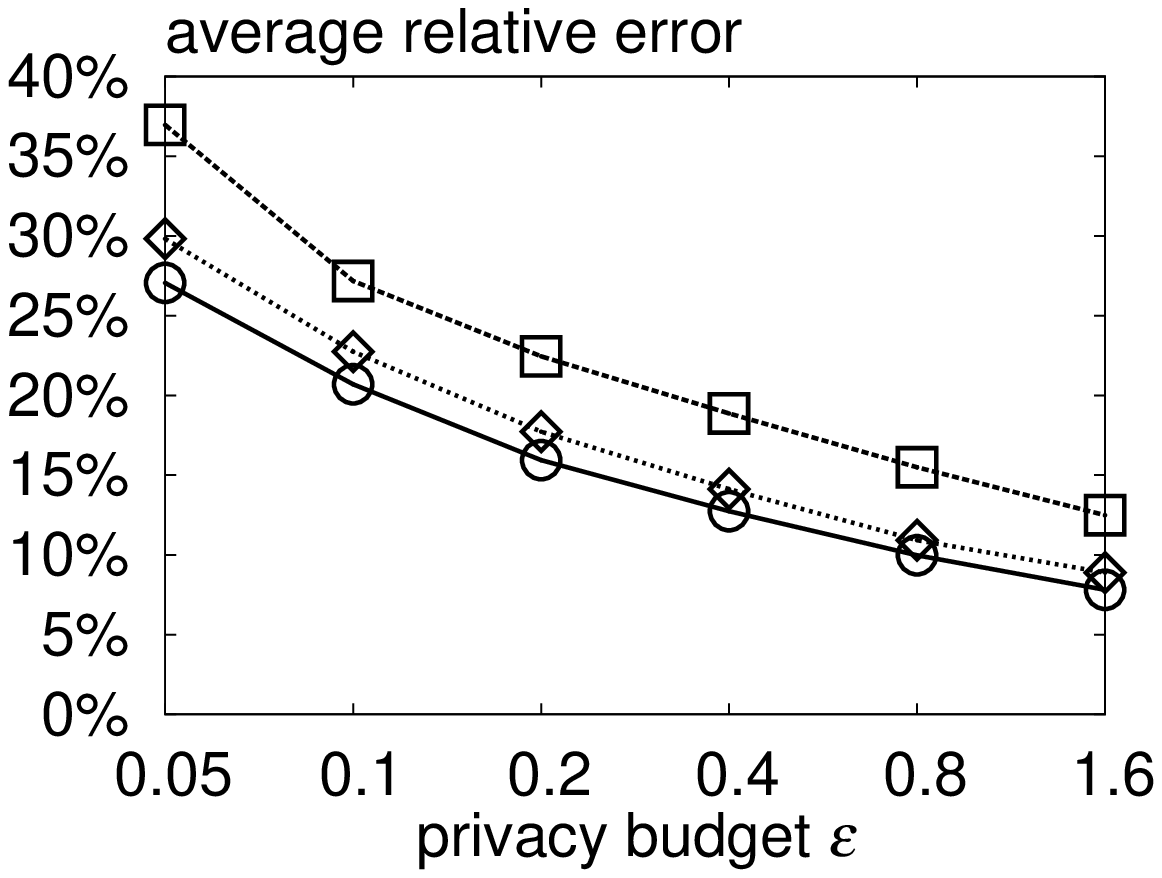} \\[0.8mm]

\hspace{-3mm} (a) {\sf road} - small queries.
&
\hspace{-2mm} (d) {\sf Gowalla} - small queries.
&
\hspace{-2mm} (g) {\sf NYC} - small queries.
&
\hspace{-2mm} (j) {\sf Beijing} - small queries. \\[2mm]

\hspace{-3mm}\includegraphics[height=31.5mm]{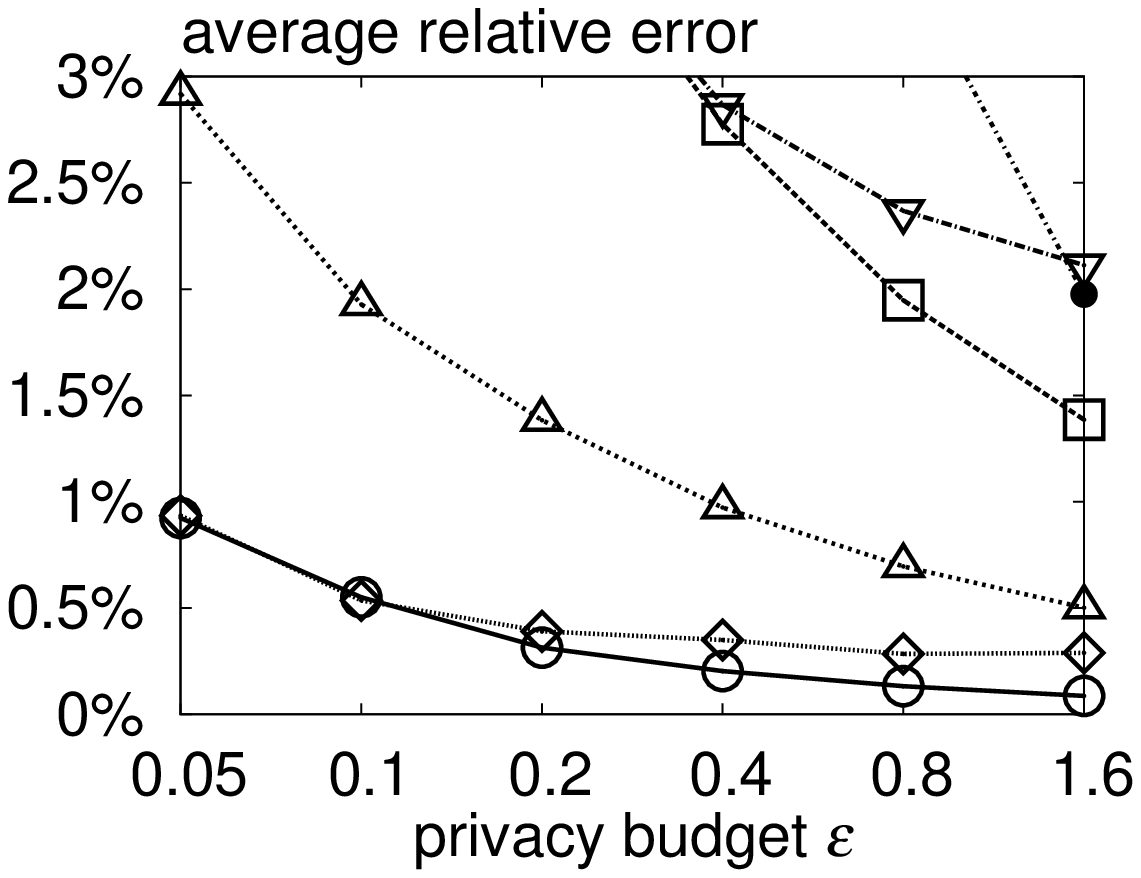}
&
\hspace{-2mm}\includegraphics[height=31.5mm]{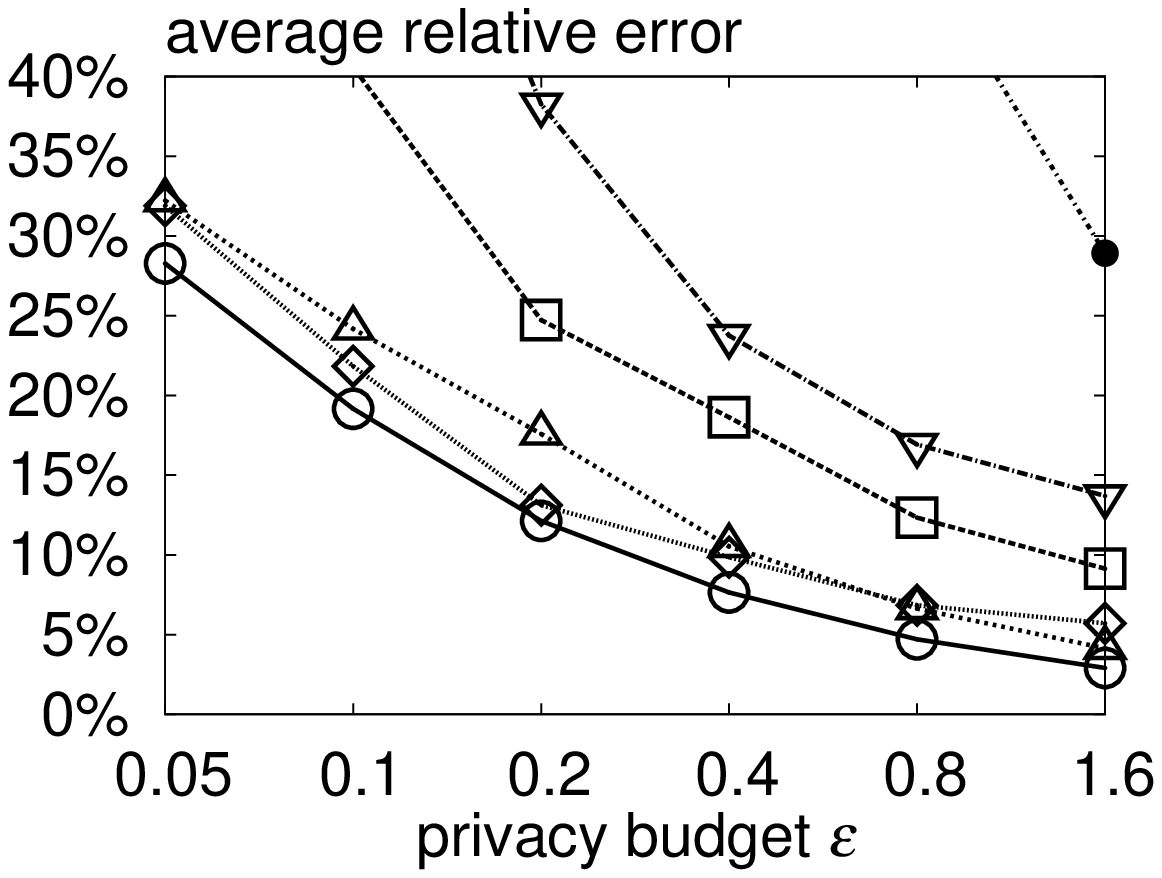}
&
\hspace{-2mm}\includegraphics[height=31.5mm]{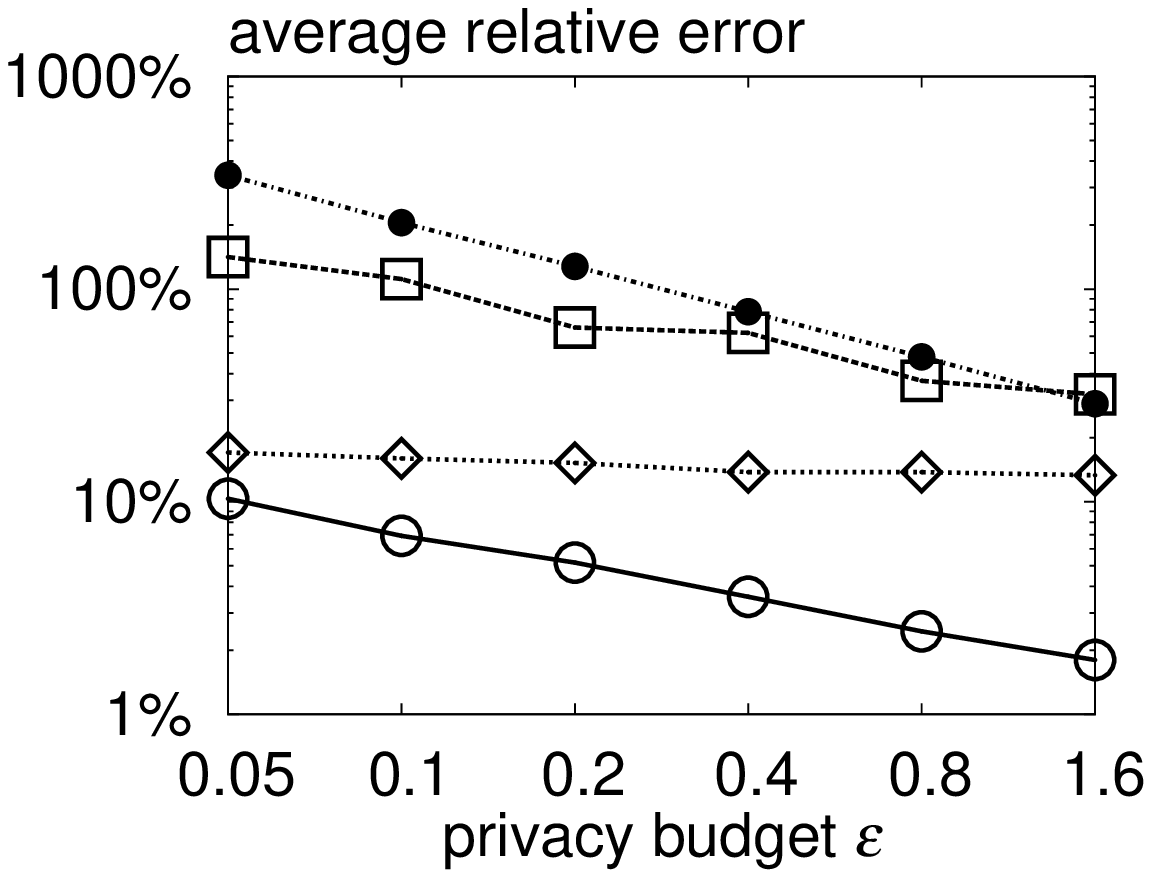}
&
\hspace{-2mm}\includegraphics[height=31.5mm]{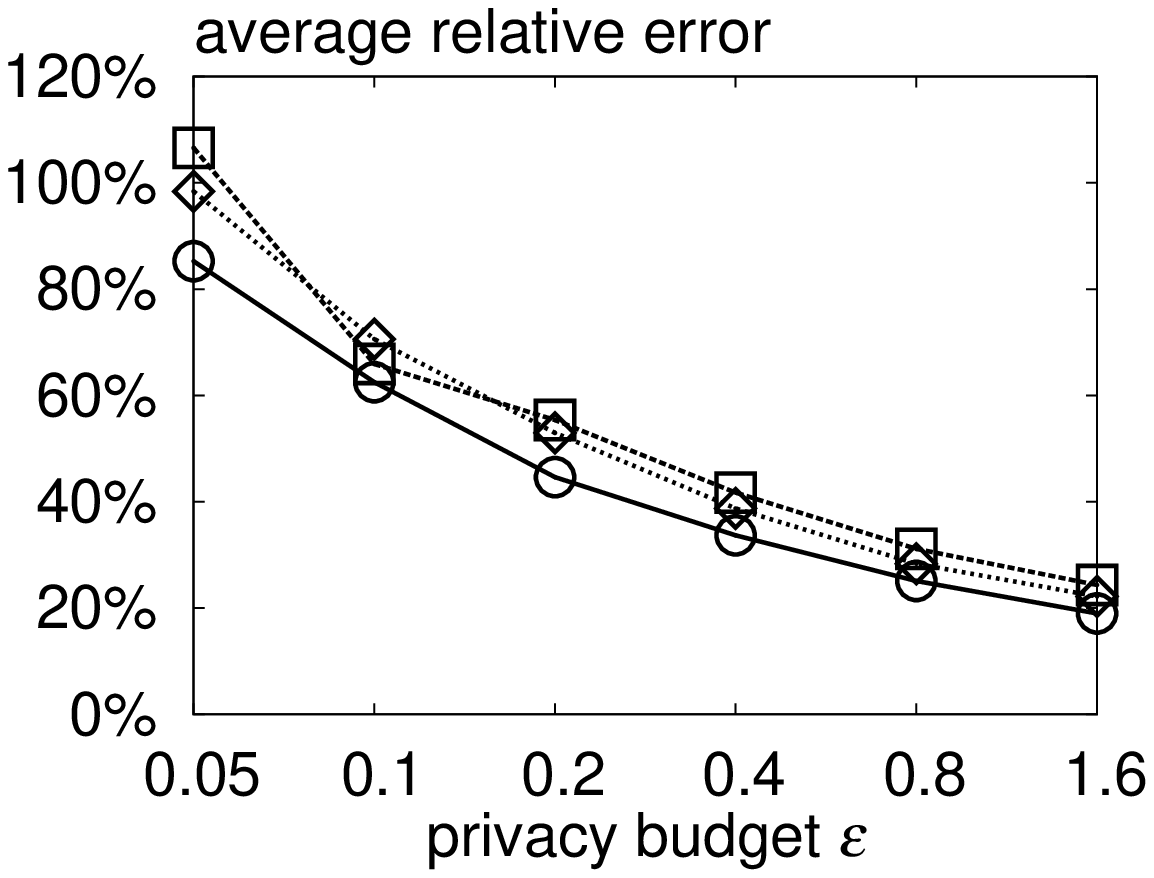}\\[0.8mm]

\hspace{-3mm} (b) {\sf road} - medium queries.
&
\hspace{-2mm} (e) {\sf Gowalla} - medium queries.
&
\hspace{-2mm}(h) {\sf NYC} - medium queries.
&
\hspace{-2mm}(k) {\sf Beijing} - medium queries. \\[2mm]

\hspace{-3mm}\includegraphics[height=31.5mm]{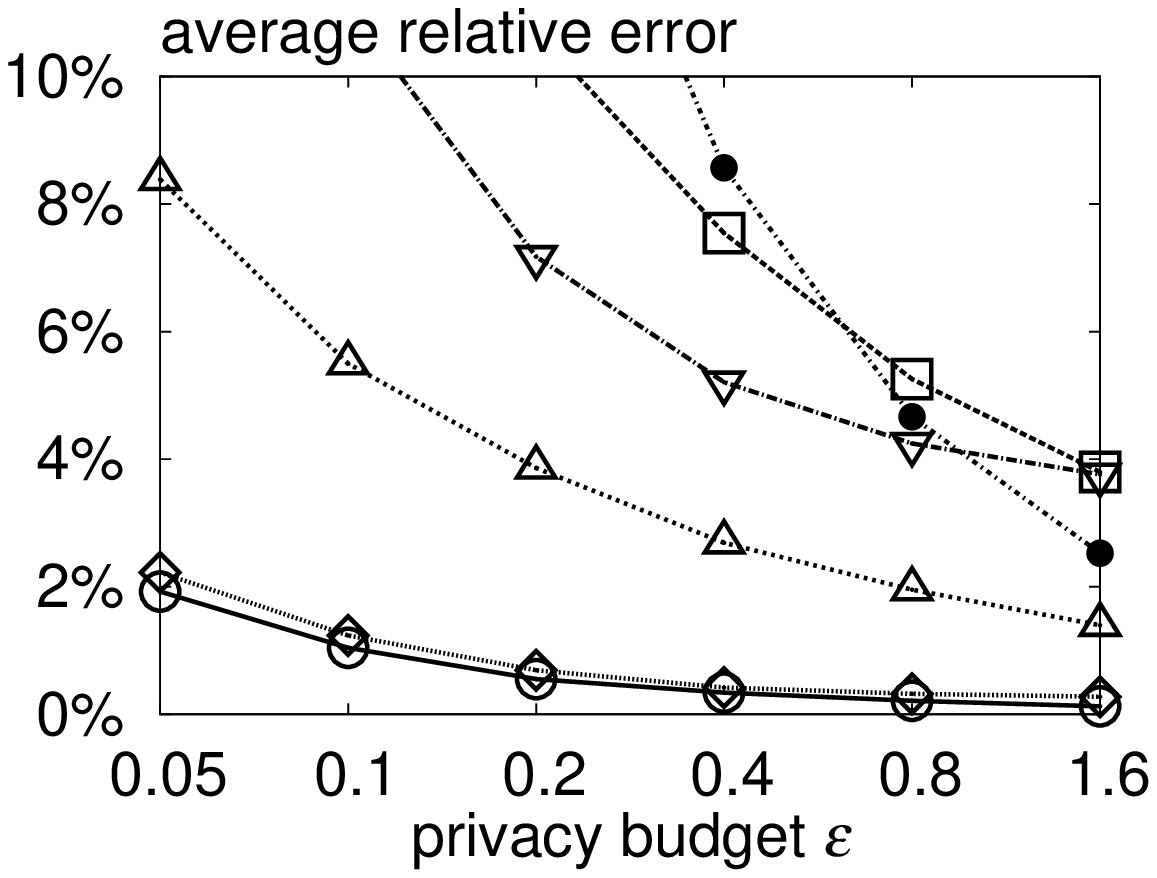}
&
\hspace{-2mm}\includegraphics[height=31.5mm]{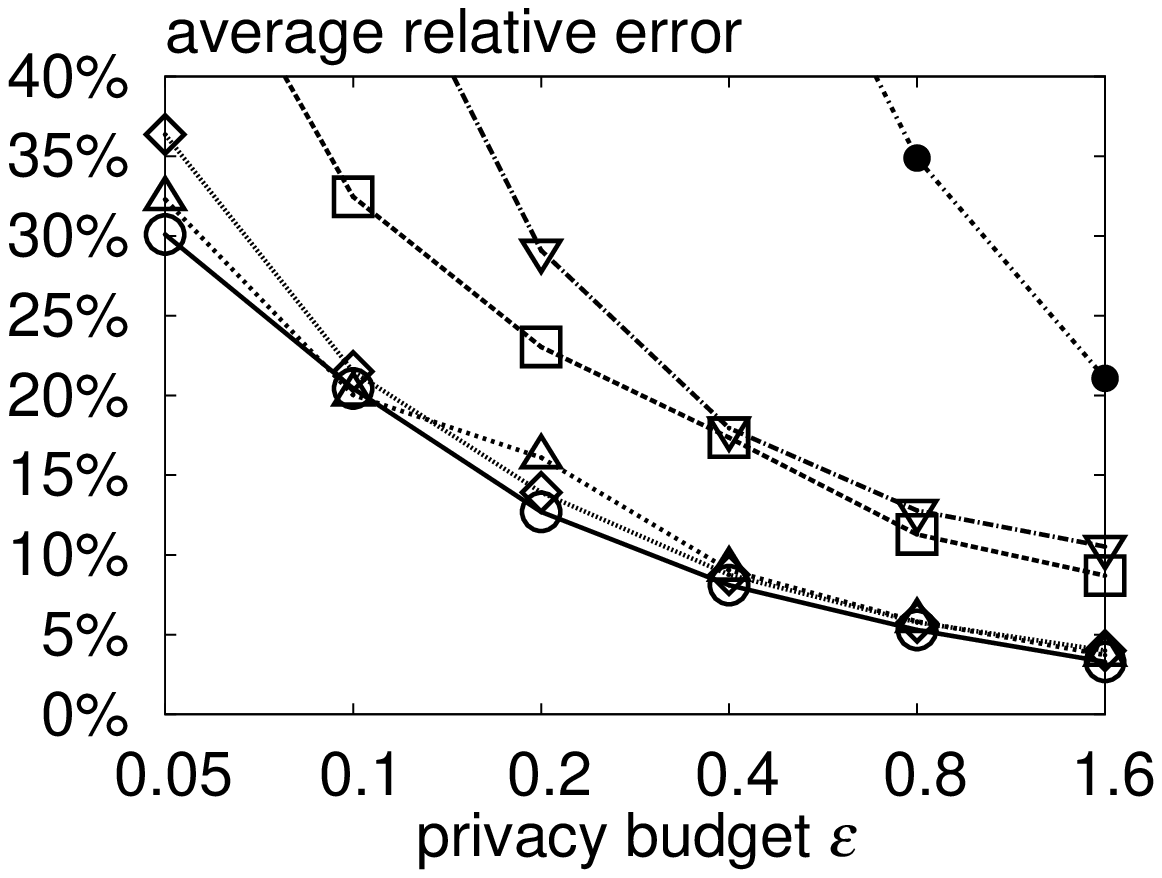}
&
\hspace{-2mm}\includegraphics[height=31.5mm]{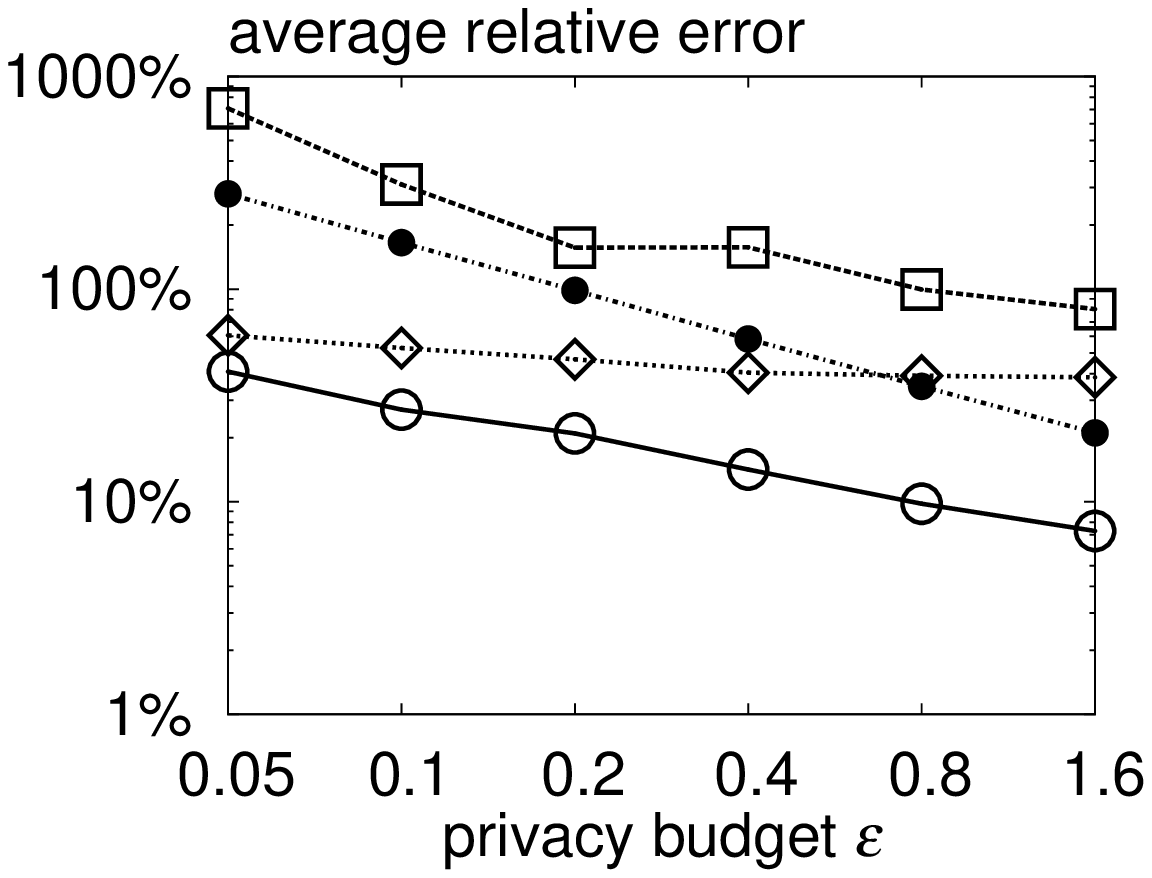}
&
\hspace{-2mm}\includegraphics[height=31.5mm]{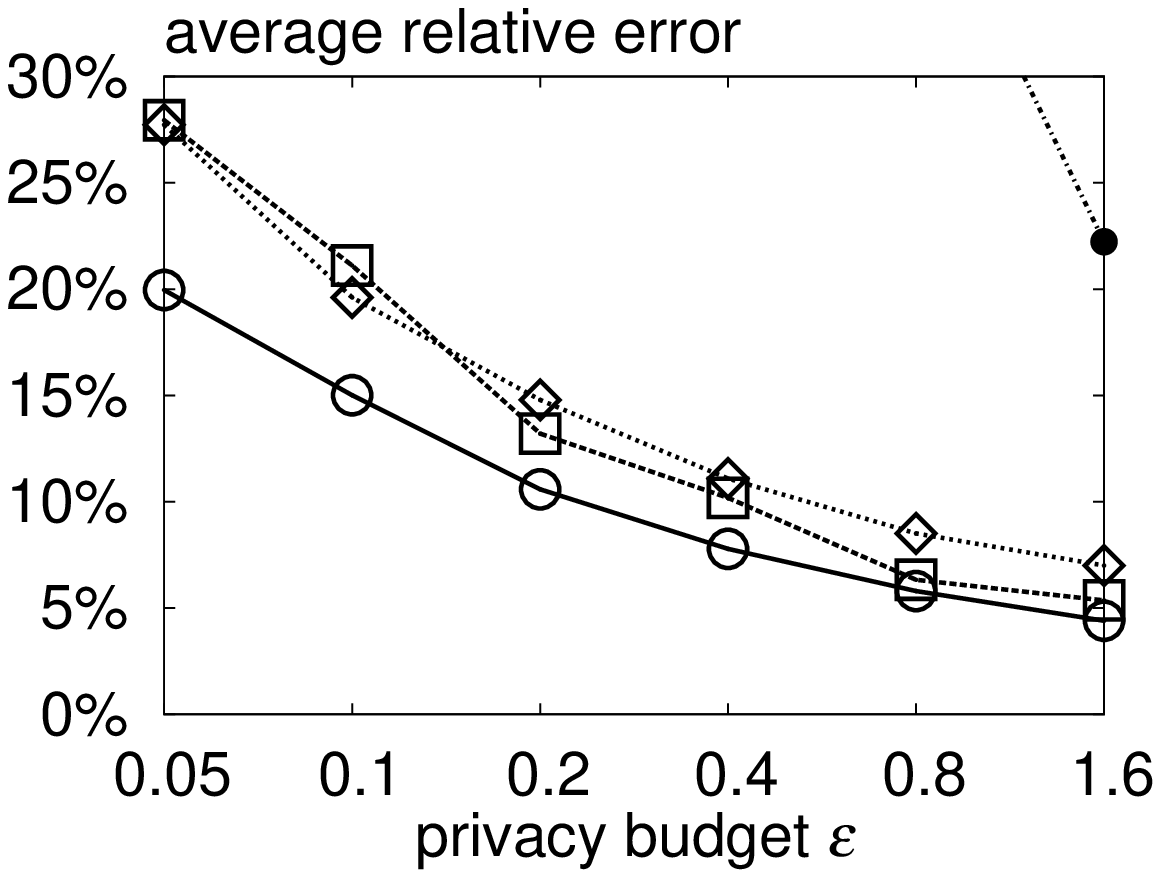} \\[0.8mm]

\hspace{-3mm} (c) {\sf road} - large queries.
&
\hspace{-2mm} (f) {\sf Gowalla} - large queries.
&
\hspace{-2mm}(i) {\sf NYC} - large queries.
&
\hspace{-2mm}(l) {\sf Beijing} - large queries.
\end{tabular}
\end{small}
\vspace{-2mm}
\caption{Results of range count queries on spatial datasets.}
\label{fig:exp:spatial}
}
\vspace{-4mm}
\end{figure*}

\vspace{-1mm}
\para{Datasets.} We make use of four real spatial datasets shown in Table~\ref{tbl:datasets:spatial}: {\sf road} \cite{CPSSY12, QYL13}, where each point represents the latitude and longitude of a road junction in the states of Washington and New Mexico; {\sf Gowalla} \cite{QYL13, SCLBJ15}, which contains check-in locations shared by users on a location-based social networking website; {\sf NYC}\footnote{http://publish.illinois.edu/dbwork/open-data/.} and {\sf Beijing}\footnote{http://research.microsoft.com/apps/pubs/?id=152883.}, which are $4$-dimensional datasets that record the pickup and drop-off locations of NYC and Beijing taxis, respectively. Figure~\ref{fig:visual} visualizes the points in {\sf road} and {\sf gowalla}, as well as the pickup locations in {\sf NYC} and {\sf Beijing}. Observe that the data distribution in {\sf road} (resp.\ {\sf NYC}) is more skewed than that in {\sf Gowalla} (resp.\ {\sf Beijing}).


\edit{\para{Methods.} We compare \tree against five state-of-the-art methods: UG \cite{QYL13, QYL13hierarchy, SCLBJ15}, AG \cite{QYL13}, Hierarchy \cite{QYL13hierarchy}, DAWA \cite{LHMW14}, and Privelet$^*$ \cite{XWG11}. UG partitions the data domain into $m^d$ grid cells of equal size, and releases a noisy count for each cell, with $m = (n\e/10)^{2/(d+2)}$ \cite{SCLBJ15}. AG is an improved version of UG that is specifically designed for two-dimensional data. It first employs a coarsened version of UG to produce a set of grid cells; after that, for each cell whose noisy count is above a threshold, AG further splits it into smaller cells and releases their noisy counts. Hierarchy utilizes a multi-level decomposition tree to generate spatial histograms, with the tree height and fanout heuristically chosen to minimize the mean squared error in answering range count queries. DAWA requires as input a workload of range count queries, and it employs the {\em matrix mechanism} \cite{LHR+10} to generate a histogram that is optimized for the given workload. Privelet$^*$ publishes multi-dimensional datasets by utilizing the Haar wavelet transformation to reduce the errors of range count queries.


DAWA and Privelet$^*$ both require that the input data should have a discrete domain. Following \cite{LHMW14}, we discretize the domain of each dataset into a uniform grid with $2^{20}$ cells before feeding it to DAWA and Privelet$^*$. The other parameters of each method (e.g., the height and fanout of the decomposition tree, and the grid granularity) are set as suggested in the original papers. For \tree, we set its fanout to $4$ (resp.\ $16$) for two-dimensional (resp.\ four-dimensional) datasets, which is standard for quadtrees.

We adopt the implementations of DAWA and Privelet$^*$ provided by their respective authors, and we implement all other methods in C++. All experiments are conducted on a windows/linux machine with a 2.4GHz CPU and 16GB main memory.}

\para{Tasks.} We apply each method to create private synopses of every dataset, and we evaluate the quality of each decomposition by the accuracy of its answers to range count queries. In particular, we construct three query sets on each dataset: {\em small}, {\em medium}, and {\em large}, each of which contains $10,000$ randomly generated range count queries. Each query in the small, medium, and large set has a region that cover $[0.01\%, 0.1\%)$, $[0.1\%, 1\%)$, and $[1\%, 10\%)$ of the data domain, respectively. Following prior work\cite{CPSSY12, QYL13}, we measure accuracy of an answer $\hat{q}(D)$ to a query $q$ by its {\em relative error}, defined as

\vspace{-4.5mm}
$$
RE\left(\hat{q}(D)\right) = \frac{\left|\hat{q}(D) - q(D)\right|}{\max\left\{q(D), \Delta\right\}}.
$$
where $\Delta$ is a smoothing factor set to $0.1\%$ of the dataset cardinality $n$ \cite{QYL13, XWG11}. We repeat each experiment $100$ times, and report the average relative error of each method for each query set. \edit{For DAWA (which is {\em query-dependent}), we allow it to generate a synopsis for each query set separately, based on a sample set of $500$ queries.}



\begin{table*}[!t]
\centering
\caption{Characteristics of sequence datasets.}
\vspace{1mm}
\begin{small}
 \begin{tabular}{|l|r|r|r|r|c|r|} 
 \hline
 {\bf Name} & $\boldsymbol{|\mathcal{I}|}$ & {\bf Cardinality} & {\bf Avg.\ sequence length} & {\bf Description} & $\Gape[1pt][0pt]{\boldsymbol{l^\top}}$ & {\bf \# of sequences with length $\boldsymbol{>l^\top}$}\\
 \hline \hline
 {\sf mooc} & $7$ & $80,362$ & $13.46$ & Users' behavior sequences on a MOOC website  & $50$ & 3,653 \\
 \hline
 {\sf msnbc} & $17$ & $989,818$ & $4.75$ & Users' web navigation histories on a news portal  & $20$ & 31,606 \\
 \hline
\end{tabular}
\end{small}
\label{tbl:datasets:sequential}
\vspace{-3mm}
\end{table*}

\edit{\para{Results.} Figure~\ref{fig:exp:spatial} illustrates the average relative error of each method on each dataset as a function of the privacy budget $\e$. On {\sf road}, \tree significantly outperforms UG, AG, Hierarchy, and Privelet$^*$, regardless of the query set used and the value of $\e$. In particular, on the large query set, the average relative error of \tree is at most $\frac{1}{4}$ (resp.\ $\frac{1}{10}$) of the error of AG (resp.\ UG and Hierarchy). This demonstrates the effectiveness of \tree in approximating the distribution of the input data. Meanwhile, AG is superior to UG and Hierarchy in all cases, which is consistent with the results in previous work \cite{QYL13}. DAWA is the only method that comes close to \tree, but its relative error is never smaller than that of \tree, and is $2$ to $3$ times higher than the latter for the small and medium query sets on {\sf road} (resp.\ small query set on {\sf Gowalla}) when $\e \ge 0.8$. Furthermore, we note that DAWA is given a sample query set in advance to optimize its query performance, whereas \tree is not given such an advantage.

On {\sf Gowalla}, \tree still consistently achieves the best results, but the performance gaps between \tree and the other methods are reduced. The reason is that the data distribution in {\sf Gowalla} is less skewed than that of {\sf road} (see Figure~\ref{fig:visual}), which makes {\sf Gowalla} easier to dealt with for all methods. DAWA incurs relatively small errors in all cases, but is noticeably inferior to \tree on the small and medium query sets.

On {\sf NYC} and {\sf Beijing}, we omit AG and Hierarchy since (i) AG is only applicable on two-dimensional data, and (ii) when applied on a four-dimensional dataset, Hierarchy produces a decomposition tree with at least 2.18 billion leaf nodes \cite{QYL13hierarchy}, which cannot fit in the main memory of our machine. As shown in Figures \ref{fig:exp:spatial}g-\ref{fig:exp:spatial}l, \tree consistently outperforms all other methods by a large margin on the highly skewed {\sf NYC}, because its tree construction mechanism enables it to effectively adapt to the skewness of the data, by growing the tree tall (resp.\ short) in the dense (resp.\ sparse) regions of the data. On the other hand, on the less skewed {\sf Beijing}, the accuracies of UG and DAWA are considerably improved. Nevertheless, \tree still incurs smaller query errors in all settings. One may notice that the error of DAWA on {\sf NYC} only decreases around $2$ times when $\e$ increases from $0.05$ to $1.6$. We find that it is caused by a ``private partitioning'' step of DAWA \cite{LHMW14}, as well as the discretization of data domain $\Omega$ that it requires. We have tried to reduce this error by using a more fine-grained discretization of $\Omega$, but it leads to an ``out of memory'' error on our machine.



In summary, \tree provides better data utility than all baselines, especially when the input dataset follows a skewed distribution. This makes \tree a more favorable approach for releasing spatial data under differential privacy.}

\subsection{Experiments on Sequence Data} \label{sec:exp-squence}

%
%
%

\para{Datasets.} We use two real sequence datasets: {\sf mooc}\footnote{https://www.kddcup2015.com/} and {\sf msnbc} \cite{CAC12,UCI}. {\sf mooc} contains $80,362$ learners' behavior sequences on a MOOC platform, and the behaviors are divided into seven categories: working on assignments, watching videos, accessing other course objects, accessing the course wiki, accessing the course forum, navigating to other part of course, and closing the web page. {\sf msnbc} consists of $989,818$ sequences of URL categories, each of which corresponds to a user's browsing history during a 24-hour period on {\it msnbc.com}. Table~\ref{tbl:datasets:sequential} shows the key statistics of {\sf mooc} and {\sf msnbc}. \edit{Note that the total number $|\mathcal{I}|$ of symbols in {\sf mooc} (resp.\ {\sf msnbc}) is not excessively large. Otherwise (e.g., when $|\mathcal{I}| > 1000$), the domain of the sequence data would be extremely sparse, in which case it is enormously difficult to publish useful information under differential privacy.}


\para{Tasks.} We consider two analytical tasks on each sequence dataset $D$. The first task is to identify the top-$k$ frequent strings in $D$, i.e., the $k$ strings that appear the largest number of times in the sequences in $D$. This task is an important primitive in sequence data mining \cite{HKP11}, and is also considered in existing work \cite{CAC12} on sequence data publishing. Following previous work \cite{CAC12}, we measure the {\em precision} of the top-$k$ strings returned by differentially private algorithm, i.e.,
\vspace{-2mm}
$$
\text{precision} = \frac{\left|K(D)\cap \mathcal{A}(D)\right|}{k},
$$
where $K(D)$ is the exact set of top-$k$ frequent strings in $D$, and $\mathcal{A}(D)$ is the set returned by algorithm $\mathcal{A}$. 

The second task is to approximate the distribution of sequence lengths in $D$. In particular, we apply \tree and other existing methods to generate synthetic sequence data from $D$. Then, we compare the distribution of sequence lengths in the synthetic data with that in $D$, and we measure their {\em total variation distance}~\cite{C09}, i.e., half of the $L_1$ distance between the two probability distributions. For each task, we repeat each experiment $100$ times and report the average measurements.

\para{Methods.} For the task of top-$k$ frequent string mining, we compare \tree against two differentially private techniques: N-gram \cite{CAC12} and EM \cite{MT07}. In particular, N-gram is the state-of-the-art solution for sequence data publishing, and it is based on a variable-length $n$-gram model. N-gram requires a pre-defined threshold $n_{max}$ on the maximum length of $n$-grams; we set $n_{max}$ = $5$, as suggested in \cite{CAC12}. Meanwhile, EM is a standard application of the exponential mechanism \cite{MT07} in our context. It first initializes a set $R$ that contains $|\I|$ string of length $1$, each of which consists of a unique symbol in $\I$. After that, it invokes the exponential mechanism $k$ times. In each invocation, it selects the most frequent string $r$ from $R$ with differential privacy, and then replaces $r$ in $R$ with $|\I|$ strings, each of which is obtained by adding a symbol to the end of $r$. The $k$ strings obtained are then returned as the result. For the task of approximating sequence length distributions, we omit EM since it is inapplicable.

Note that \tree, N-gram, and EM all require that the maximum sequence length in the input data is bounded by a constant $l^\top$ that is not excessively large (see Section~\ref{sec:markov-tree} for a discussion on the necessity of $l^\top$). Following previous work \cite{CAC12}, we set $l^\top$ to be roughly the $95\%$ quantile of the sequence lengths in the input data, i.e., only around $5\%$ sequences are truncated (see Table~\ref{tbl:datasets:sequential}). To illustrate the effects of truncation, we also include in our experiments a baseline approach dubbed {\em Truncate}. This approach directly answers all queries on the truncated dataset, without any privacy assurance.

\begin{figure}[!t]
\begin{small}
\begin{tabular}{c@{\hspace{-0.5mm}}:c}
\multicolumn{2}{c}{ \vspace{-0mm}\hspace{-5mm} \includegraphics[height=3.8mm]{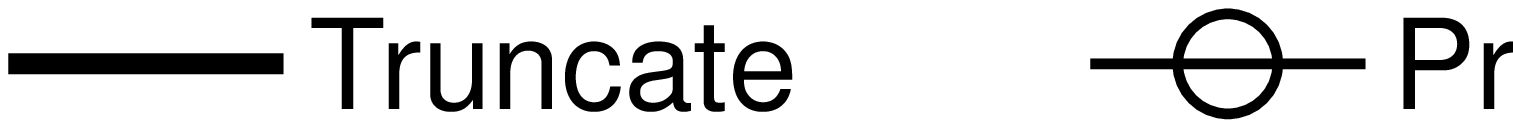}} \\

\hspace{-3mm}\includegraphics[height=31mm]{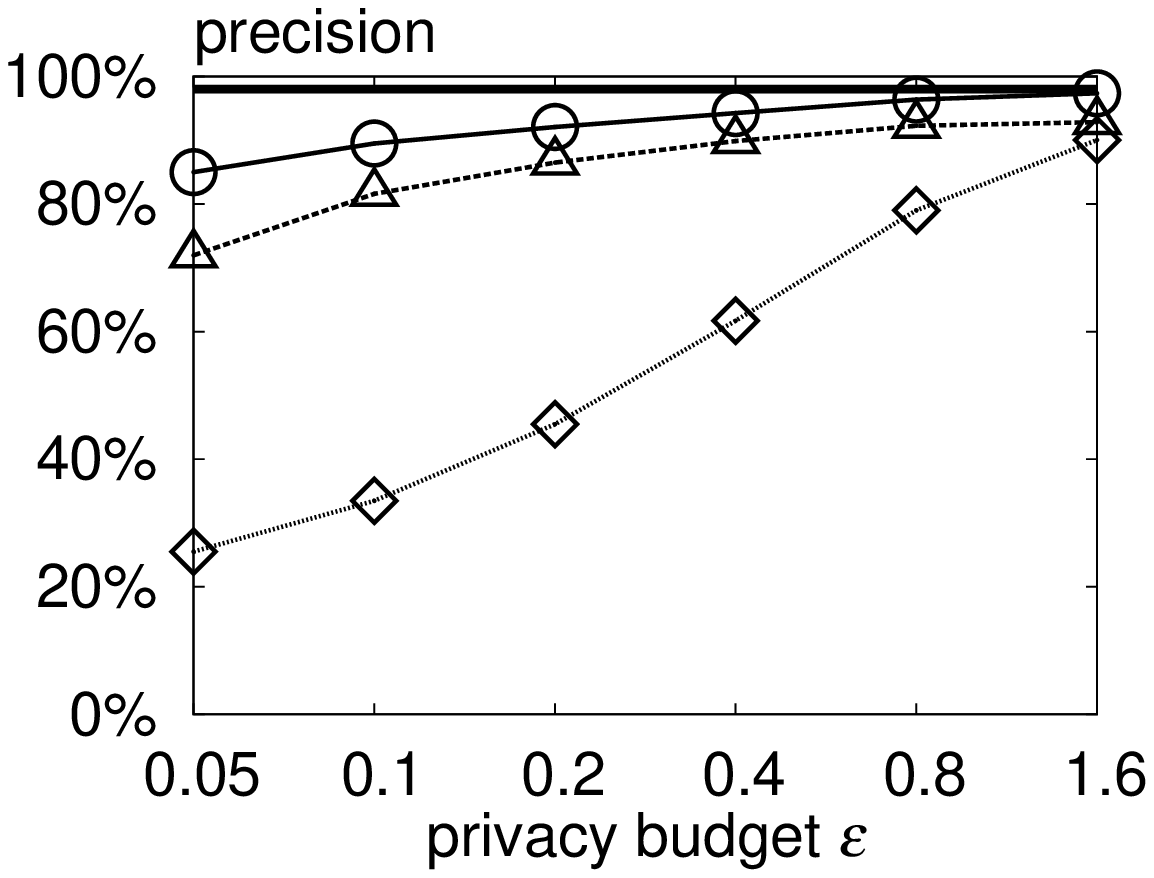}
&
\hspace{-2mm}\includegraphics[height=31mm]{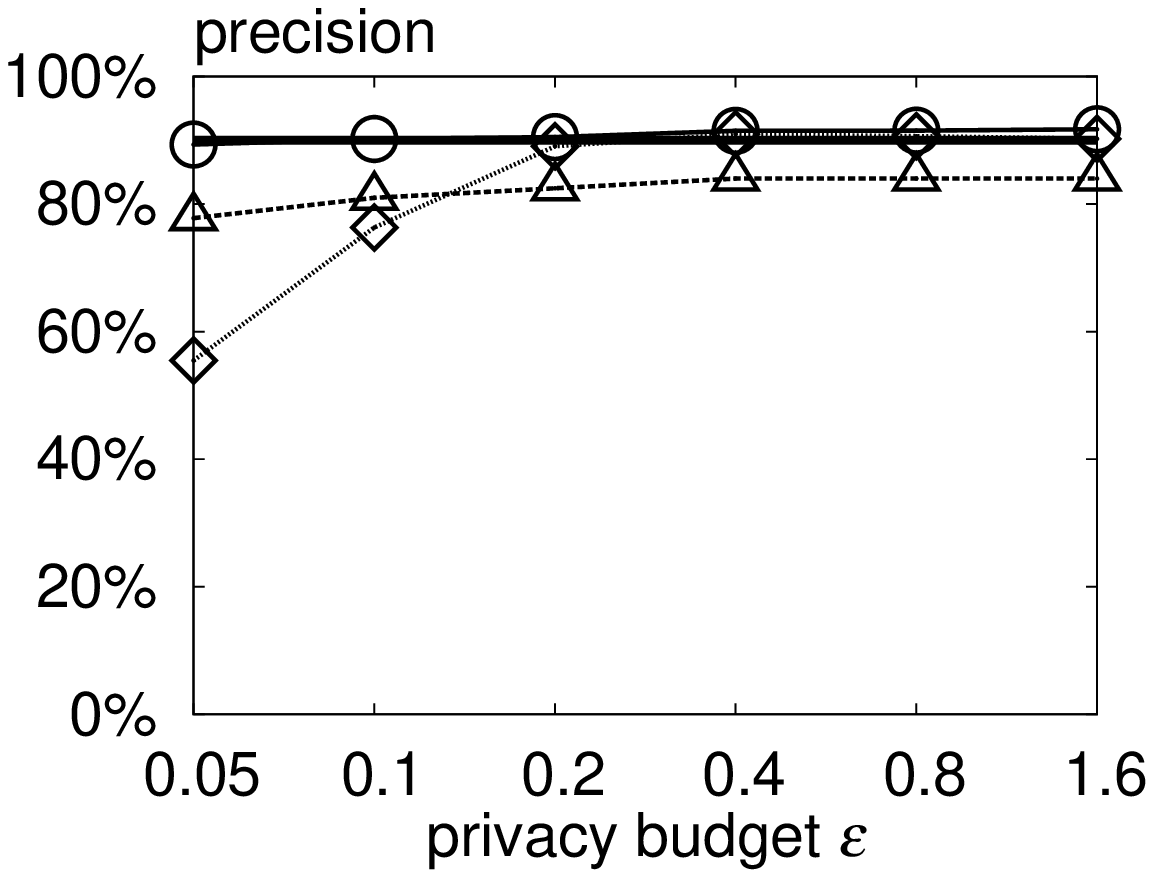}\\

\hspace{-3mm}(a) {\sf mooc} - top$50$. & \hspace{-2mm}(d) {\sf msnbc} - top$50$. \\[2mm]

\hspace{-3mm}\includegraphics[height=31mm]{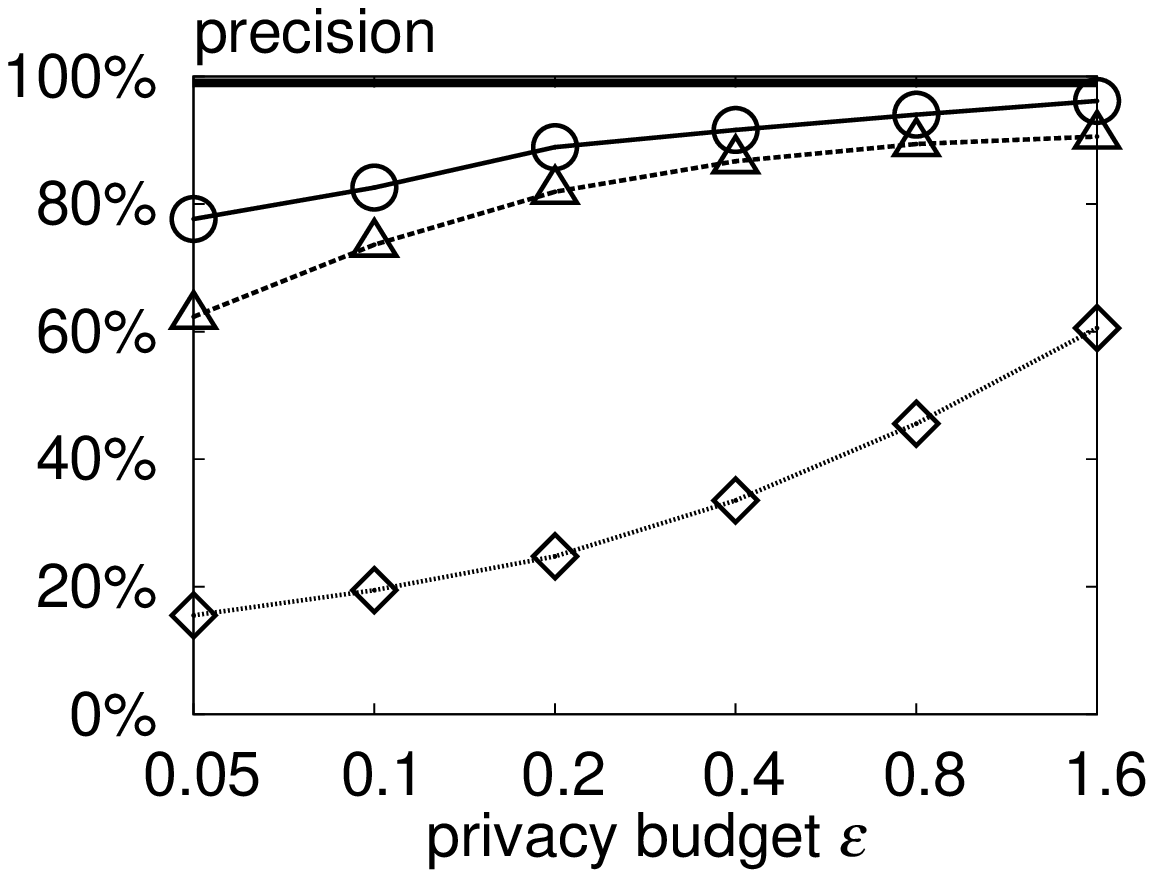}
&
\hspace{-2mm}\includegraphics[height=31mm]{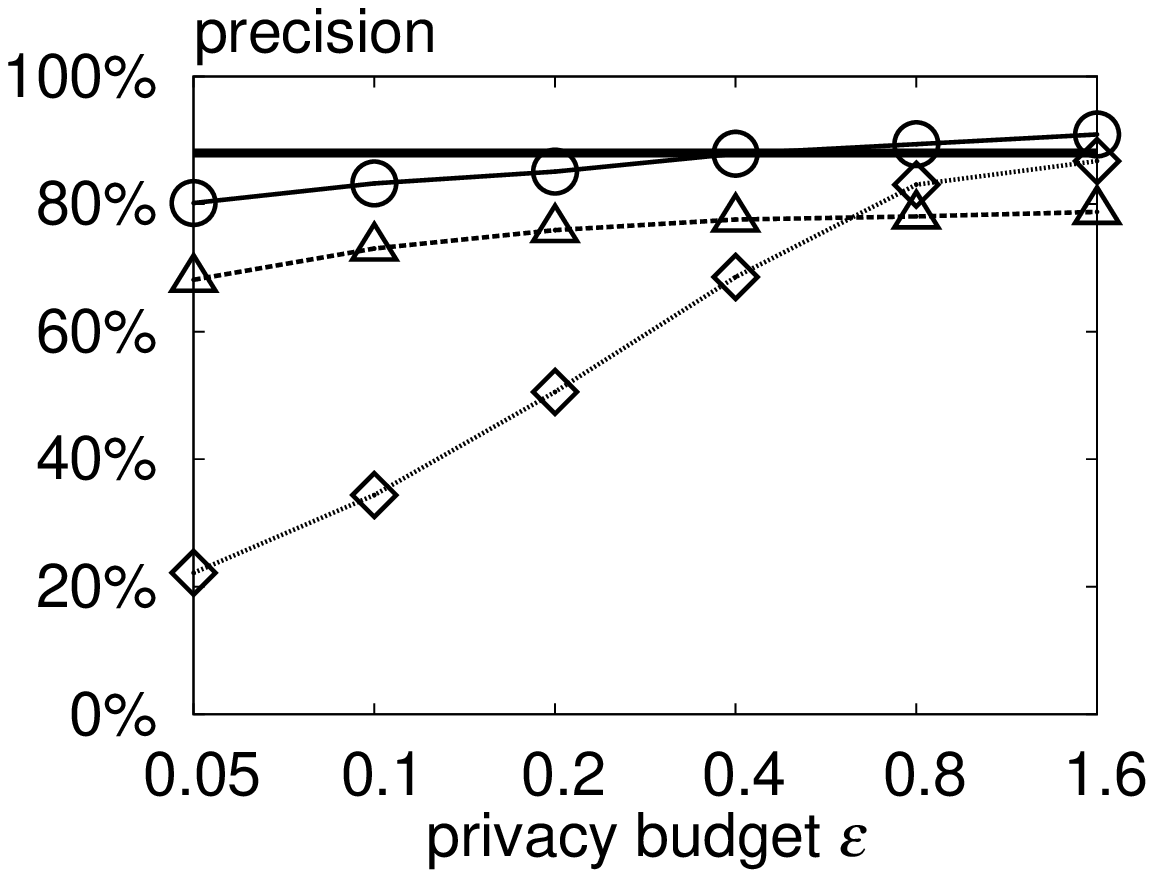}\\

\hspace{-3mm}(b) {\sf mooc} - top$100$. & \hspace{-2mm}(e) {\sf msnbc} - top$100$. \\[2mm]

\hspace{-3mm}\includegraphics[height=31mm]{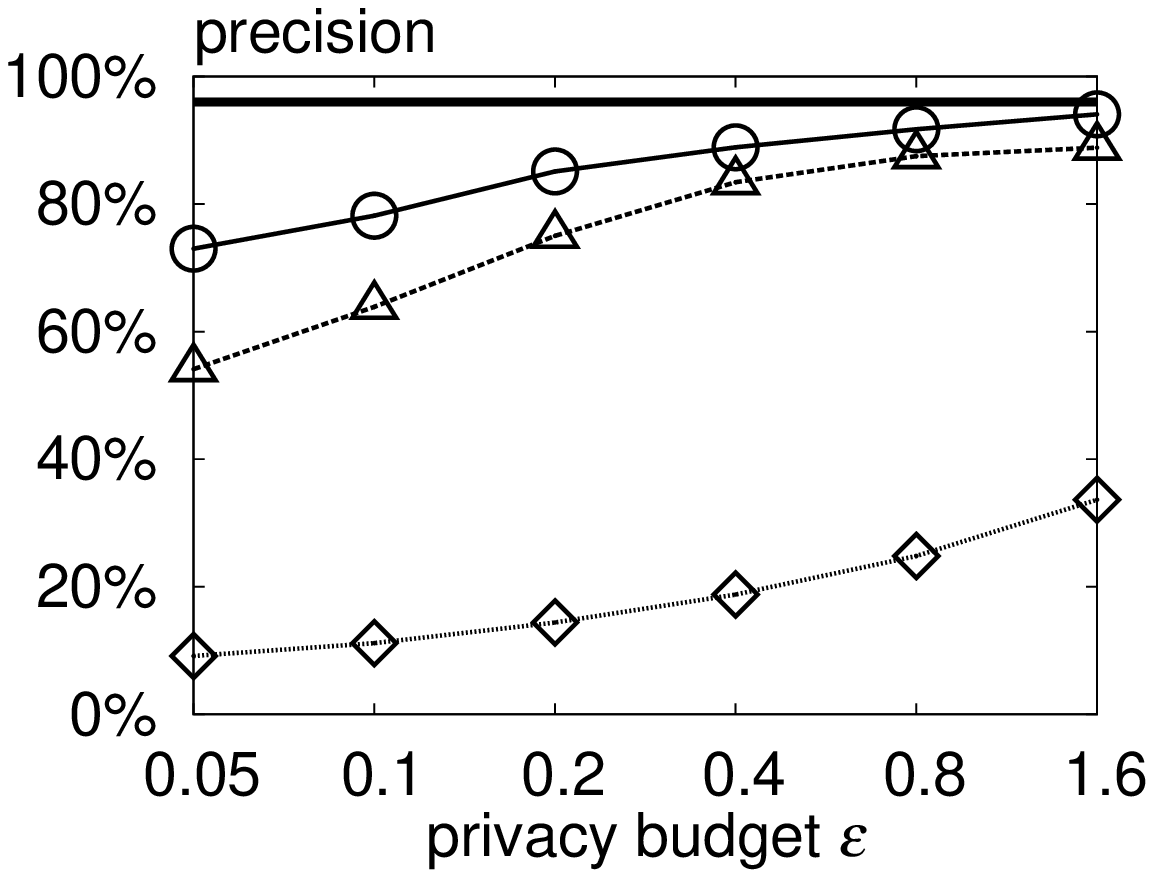}
&
\hspace{-2mm}\includegraphics[height=31mm]{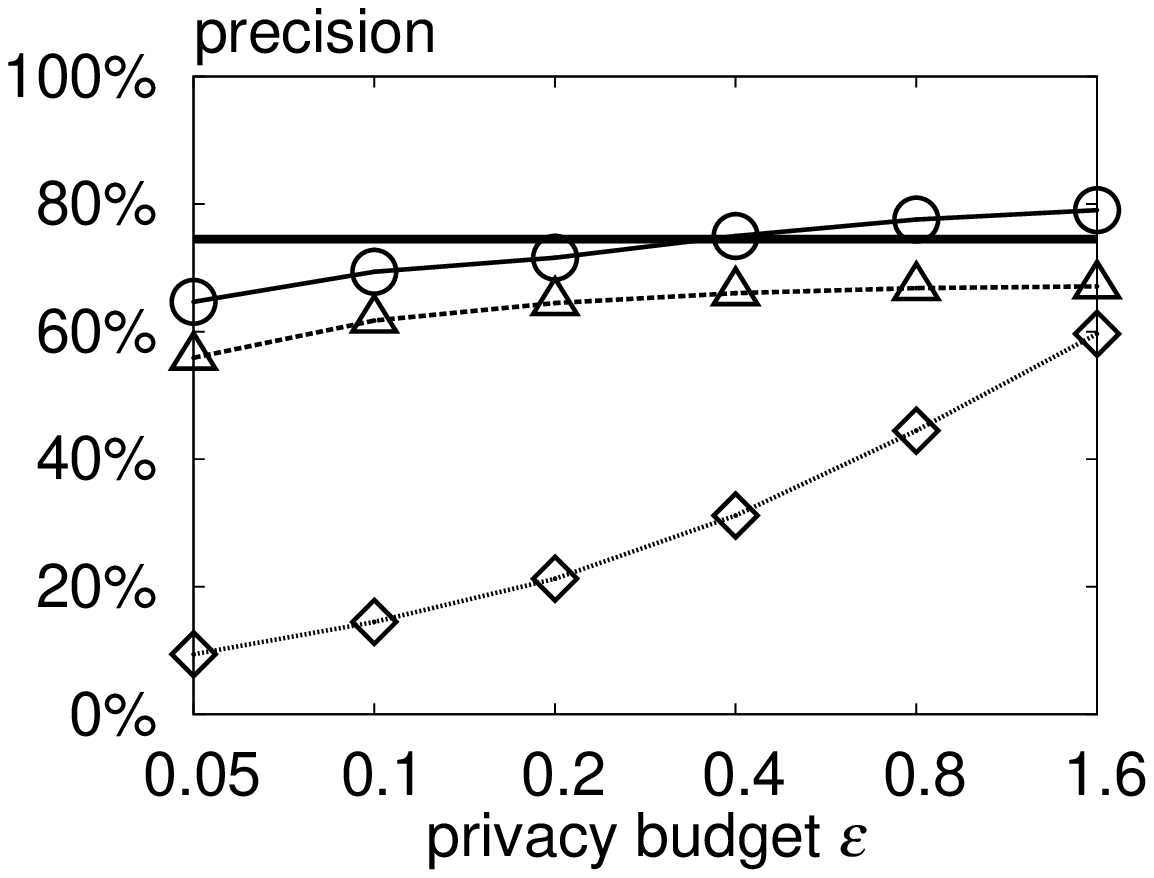} \\

\hspace{-3mm}(c) {\sf mooc} - top$200$.  & \hspace{-2mm}(f) {\sf msnbc} - top$200$.
\end{tabular}
\end{small}
\vspace{-2mm}
\caption{Results of top-$\boldsymbol{k}$ frequent string mining.}
\label{fig:exp:topk}
\end{figure}

\para{Results.} Figure~\ref{fig:exp:topk} shows the precision of each method (for top-$k$ string mining) as a function of the privacy budget $\e$. The precision of Truncate remains unchanged for all $\e$, since it does not enforce differential privacy. Among the differentially private methods, \tree consistently outperforms N-gram and EM, in most cases by a large margin. Furthermore, in Figure~\ref{fig:exp:topk}d-\ref{fig:exp:topk}f, \tree has an even higher precision than Truncate when $\e\ge0.8$. The reason is that the Markov model adopted by \tree is able to {\em recover} some information that is lost due to the truncation of sequences. For example, suppose that the string $aa$ appears in $5$ sequences in a dataset $D$ and, in each appearance, it is immediately followed by a symbol $b$. Assume that one of those $5$ sequences (denoted as $s$) is truncated, and its suffix $aab$ becomes $aa$ after the truncation. In that case, the Markov model can be used able to accurately recover the truncated symbol of $s$, because, based on the truncated data, it would predict that the ``next symbol'' after $aa$ is always $b$. Intuitively, such recovering of information is more effective when the amount of noise in the Markov model is small, which explains why \tree outperforms Truncate only when $\e$ is large. In contrast, N-gram never outperforms Truncate, and its precision is lower than that of \tree by more than $10\%$ in most settings. In addition, EM yields unattractive precision in almost all cases. Its accuracy degrades with the increases of $k$, since a larger $k$ requires it to inject more noise into the selection procession of top-$k$ frequent strings.

In the last set of experiments, we evaluate the accuracy of the sequence length distribution in the synthetic sequence data generated by each method. Figure~\ref{fig:exp:length} illustrates the total variation distance of each sequence length distribution. Observe that \tree incurs a small error comparable to that of Truncate, especially when $\e \ge 0.2$. In contrast, N-gram entails an enormous error in all cases. Based on the results in Figures \ref{fig:exp:topk} and \ref{fig:exp:length}, we conclude that \tree is a more preferable solution than N-gram to modeling sequential data under $\e$-differential privacy.

\begin{figure}[t]
\begin{small}
\begin{tabular}{cc}
\multicolumn{2}{c}{ \vspace{-0mm}\hspace{-3mm} \includegraphics[height=3.8mm]{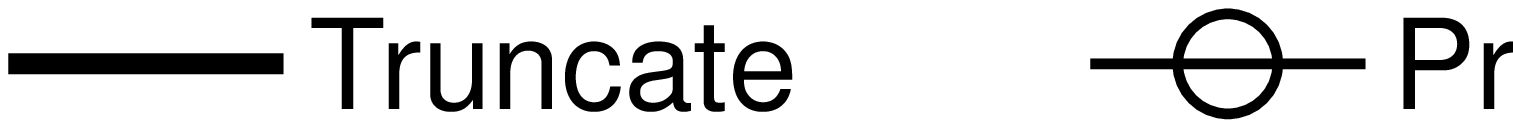}} \\

\hspace{-3mm}\includegraphics[height=31mm]{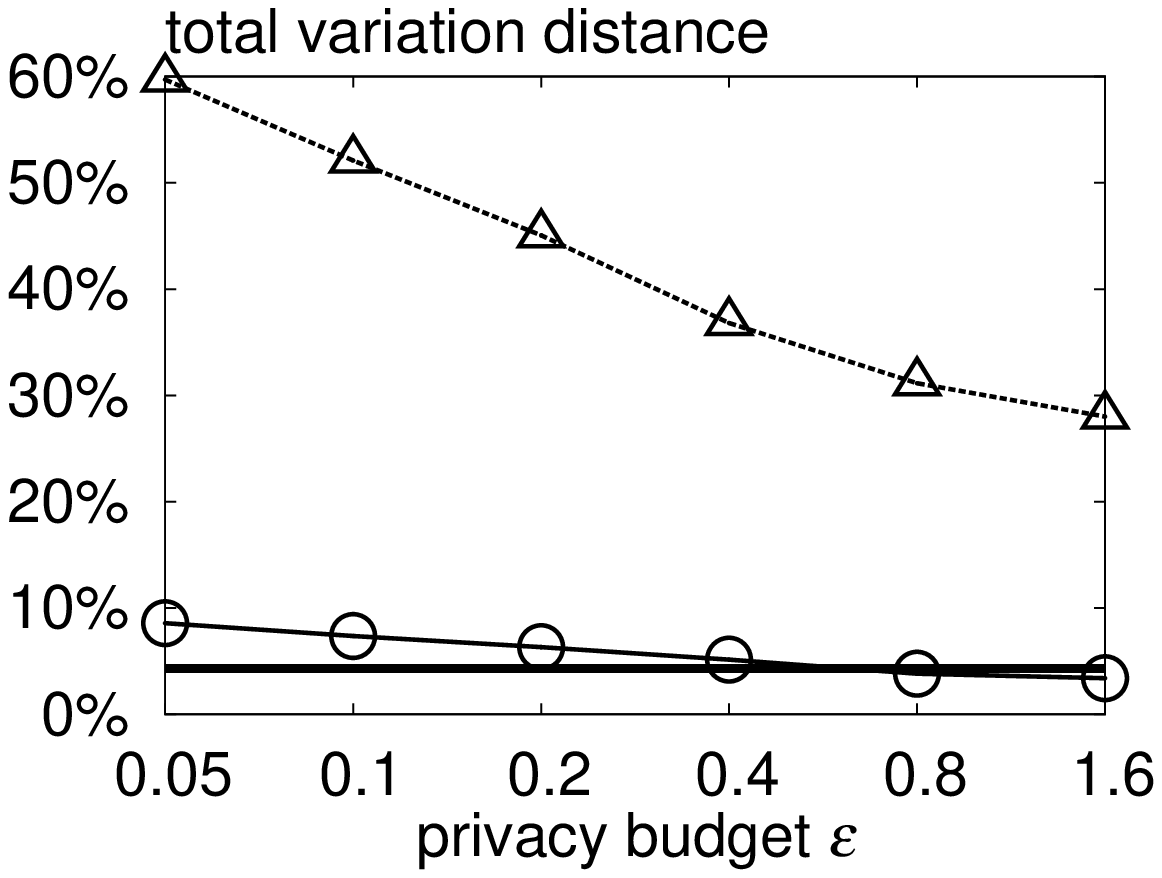}
&
\hspace{-5mm}\includegraphics[height=31mm]{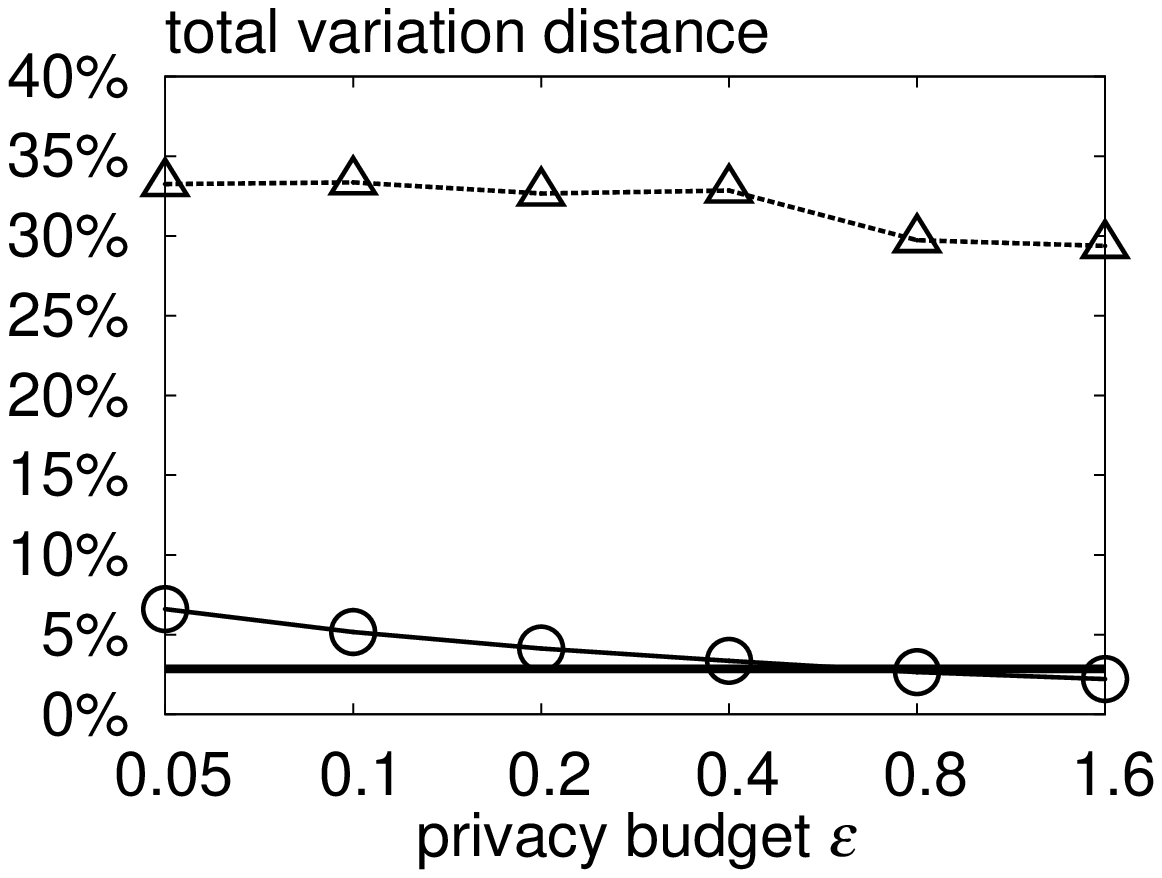} \\

\hspace{-3mm}(a) {\sf mooc} & \hspace{-5mm}(b) {\sf msnbc} \\
\end{tabular}
\end{small}
\vspace{-2mm}
\caption{Errors of sequence length distributions.}
\label{fig:exp:length}
\end{figure}

\balance

\section{Additional Related Work}\label{sec:related}

In Sections \ref{sec:alg:first}, \ref{sec:markov-compare}, and \ref{sec:exp}, we have introduced the states-of-the-art solutions \cite{CPSSY12,QYL13,SCLBJ15,CAC12,XWG11,QYL13hierarchy,LHMW14} for publishing spatial and sequence data under differential privacy. Besides those solutions, there are a few other methods for private modeling of spatial and sequence data. In particular, Xiao et al.\ \cite{XXY10} present a spatial decomposition algorithm based on the $k$-d tree \cite{b75}. It first imposes a uniform grid over the data domain, and then construct a private $k$-d tree over the cells in the grid. This method, however, is shown to be inferior to the UG and AG methods tested in our experiments, in terms of data utility \cite{QYL13}. Chen et al.\ \cite{CFDS12} consider the publication of sequence data under differential privacy, and propose an algorithm that releases a prefix tree of sequences to support count queries and frequent string mining. Nevertheless, subsequent work by Chen et al.\ \cite{CAC12} shows that the prefix-based method is considerably outperformed by the N-gram approach in our experiment.


In addition, there is a long line of research on processing aggregate queries in a differentially private manner. Specifically, Barak et al.\ \cite{BCD+07} investigate the publication of {\em marginals} (i.e., projections of a dataset on subsets of its dimensions), and propose a solution based on the Fourier transform. Ding et al.\ \cite{DWHL11} publish multiple data cubes with both privacy and consistency guarantees. The {\em matrix mechanism} of Li and Miklau \cite{LM12,LM13} and follow-up approaches \cite{HLM12,YZWXYH12,YCPS13,LHMW14} take into account a query workload, and aim to release a version of the data that maximizes the overall accuracy of the workload. DAWA \cite{LHMW14} is the most advanced method among these approaches, but as shown in our experiments, it is outperformed by \tree in terms of the relative errors of range count queries on spatial data.


Moreover, there exists extensive work that addresses numerous other tasks under differential privacy, such as regressions~\cite{NRS07,Smith11,CMS11,KST12,ZXYZW13,ZCPSX14},  clusterings~\cite{SCLBJ15}, decision trees~\cite{FS10}, recommendation systems~\cite{MM09}, time-series data analysis~\cite{RN10}, combinatorial optimizations~\cite{TGLMR10}, frequent itemset mining~\cite{LQSC12}, and graph queries~\cite{KRSY14,ZCPSX15,CZ13,LM14}. Finally, recent research has also studied the adoption of differential privacy in various systems \cite{RSKSW10,CRFG12,NH12}.

\section{Concluding Remarks}
In this paper, we study the problem of hierarchical decomposition under differential privacy, and address the central dilemma of choosing the maximum height $h$ of the decomposition tree. We show that the constraint on $h$ can be removed by introducing a carefully controlled bias in deciding when a node should be split. Based on this result, we propose \tree, a general approach for hierarchical decomposition on private data, and we showcase its applications on spatial and sequence data release. Our experimental results demonstrate that \tree significantly outperforms the states of the art in terms of data utility. For future work, we plan to extend the idea behind \tree to other problems that are based on a lattice-model instead of a tree-model, such as frequent itemset mining.

\bibliographystyle{abbrv}
\bibliography{tree}

\appendix

\section{Additional Analysis on SVT} \label{appen:svt}
We continue our discussions on the sparse vector techniques (SVTs), focusing on two existing variants that we referred to as the {\em vanilla SVT} \cite{H11} and {\em reduced SVT} \cite{DR13}, respectively. We will also introduce a new variant, namely, {\em improved SVT}, which is an improvement over the reduced SVT.

\begin{algorithm}[t]
\caption{\label{alg:svt-old} {\bf VanillaSVT} ($D$, $Q = \left\{q_1, q_2, \ldots\right\}$, $\theta$, $\lambda$, $t$)}
\BlankLine
    initialize a counter: $cnt = 0$\;
    compute a noisy version of $\theta$: $\hat{\theta} = \theta + \Lap\left(\lambda\right)$\;
    \For{$i  = 1, 2, \ldots$}
    {
        compute a noisy version of $q_i(D)$: $\hat{q_i}(D) = q_i(D) + \Lap\left(t \cdot \lambda\right)$\;
        \If{$\hat{q_i}(D) > \hat{\theta}$}
        {
            {\bf output} $o_i = \hat{q_i}(D)$ and {\bf continue}\;
            increase $cnt$ by 1\;
            \If{$cnt \ge t$}
            {
                \Return\;
            }
        }
        \Else
        {
            {\bf output} $o_i = \bot$ and {\bf continue}\;
        }
    }
    \Return\;
\end{algorithm}

\header
{\bf Analyzing the Vanilla SVT.} Algorithm~\ref{alg:svt-old} presents a generic version of the vanilla SVT. The algorithm takes as input five parameters: (i) a dataset $D$, (ii) a sequence $Q$ of counting queries, each of which has sensitivity $1$, (iii) a threshold $\theta$, (iv) a noise scale $\lambda$, and (v) a positive integer $t$. The algorithm is largely identical to the binary SVT (Algorithm~\ref{alg:svt-binary}), except for three relatively minor changes. First, instead of outputting a binary variable for each query $q_i$ to indicate whether the noisy query answer $\hat{q}_i(D)$ is above the noisy threshold $\hat{\theta}$, the vanilla SVT directly outputs $\hat{q}_i(D)$ if it is larger than $\hat{\theta}$, and it outputs a placeholder $\bot$ otherwise. Second, the vanilla SVT maintains a counter $cnt$ of the number of noisy query answers that have been directly output; whenever the counter reaches $t$, the algorithm terminates. In other words, at most $t$ noisy answers are reported. Third, in each noisy answer $\hat{q}_i(D)$, the vanilla SVT injects Laplace noise of scale $t \cdot \lambda$ (instead of $\lambda$), which is in accordance with the total number of noisy answers that might be released. Previous work \cite{H11} makes the following claim about the privacy guarantee of the vanilla SVT.
\begin{claim}\label{clm:svt-old}
Algorithm~\ref{alg:svt-old} satisfies $\e$-differential privacy if $\lambda \! \ge \! \frac{2}{\e}$.
\end{claim}

In the following, we invalidate Claim~\ref{clm:svt-old} with a counter-example similar to the one used in the proof of Lemma~\ref{lmm:svt-binary}. Consider three datasets $D_1=\left\{a, b\right\}$, $D_2=\left\{a,a, b\right\}$ and $D_3=\left\{a, a\right\}$, among which $(D_1, D_2)$ and $(D_2, D_3)$ are two pairs of neighboring datasets. Let $q_a$ (resp.\ $q_b$) be a query that asks for the number of $a$ (reps.\ $b$) in a dataset, e.g., $q_a(D_1) = 1$. Let $Q$ consist of $k$ queries, with the first $k-1$ queries being $q_a$ and the last query being $q_b$.

Suppose that we invoke Algorithm~\ref{alg:svt-old} on $D_1,D_2,D_3$, respectively, with $Q$, a noise scale $\lambda$, a threshold $\theta = 0$ and $t = 1$. Let $E$ be the event that the Algorithm~\ref{alg:svt-old} outputs the placeholder $\bot$ for the first $k-1$ queries but returns a noisy answer $1$ for the last query. That is, $o_i = \bot$ for any $i \in [1, k-1]$, and $o_k = 1$. If Algorithm~\ref{alg:svt-old} satisfies $\e$-differential privacy, then
$$
\frac{\Pr[D_1 \rightarrow E]}{\Pr[D_3\rightarrow E]} \; = \; \frac{\Pr[D_1 \rightarrow E]}{\Pr[D_2\rightarrow E]} \cdot \frac{\Pr[D_2 \rightarrow E]}{\Pr[D_3\rightarrow E]} \; \le \; e^{2\e}.
$$

Recall that Algorithm~\ref{alg:svt-old} outputs the noisy answer $\hat{q}(D)$ only if it is larger than the noisy threshold $\hat{\theta}$. Therefore,

\begin{algorithm}[h]
\caption{\label{alg:svt-reduced} {\bf ReducedSVT} ($D$, $Q = \left\{q_1, q_2, \ldots\right\}$, $\theta$, $\lambda$, $t$)}
\BlankLine
    initialize a counter: $cnt = 0$\;
    compute a noisy version of $\theta$: $\hat{\theta} = \theta + \Lap\left(t \cdot \lambda\right)$\;
    \For{$i  = 1, 2, \ldots$}
    {
        compute a noisy version of $q_i(D)$: $\hat{q_i}(D) = q_i(D) + \Lap\left(t \cdot \lambda\right)$\;
        \If{$\hat{q_i}(D) > \hat{\theta}$}
        {
            {\bf output} $o_i = 1$ and {\bf continue}\;
            update the noisy threshold: $\hat{\theta} = \theta + \Lap\left(t \cdot \lambda\right)$\;
            increase $cnt$ by 1\;
            \If{$cnt \ge t$}
            {
                \Return\;
            }
        }
        \Else
        {
            {\bf output} $o_i = 0$ and {\bf continue}\;
        }
    }
    \Return\;
\end{algorithm}

\begin{algorithm}[h]
\caption{\label{alg:svt-improved} {\bf ImprovedSVT} ($D$, $Q = \left\{q_1, q_2, \ldots\right\}$, $\theta$, $\lambda$, $t$)}
\BlankLine
    initialize a counter: $cnt = 0$\;
    compute a noisy version of $\theta$: $\hat{\theta} = \theta + \Lap\left(\lambda\right)$\;
    \For{$i  = 1, 2, \ldots$}
    {
        compute a noisy version of $q_i(D)$: $\hat{q_i}(D) = q_i(D) + \Lap\left(t \cdot \lambda\right)$\;
        \If{$\hat{q_i}(D) > \hat{\theta}$}
        {
            {\bf output} $o_i = 1$ and {\bf continue}\;
            increase $cnt$ by 1\;
            \If{$cnt \ge t$}
            {
                \Return\;
            }
        }
        \Else
        {
            {\bf output} $o_i = 0$ and {\bf continue}\;
        }
    }
    \Return\;
\end{algorithm}

\begin{small}
\begin{align}
& \frac{\Pr[D_1  \rightarrow E]}{\Pr[D_3\rightarrow E]} \nonumber \\
& = \frac{\int_{-\infty}^1 \Pr[\hat{\theta} = x] \cdot \big(\Pr[\hat{q_a}(D_1)\le x]\big)^{k-1} \cdot \Pr[\hat{q_b}(D_1) = 1] dx}
{\int_{-\infty}^1 \Pr[\hat{\theta} = x] \cdot \big(\Pr[\hat{q_a}(D_3)\le x]\big)^{k-1} \cdot \Pr[\hat{q_b}(D_3) = 1] dx} \nonumber \\
& = \frac{\int_{-\infty}^1 \Pr[\hat{\theta} = x] \cdot \big(\Pr[\Lap(\lambda)\le x-1]\big)^{k-1} \cdot \Pr[\Lap(\lambda) = 0] dx}
{\int_{-\infty}^1 \Pr[\hat{\theta} = x] \cdot \big(\Pr[\Lap(\lambda)\le x-2]\big)^{k-1} \cdot \Pr[\Lap(\lambda) = 1] dx} \nonumber
\end{align}
\end{small}

\vspace{-2mm}
\noindent
Note that in the above equation, we restrict the range of the noisy threshold $\hat{\theta}$ to $(-\infty, 1)$. This is because $\hat{\theta}$ has to be smaller than all noisy answers output by Algorithm~\ref{alg:svt-old}, i.e., $\hat{\theta} < o_{k} = 1$. (Previous work \cite{H11} overlooks this issue and considers $\hat{\theta}$ independent of the noisy answers.)

Consider any $\hat{\theta} < 1$. By Equation~\eqref{eqn:pre-laplace},
$$\frac{\Pr[\hat{q_b}(D_1) = 1]}{\Pr[\hat{q_b}(D_3) = 1]} \; = \; \frac{\Pr[\Lap(\lambda) = 0]}{\Pr[\Lap(\lambda) = 1]} \; = \; e^{1/\lambda}.$$
In addition,
$$\frac{\Pr[\Lap(\lambda) \le x-1]}{\Pr[\Lap(\lambda) \le x-2]} = e^{1/\lambda}.$$
It follows that

\vspace{-2mm}
\begin{small}
\begin{align*}
& \frac{\Pr[D_1  \rightarrow E]}{\Pr[D_3\rightarrow E]} \nonumber \\
& = \frac{\int_{-\infty}^1 \Pr[\hat{\theta} = x] \cdot \big(\Pr[\Lap(\lambda)\le x-1]\big)^{k-1} \cdot \Pr[\Lap(\lambda) = 0] dx}
{\int_{-\infty}^1 \Pr[\hat{\theta} = x] \cdot \big(\Pr[\Lap(\lambda)\le x-2]\big)^{k-1} \cdot \Pr[\Lap(\lambda) = 1] dx} \nonumber \\
& = e^{\frac{k}{\lambda}}.
\end{align*}
\end{small}

\vspace{-3mm}
\noindent
Therefore, $\frac{\Pr[D_1  \rightarrow E]}{\Pr[D_3\rightarrow E]} > e^{2\e}$ whenever $\lambda < k/2\e$. This indicates that, in the worst case, Algorithm~\ref{alg:svt-old} requires Laplace noise of scale $t\lambda = \Omega(\frac{t\cdot k}{\e})$ (instead of $\frac{2 t}{\e}$) in each noisy answer to ensure $\e$-differential privacy, where $k$ is the total number of queries and $t$ is the total number of noisy answers returned.

\header
{\bf Analyzing the Reduced SVT.} Algorithm~\ref{alg:svt-reduced} shows the pseudo-code of the reduced SVT \cite{DR13}. The input of the algorithm is the same as that of the vanilla SVT \cite{H11}, while the other parts are almost identical with only two differences: (i) when a noisy answer $\hat{q_i}(D)$ is larger than the noisy threshold $\hat{\theta}$, the reduced SVT outputs a binary value $o_i = 1$; otherwise, it outputs $o_i = 0$, and (ii) the scale of noise added to the threshold $\theta$ is set to $t\cdot \lambda$ instead of $\lambda$, and it re-generates the noisy threshold each time after outputting an $o_i=1$. Dwork and Roth \cite{DR13} prove that Algorithm~\ref{alg:svt-reduced} satisfies $\e$-differential privacy when $\lambda \ge \frac{2}{\e}$. That is, the scale of Laplace noise in each noisy answer $\hat{q}_i(D)$ should be at least $\frac{2 t}{\e}$.

We observe that the reduced SVT can be improved, by reducing the amount of noise added to the threshold $\theta$. In particular, instead of using multiple noisy versions of $\theta$ with noise scale set to $t \cdot \lambda$, we may use a {\em single} noisy version of $\theta$ throughout the algorithm, with noise scale set to $\lambda$ instead. Algorithm~\ref{alg:svt-improved} presents the pseudo-code of the revised algorithm, which we refer to as the {\em improved SVT}. Compared with the reduced SVT, the improved SVT yields more accurate results since it uses a more accurate version of $\theta$ to decide whether a query answer is larger than the threshold. The following lemma shows that the improved SVT provides the same privacy guarantee as the reduced SVT.
\begin{lemma}\label{lmm:svt-improved}
Algorithm~\ref{alg:svt-improved} satisfies $\e$-differential privacy if $\lambda\ge \frac{2}{\e}$.
\end{lemma}

Notice that both the reduced SVT and the improved SVT require adding Laplace noise of scale $\frac{2 t}{\e}$ into each query answer. As such, neither of them yields a competitive solution for the hierarchical decomposition problem that we consider. For example, suppose that we construct a private quadtree by using the improved SVT to choose the nodes that should be split, i.e., the nodes $v$ whose point counts $c(v)$ are larger than a given threshold $\theta$. In that case, we would have to choose an appropriate value for $t$ (i.e., the maximum number of nodes that we can split), which is rather difficult without prior knowledge of the input data. In addition, even if we are able to select an appropriate $t$, we still need to inject Laplace noise of scale $\frac{2 t}{\e}$ into $c(v)$ when we decide whether $v$ should be split. In contrast, \tree only requires Laplace noise of scale $\Theta(\frac{1}{\e})$. Therefore, the reduced SVT and the improved SVT are both less favorable than \tree for hierarchical decomposition.

\section{Proofs} \label{appen:proof}

\para{Proof of Lemma~\ref{lmm:alg-bound}.} By Equations \eqref{eqn:pre-laplace} and \eqref{eqn:alg-risk}, for any $x$,
\begin{equation*}
\rho(x) = \ln\left(\frac{\int_{\theta - x}^{+\infty} \frac{1}{2\lambda}\exp\left(-\frac{|y|}{\lambda}\right)dy}{\int_{\theta+1 - x}^{+\infty} \frac{1}{2\lambda}\exp\left(-\frac{|y|}{\lambda}\right)dy}\right) \le \frac{1}{\lambda}.
\end{equation*}
Recall that $\rho^\top(x) = 1/\lambda$ when $c(v) < \theta + 1$. Therefore, $\rho(x) \le \rho^\top(x)$ holds if $x < \theta + 1$.

Now consider that $x \ge \theta + 1$. In that case,
\begin{equation*}
\rho(x) = \ln\left(\frac{1 - \frac{1}{2}\exp\left(\frac{\theta - x}{\lambda}\right)}{1 - \frac{1}{2}\exp\left(\frac{\theta + 1 - x}{\lambda}\right)}\right).
\end{equation*}
For convenience, we let $\alpha = \frac{1}{2}\exp\left(\frac{\theta + 1 - x}{\lambda}\right)$, and define a function $f$ of $\alpha$ as follows:

\vspace{-5mm}
\begin{align*}
f(\alpha) & =  \rho(x) - \frac{1}{\lambda} \exp\left(\frac{\theta + 1 - x}{\lambda}\right) \\
& = \ln\left(\frac{1 - \alpha e^{-1/\lambda}}{1 - \alpha}\right) - \frac{2\alpha}{\lambda}.
\end{align*}
Observe that $\alpha \in (0, 1/2]$ whenever $x \ge \theta + 1$. Therefore, we can prove the lemma by showing that $f(\alpha) \le 0$ for all $\alpha \in (0, 1/2]$. For this purpose, we first compute the second derivative of $f$ with respect to $\alpha$:

\vspace{-5mm}
\begin{equation*}
f''(\alpha) \quad = \quad (e^{1/\lambda} - 1) \cdot \frac{e^{1/\lambda} + 1 - 2\alpha}{(1-\alpha)^2 \cdot (e^{1/\lambda} - \alpha)^2}.
\end{equation*}
Given that $e^{1/\lambda} -1  > 0$ and $e^{1/\lambda} + 1 - 2\alpha \ge e^{1/\lambda} > 0$, we have $f''(\alpha) > 0$. This indicates that
\begin{equation*}
\max_{\alpha \in (0, 1/2]} f(\alpha) = \max\{ f(0), f(1/2)\}.
\end{equation*}
Meanwhile, $f(0) = 0$, and
\begin{align*}
f(1/2) & =  \ln(2 - e^{-1/\lambda}) - \frac{1}{\lambda} \\
& =  \ln\Big(e^{1/\lambda} - \left(e^{1/2\lambda} - e^{-1/2\lambda}\right)^2\Big) -\frac{1}{\lambda} \\
& <  \ln(e^{1/\lambda}) - \frac{1}{\lambda} = 0.
\end{align*}
Therefore, $\max_{\alpha \in (0, 1/2]} f(\alpha) \le 0$, which proves the lemma. \done

\para{Proof of Theorem~\ref{thm:alg:privacy}.} The theorem directly follows from the privacy analysis in Section~\ref{sec:alg:privtree}. \done

\para{Proof of Lemma~\ref{lmm:alg-converge}.} Let $V$ be the set of all possible nodes in a quadtree built on $D$. We divide the nodes in $V$ into three subsets: (i) the set $V_1$ of nodes that appear as non-leaf nodes in $\mathcal{T^*}$, (ii) the set $V_2$ of the nodes that appear as leaves in $\mathcal{T^*}$, and (iii) the set $V_3$ of the nodes that do not appear in $\mathcal{T^*}$. Let $g(S)$ be the expected number of nodes in a set $S$ that appear in $\mathcal{T}$. Then, we have
\begin{align*}
\mathbb{E}[|\mathcal{T}|] \; = \; & g(V_1) + g(V_2) + g(V_3) \\
\; \le \; & |V_1| + |V_2| + g(V_3) \; = \; |\mathcal{T^*}| + g(V_3).
\end{align*}
Therefore, the lemma can be proved by showing $g(V_3) \le |\mathcal{T^*}|$.

Observe that each node in $V_3$ must be the descendant of a node in $V_2$, i.e., a node that appears as a leaf in $\mathcal{T^*}$. Therefore, we can divide the nodes in $V_3$ into $|V_2|$ subsets, such that all nodes in the same subset are descendants of the same node in $V_2$. Consider any such subset $S$ that corresponds to a node $v \in V_2$. Given that $v$ appears as a leaf in $\mathcal{T^*}$, we have $c(v) \le \theta$. In addition, since $|\mathcal{T^*}| > 1$, $\depth(v) \ge 1$ holds. Therefore,
\begin{equation*}
c(v) - \depth(v)\cdot \delta \:\: \le \:\: c(v) - \delta \:\: \le \:\: \theta - \delta.
\end{equation*}
By Equation~\eqref{eqn:alg-bv}, we have $b(v) = \theta - \delta$. Furthermore, for any $v' \in S$, we have $c(v') \le c(v)$, which also leads to $b(v') = \theta - \delta$. Therefore, $v$ and $v'$ have the same probability $p_s$ to be split. Given $\delta = \lambda \cdot \ln \beta$, we have
\begin{align*}
p_s \:\: = \:\: &  \Pr[\theta - \delta + \Lap(\lambda)>\theta] \:\: = \:\: \Pr[\Lap(\lambda)>\delta] \\
= \:\: &  \int_{\lambda\cdot\ln\beta}^{\infty} \frac{1}{2\lambda}\exp\left(-\frac{|y|}{\lambda}\right)dy \:\: = \:\: \frac{1}{2\beta}.
\end{align*}

Assume that $v$ appears in $\mathcal{T}$. Then, given that $v$ is split with $\frac{1}{2 \beta}$ probability, each child of $v$ has $\frac{1}{2 \beta}$ probability to appear in $\mathcal{T}$. Therefore, in expectation, the number of $v$'s children that appear in $\mathcal{T}$ should equal $\beta \cdot \frac{1}{2 \beta} = \frac{1}{2}$. In general, for any $i \ge 1$, there exist $\beta^i$ nodes $v' \in S$ with $\depth(v') - \depth(v) = i$, and each such $v'$ has $(\frac{1}{2 \beta})^i$ probability to appear in $\mathcal{T}$. Hence, the expected number of nodes in $S$ that appear in $\mathcal{T}$ is
\begin{align*}
g(S) \:\: = \:\: \sum_{i = 1}^{+\infty} \beta^i \cdot \left(\frac{1}{2 \beta}\right)^i \:\: = \:\: \sum_{i = 1}^{+\infty} \frac{1}{2^i} \:\: = \:\: 1.
\end{align*}
Since each $S$ uniquely corresponds to a node in $V_2$, we have
\begin{equation*}
g(V_3) \:\: = \:\: |V_2| \cdot g(S) \:\: = |V_2| \:\: \le \:\: |\mathcal{T^*}|.
\end{equation*}
Therefore, the lemma is proved. \done

\para{Proof of Corollary~\ref{coro:alg:privacy}.} The corollary follows from Theorem~\ref{thm:alg:privacy} when $\gamma = \ln \beta$. \done

\para{Proof of Lemma~\ref{lmm:newc:monotone}.} Let $u$ and $v$ be two nodes in $\mathcal{T}$, such that $u$ is the parent of $v$. Then, $\hist(v)[x] \le \hist(u)[x]$ holds for every symbol $x\in\I\cup\{\&\}$. Let $x_v$ (resp.\ $x_u$) be the symbol that has the largest count in $\hist(v)$ (resp.\ $\hist(u)$). We have
\begin{align*}
c(v) \:\: =  \:\: & \|\hist(v)\|_1 - \hist(v)[x_v] \:\: \le \:\: \|\hist(v)\|_1 - \hist(v)[x_u] \\
= \:\: & \sum_{x \ne x_u} \hist(v)[x]  \:\: \le  \:\: \sum_{x \ne x_u} \hist(u)[x] \\
= \:\: & \|\hist(u)\|_1 - \hist(u)[x_u] \:\: = \:\: c(u).
\end{align*}
Therefore, $c(\cdot)$ is monotonic. \done

\para{Proof of Theorem~\ref{thrm:markov:privtree}.} Let $D$ and $D'$ be two neighboring datasets, such that $D$ is obtained by inserting a sequence $s$ into $D'$. Assume that $s = \$x_1\ldots x_{l}$, where $x_i\in I\cup\{\&\}$ for $i\in[1,l]$. To facilitate our proof, we define $l$ datasets $D_1, D_2, \ldots, D_l$, such that $D_i = D' \cup \{s_i\}$ and $s_i = \$x_1x_2\ldots x_i$ is the length-$i$ prefix of $s$ ended at symbol $x_i$. Observe that $D_l = D$. For convenience, we define $D_0 =  D'$.

In the following, we will prove that for any $i \in [1, l]$ and any output $\mathcal{T}$ of the modified \tree,
\begin{equation} \label{eqn:markov-proof-1}
-\frac{\e}{l^\top} \; \le \; \ln\left( \! \frac{\Pr[D_i\rightarrow \mathcal{T}]}{\Pr[D_{i-1}\rightarrow \mathcal{T}]} \! \right) \; \le \; \frac{\e}{l^\top},
\end{equation}
where $\Pr[D_i \rightarrow \mathcal{T}]$ denotes the probability that \tree outputs $\mathcal{T}$ given $D_i$. This would prove the theorem because, given that $l \le l^\top$,
$$
\ln\left( \! \frac{\Pr[D\rightarrow \mathcal{T}]}{\Pr[D'\rightarrow \mathcal{T}]} \! \right) \; = \; \sum_{i=1}^{l}\ln\left( \! \frac{\Pr[D_i\rightarrow \mathcal{T}]}{\Pr[D_{i-1}\rightarrow \mathcal{T}]} \! \right) \; \in \; [-\e, \e].
$$

Observe that $D_i$ can be obtained by appending a symbol $x_i$ to the end of the sequence $s_{i-1}$ in $D_{i-1}$. Therefore, when we change the input data from $D_{i-1}$ to $D_i$, the only changes in the PST are the histogram counts that $x_i$ contributes to. Observe that if $x_i$ contributes to the prediction histogram $hist(v)$ of a node $v$, then $\dom(v)$ must be a suffix of $s_{i-1}$, and the only possible change in $\hist(v)$ is that $\hist(v)[x_i]$ would be increased by one. Then, by the definition of the PST, all of those nodes $v$ should form a path from the root of the PST to a leaf. In addition, by Equation~\eqref{eqn:markov-newc}, the score $c(v)$ of each of those nodes $v$ is changed by at most one. In that case, we can prove Equation~\eqref{eqn:markov-proof-1} by reusing the analysis in the proof of Theorem~\ref{thm:alg:privacy}.

To explain, recall that the correctness of Theorem~\ref{thm:alg:privacy} only replies on two conditions. First, the score $c(v)$ of each node $v$ is monotonic. Second, when we change the input data, all of the nodes affected should form a path from the root of the decomposition tree to a leaf, and the score of each of those nodes should change by at most one. Notice that all three conditions are satisfied when we change the input of the modified \tree from $D_{i-1}$ to $D_i$. Combining this with the fact that the modified \tree uses a noise scale that is $l^\top$ times that of Algorithm~\ref{alg:privtree}, it can be verified that Equation~\eqref{eqn:markov-proof-1} holds. Therefore, the theorem is proved. \done

\para{Proof of Theorem~\ref{thrm:markov:post}.}
Let $\mathcal{T}$ be the output of the modified \tree. Let $D$ and $D'$ be two neighboring datasets, such that $D$ is obtained by inserting a sequence $s$ into $D'$. Assume that $s = \$x_1\ldots x_{l}$, where $x_i\in I\cup\{\&\}$ for $i\in[1,l]$. Observe that each symbol $x_i$ in $s$ contributes to the prediction histograms of the nodes whose predictor strings $\dom(\cdot)$ are suffixes of $\$x_1\ldots x_{i-1}$. By the definition of PSTs, these nodes form a path from the root of $\mathcal{T}$ to a leaf. This indicates that each $x_i$ gets counted in the histogram of {\em one} leaf node only. Taking into account $l \le l^\top$, it follows that the sensitivity of releasing the histogram counts of all leaf nodes in $\mathcal{T}$ is $l^\top$. By the property of the Laplace mechanism, the postprocessing step ensures $\e$-differential privacy if $\lambda\ge \frac{l^\top}{\e}$.
\done

\para{Proof of Lemma~\ref{lmm:svt-binary}.} Consider three datasets $D_1=\left\{a,b\right\}$, $D_2=\left\{a,b,b\right\}$, and $D_3=\left\{b,b\right\}$, where each tuple is either $a$ or $b$. Observe that $D_1$ is a neighboring dataset of $D_2$, while $D_2$ is a neighboring dataset of $D_3$. Let $q_a$ (resp.\ $q_b$) be a query that asks for the number of $a$ (resp.\ $b$) in a dataset. Let $Q$ be a sequence of $k$ queries, such that first $k/2$ queries are all $q_a$, and the remaining $k/2$ queries are all $q_b$.

Suppose that we invoke Algorithm~\ref{alg:svt-binary} on $D_1, D_2, D_3$, respectively, with $Q$, a noise scale $\lambda$, and a threshold $\theta = 1$. Let $E$ be the event that Algorithm~\ref{alg:svt-binary} outputs $1$ for the first $k/2$ queries in $Q$, and $0$ for the remaining $k/2$ queries. In addition, let $\Pr[D\rightarrow E]$ denote the probability that $E$ occurs when the input dataset is $D$. If Algorithm~\ref{alg:svt-binary} satisfies $\e$-differential privacy, then
\begin{equation*}
\frac{\Pr[D_1 \rightarrow E]}{\Pr[D_2\rightarrow E]} \le e^{\e}, \qquad \frac{\Pr[D_2 \rightarrow E]}{\Pr[D_3\rightarrow E]} \le e^{\e}.
\end{equation*}
This indicates that
\begin{equation} \label{eqn:svt-proof-1}
\frac{\Pr[D_1 \rightarrow E]}{\Pr[D_3\rightarrow E]} \:\: = \:\: \frac{\Pr[D_1 \rightarrow E]}{\Pr[D_2\rightarrow E]} \cdot \frac{\Pr[D_2 \rightarrow E]}{\Pr[D_3\rightarrow E]} \:\: \le \:\: e^{2\e}.
\end{equation}
In what follows, we prove the lemma by showing that Equation~\eqref{eqn:svt-proof-1} does not hold when $\lambda \le k/4\e$.

Recall that Algorithm~\ref{alg:svt-binary} generates a noisy threshold $\hat{\theta}$, and outputs $1$ for a query $q$ only when its noisy answer $\hat{q}(D)$ is larger than $\hat{\theta}$. Therefore,
\begin{small}
\begin{align}
& \frac{\Pr[D_1  \rightarrow E]}{\Pr[D_3\rightarrow E]} \nonumber \\
& = \frac{\int_{-\infty}^\infty \Pr[\hat{\theta} = x] \cdot \big(\Pr[\hat{q_a}(D_1)>x] \cdot \Pr[\hat{q_b}(D_1)\le x]\big)^{\frac{k}{2}} dx}
{\int_{-\infty}^\infty \Pr[\hat{\theta} = x] \cdot \big(\Pr[\hat{q_a}(D_3)>x] \cdot \Pr[\hat{q_b}(D_3)\le x]\big)^{\frac{k}{2}} dx} \nonumber \\
& = \frac{\int_{-\infty}^\infty \Pr[\hat{\theta} = x] \cdot \big(\Pr[\Lap(\lambda)>x-1] \cdot \Pr[\Lap(\lambda)\le x-1]\big)^{\frac{k}{2}} dx}
{\int_{-\infty}^\infty \Pr[\hat{\theta} = x] \cdot \big(\Pr[\Lap(\lambda)>x] \cdot \Pr[\Lap(\lambda)\le x-2]\big)^{\frac{k}{2}} dx} \nonumber
\end{align}
\end{small}

\vspace{-1mm}
\noindent
Consider any $x \in (-\infty, +\infty)$. If $x > 1$, then
\begin{equation*}
\Pr[\Lap(\lambda) > x-1] \:\: = \:\: \frac{1}{2}e^\frac{1-x}{\lambda} \:\: = \:\: e^\frac{1}{\lambda} \cdot \Pr[\Lap(\lambda) > x],
\end{equation*}

\noindent
and $\Pr[\Lap(\lambda) \le x-1] > \Pr[\Lap(\lambda) \le x-2]$. This leads to

\vspace{-3mm}
\begin{align} \label{eqn:svt-proof-3}
& \Pr[\Lap(\lambda)>x-1] \cdot \Pr[\Lap(\lambda)\le x-1] \nonumber \\
& \ge \; e^\frac{1}{\lambda} \cdot \Pr[\Lap(\lambda)>x] \cdot \Pr[\Lap(\lambda)\le x-2]
\end{align}
Meanwhile, if $x \le 1$, then
\begin{equation*}
\Pr[\Lap(\lambda) \le x-1] \:\: = \:\: \frac{1}{2}e^\frac{x-1}{\lambda} \:\: = \:\: e^\frac{1}{\lambda} \cdot \Pr[\Lap(\lambda) \le x - 2],
\end{equation*}
and $\Pr[\Lap(\lambda) > x-1] > \Pr[\Lap(\lambda) > x]$. In that case, Equation~\eqref{eqn:svt-proof-3} still holds.

Given that Equation~\eqref{eqn:svt-proof-3} holds for all $x \in (-\infty, \infty)$, we have

\vspace{-2mm}
\noindent
\begin{small}
\begin{align*}
& \frac{\Pr[D_1  \rightarrow E]}{\Pr[D_3\rightarrow E]} \nonumber \\
& = \frac{\int_{-\infty}^\infty \Pr[\hat{\theta} = x] \cdot \big(\Pr[\Lap(\lambda)>x-1] \cdot \Pr[\Lap(\lambda)\le x-1]\big)^{\frac{k}{2}} dx}
{\int_{-\infty}^\infty \Pr[\hat{\theta} = x] \cdot \big(\Pr[\Lap(\lambda)>x] \cdot \Pr[\Lap(\lambda)\le x-2]\big)^{\frac{k}{2}} dx} \nonumber \\
& > \left(e^\frac{1}{\lambda}\right)^\frac{k}{2} = e^\frac{k}{2\lambda}.
\end{align*}
\end{small}

\vspace{-2mm}
\noindent
Therefore, $\frac{\Pr[D_1  \rightarrow E]}{\Pr[D_3\rightarrow E]} > e^{2\e}$ when $\lambda \le k/4\e$, which proves the lemma. \done

\para{Proof of Lemma~\ref{lmm:svt-improved}.}
Let $D$ and $D'$ be any two neighboring databases and $E=\left\{o_1, o_2, \ldots \right\}$ be any output returned by Algorithm~\ref{alg:svt-improved}. To prove the lemma, we shall show that $\frac{\Pr[D  \rightarrow E]}{\Pr[D'\rightarrow E]}$ is always upper bounded by $e^\e$ when $\lambda\ge 2/\e$.

Let $\mathbf{1}$ (resp. $\mathbf{0}$) denote the set of queries whose corresponding output in $E$ equals $1$ (resp. $0$). As the algorithm terminates when the counter of the number of $o_i=1$ reaches $t$,  we know that the size of $\mathbf{1}$ should be no larger than $t$.
Let $f(x)$  be the probability that the Laplace noise $\Lap(t\cdot\lambda)$ is larger than  $x$, i.e., $f(x)=\Pr[\Lap(t\cdot\lambda) > x]$. Similarly, we define $g(x)$ as the probability of being no larger than $x$, that is, $g(x) = \Pr[\Lap(t\cdot\lambda) \le x]$. Then, we revise $\frac{\Pr[D  \rightarrow E]}{\Pr[D'\rightarrow E]}$ as follows:

\vspace{-2mm}
\begin{small}
\begin{align} \label{eqn:svt-proof-4}
& \frac{\Pr[D  \rightarrow E]}{\Pr[D'\rightarrow E]} \nonumber \\
& = \frac{\int_{-\infty}^\infty \Pr[\hat{\theta} = x] \cdot \prod_{q\in \mathbf{1}}\Pr[\hat{q}(D) > x] \cdot \prod_{q\in \mathbf{0}}\Pr[\hat{q}(D) \le x]dx}
{\int_{-\infty}^\infty \Pr[\hat{\theta} = x] \cdot \prod_{q\in \mathbf{1}}\Pr[\hat{q}(D') > x] \cdot \prod_{q\in \mathbf{0}}\Pr[\hat{q}(D') \le x]dx} \nonumber \\
& = \frac{\int_{-\infty}^\infty \Pr[\hat{\theta} = x] \cdot \prod_{q\in \mathbf{1}}f (x - q(D)) \cdot \prod_{q\in \mathbf{0}} g(x-q(D)) dx}
{\int_{-\infty}^\infty \Pr[\hat{\theta} = x] \cdot \prod_{q\in \mathbf{1}}f (x - q(D')) \cdot \prod_{q\in \mathbf{0}} g(x-q(D')) dx}.
\end{align}
\end{small}

\vspace{-3mm}
To figure out the upper bound of Equation~\eqref{eqn:svt-proof-4}, we first consider an easier case when $D$ is obtained by adding a tuple into $D'$. This implies that $q(D')\le q(D) \le q(D')+1$ holds for any counting query $q \in Q$. Given $q(D')\le q(D)$ and the fact that $g(x)$ is a monotonically increasing function of $x$, we have $g(x-q(D))\le g(x-q(D'))$. Moreover, we can remove the common factor $\Pr[\hat{\theta}=x]$ from both the numerator and denominator of \eqref{eqn:svt-proof-4}. Thus, the equation is upper bounded by:
\begin{equation}
\frac{\Pr[D  \rightarrow E]}{\Pr[D'\rightarrow E]} \le \frac{\int_{-\infty}^\infty \prod_{q\in \mathbf{1}}f (x - q(D))  dx}
{\int_{-\infty}^\infty \prod_{q\in \mathbf{1}}f (x - q(D'))  dx}. \nonumber
\end{equation}
Since $q(D)\le q(D')+1$ and $f(x)$ is monotonically decreasing on $x$, it is easy to have $f(x-q(D))\le f(x-q(D')-1)$. Combined with the property of Laplace noise, we have
$$
f(x-q(D))\le f(x-q(D')-1)\le e^{\frac{1}{t\lambda}} \cdot f(x-q(D')),
$$
for any query $q\in \mathbf{1}$.
Furthermore, given the fact that the size of $\mathbf{1}$ is at most $t$, \eqref{eqn:svt-proof-4} is upper bounded by $\exp(t\cdot \frac{1}{t\lambda}) = \exp(\frac{1}{\lambda})$.\vspace{1mm}

Now we consider the other case when $D'$ is obtained by inserting a tuple into $D$. As $D'$ is a larger dataset, we have $q(D)\le q(D') \le q(D)+1$.
In this case, the $q\in \bf{0}$ terms could not be trivially eliminated, as $q(D)$ is no longer larger than $q(D')$.
Fortunately, we manage to prove by utilizing the inequality $q(D') \le q(D)+1$. In particular, we shift the x-axis by $1$ and replace $x$ with $x+1$ in the denominator of Equation~\eqref{eqn:svt-proof-4}:

\vspace{-3mm}
\begin{small}
\begin{align}
& \frac{\Pr[D  \rightarrow E]}{\Pr[D'\rightarrow E]} = \nonumber \\
&  \frac{\int_{-\infty}^\infty \Pr[\hat{\theta} = x] \cdot \prod_{q\in \mathbf{1}}f (x - q(D)) \cdot \prod_{q\in \mathbf{0}} g(x-q(D)) dx}
{\int_{-\infty}^\infty \Pr[\hat{\theta} = x+1] \cdot \prod_{q\in \mathbf{1}}f (x - q(D')+1) \cdot \prod_{q\in \mathbf{0}} g(x-q(D')+1) dx}. \nonumber
\end{align}
\end{small}

\vspace{-2mm}
\noindent
Combining $q(D')-1 \le q(D)$ with the monotonicity of $g$, we again eliminate the $q\in\bf{0}$ terms from the upper bound. For other parts of the expression, we have
$$
\Pr[\hat{\theta} = x] \le e^{\frac{1}{\lambda}} \cdot \Pr[\hat{\theta} = x+1],
$$
and
$$
f(x-q(D)) \le f(x-q(D')) \le e^{\frac{1}{t\lambda}} \cdot f(x-q(D')+1)
$$
for any query $q\in \bf{1}$.
Provided that the size of $\bf{1}$ is at most $t$, the expression is upper bounded by $\exp(\frac{1}{\lambda} + t\cdot \frac{1}{t\lambda}) = \exp(\frac{2}{\lambda})$. Therefore, the lemma is proved. \done

\edit{
\section{Additional Experiments} \label{appen:add-exp}

\begin{figure*}[!t]
\centering
\edit{
\begin{small}
\begin{tabular}{c@{\hspace{-0.5mm}}:c@{\hspace{-0.5mm}}:c@{\hspace{-0.5mm}}:c}
\multicolumn{4}{c}{ \vspace{-0mm}\hspace{-5mm} \includegraphics[height=4mm]{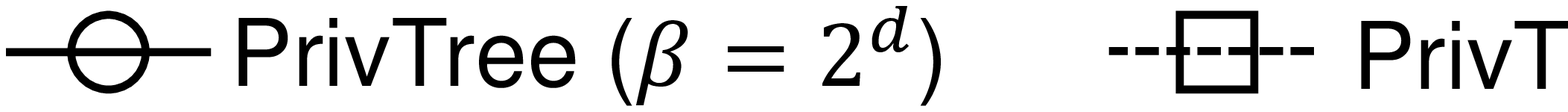}} \\

\hspace{-3mm}\includegraphics[height=31.5mm]{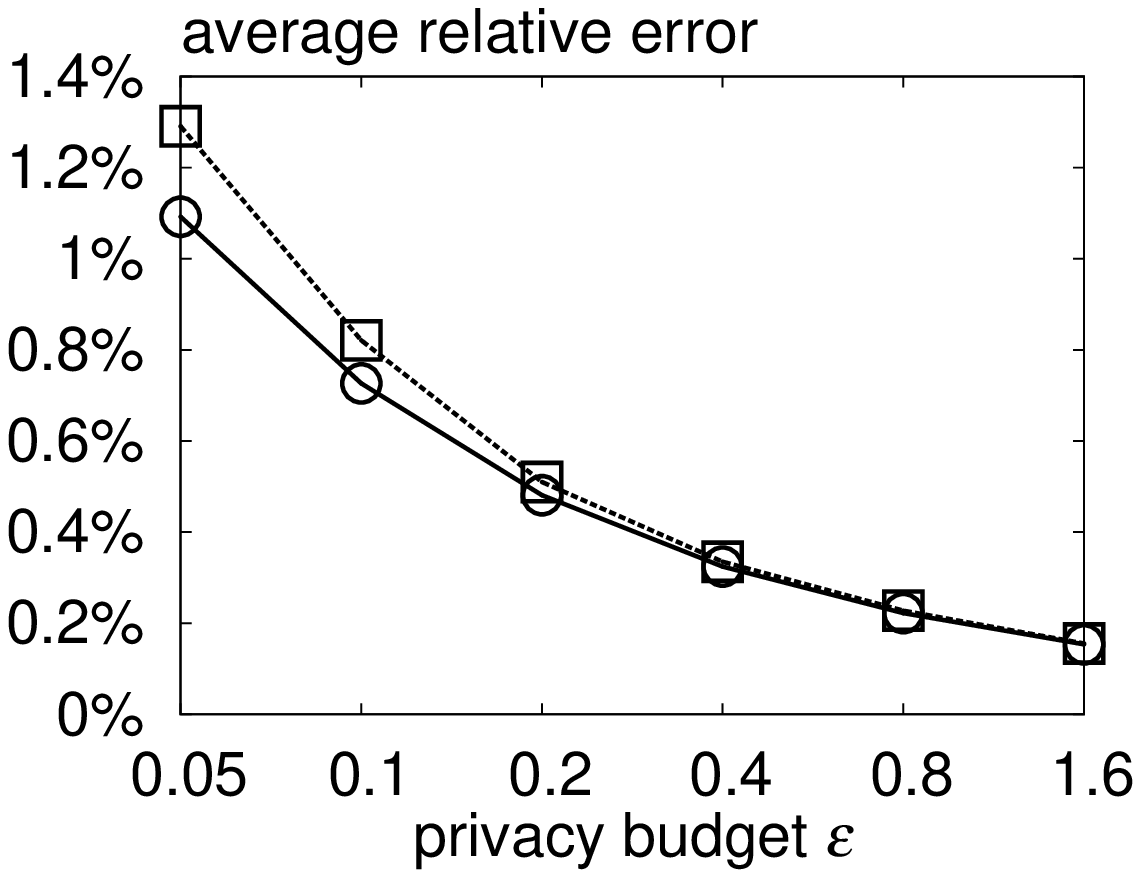}
&
\hspace{-1mm}\includegraphics[height=31.5mm]{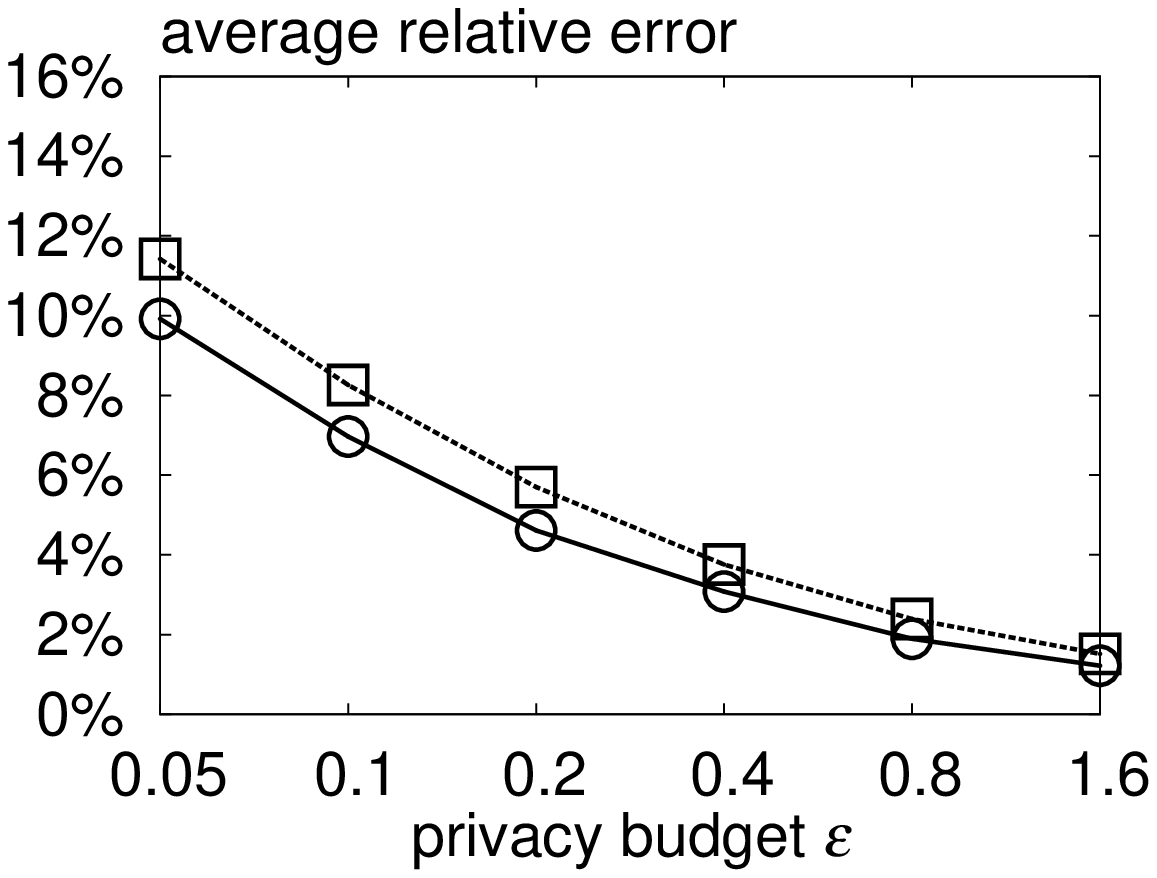}
&
\hspace{-1mm}\includegraphics[height=31.5mm]{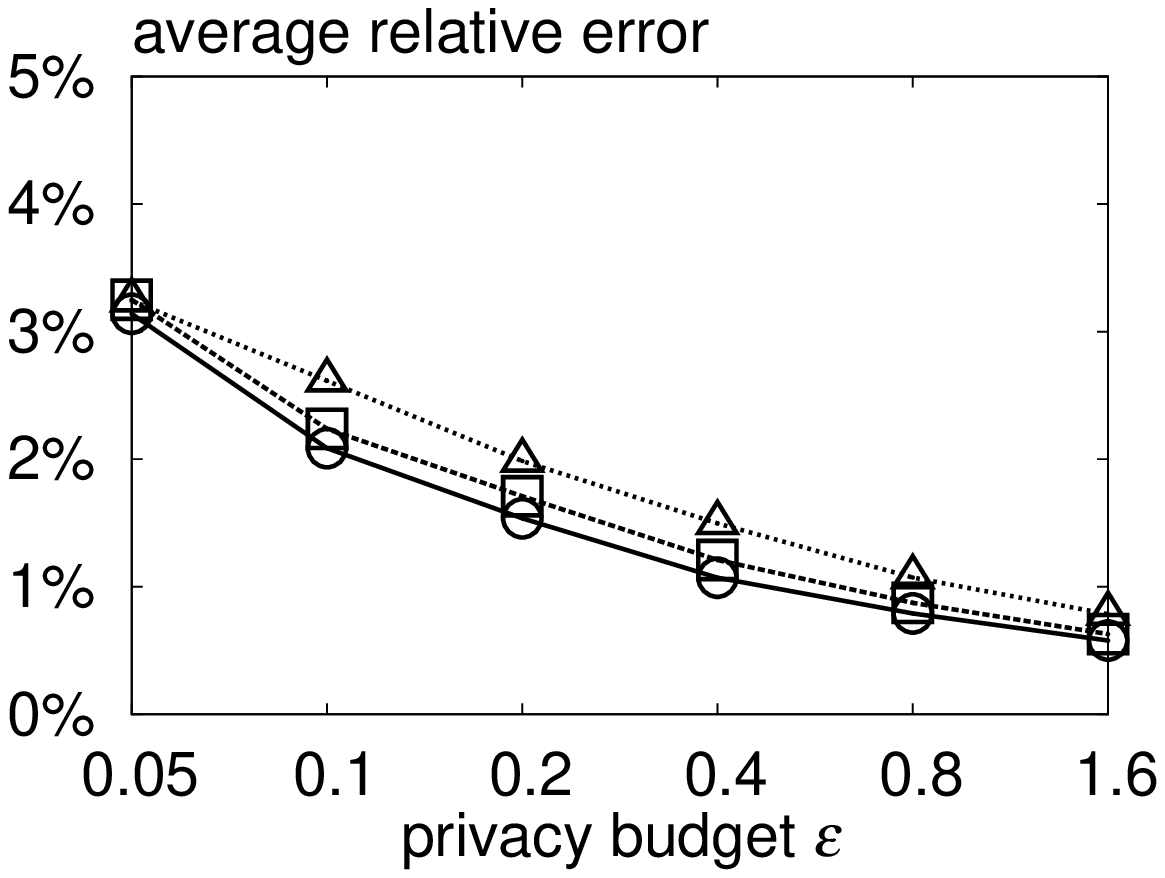}
&
\hspace{-1mm}\includegraphics[height=31.5mm]{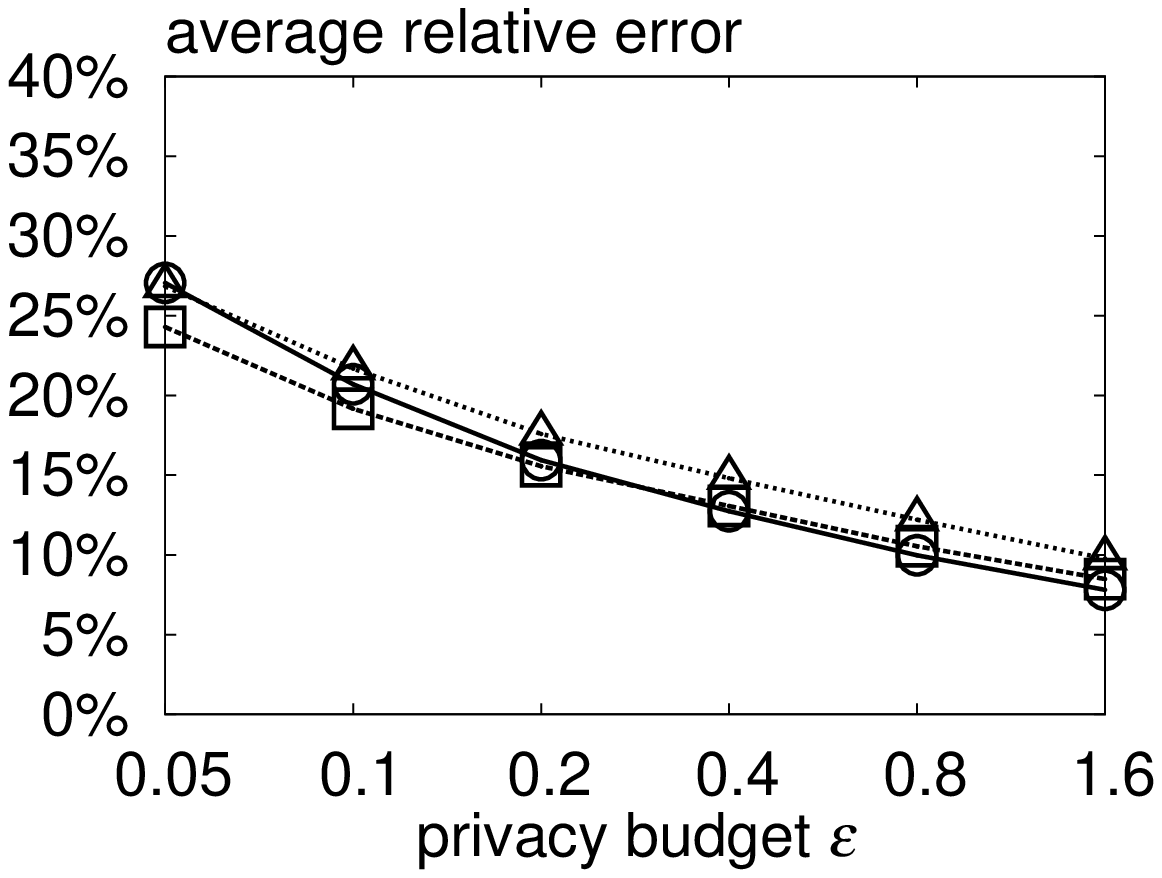} \\[0.8mm]

\hspace{-3mm} (a) {\sf road} - small queries.
&
\hspace{-2mm} (d) {\sf Gowalla} - small queries.
&
\hspace{-2mm} (g) {\sf NYC} - small queries.
&
\hspace{-2mm} (j) {\sf Beijing} - small queries. \\[2mm]

\hspace{-3mm}\includegraphics[height=31.5mm]{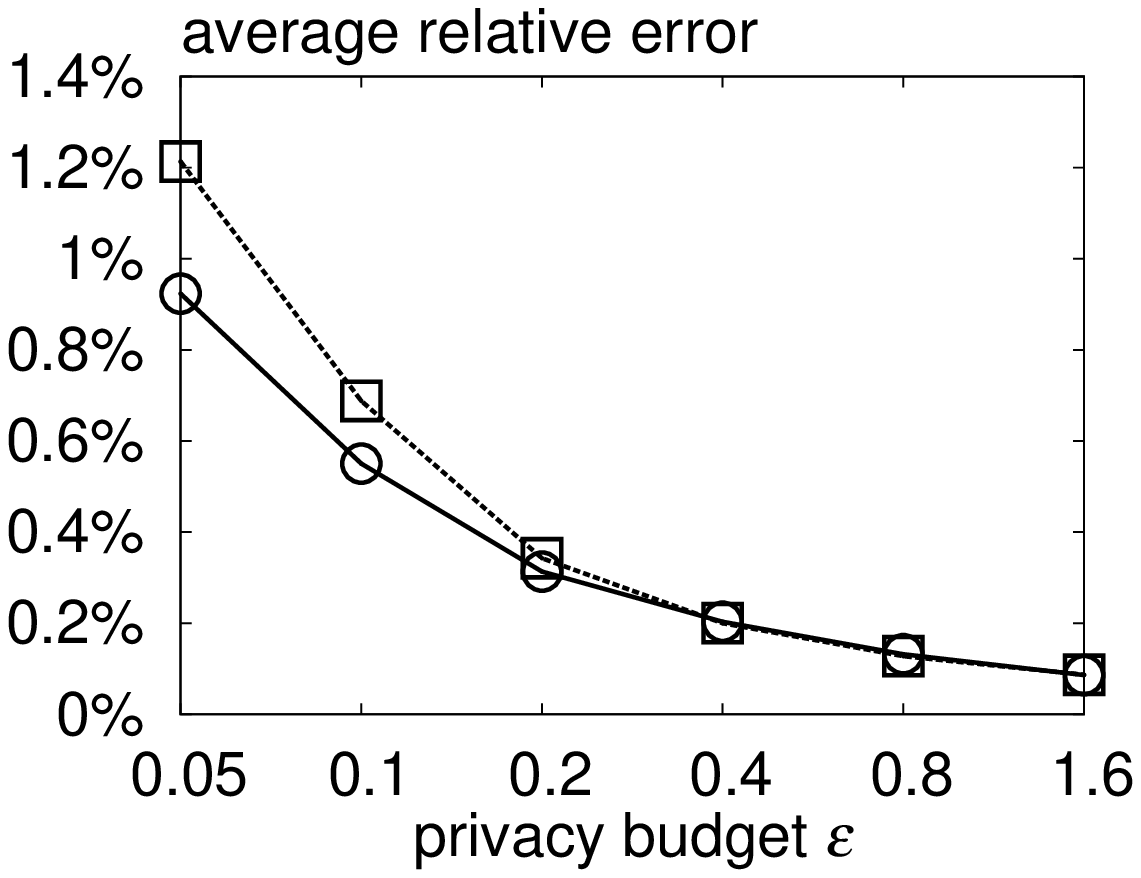}
&
\hspace{-2mm}\includegraphics[height=31.5mm]{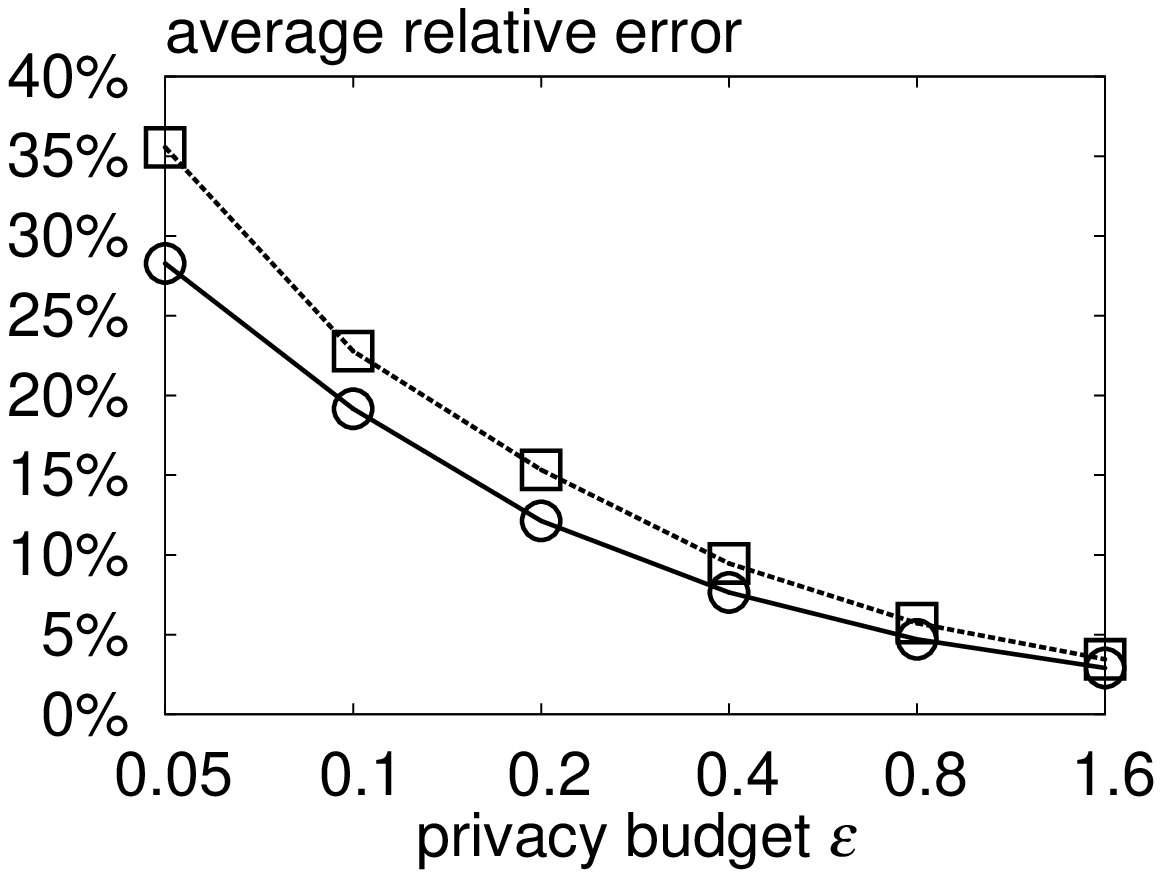}
&
\hspace{-2mm}\includegraphics[height=31.5mm]{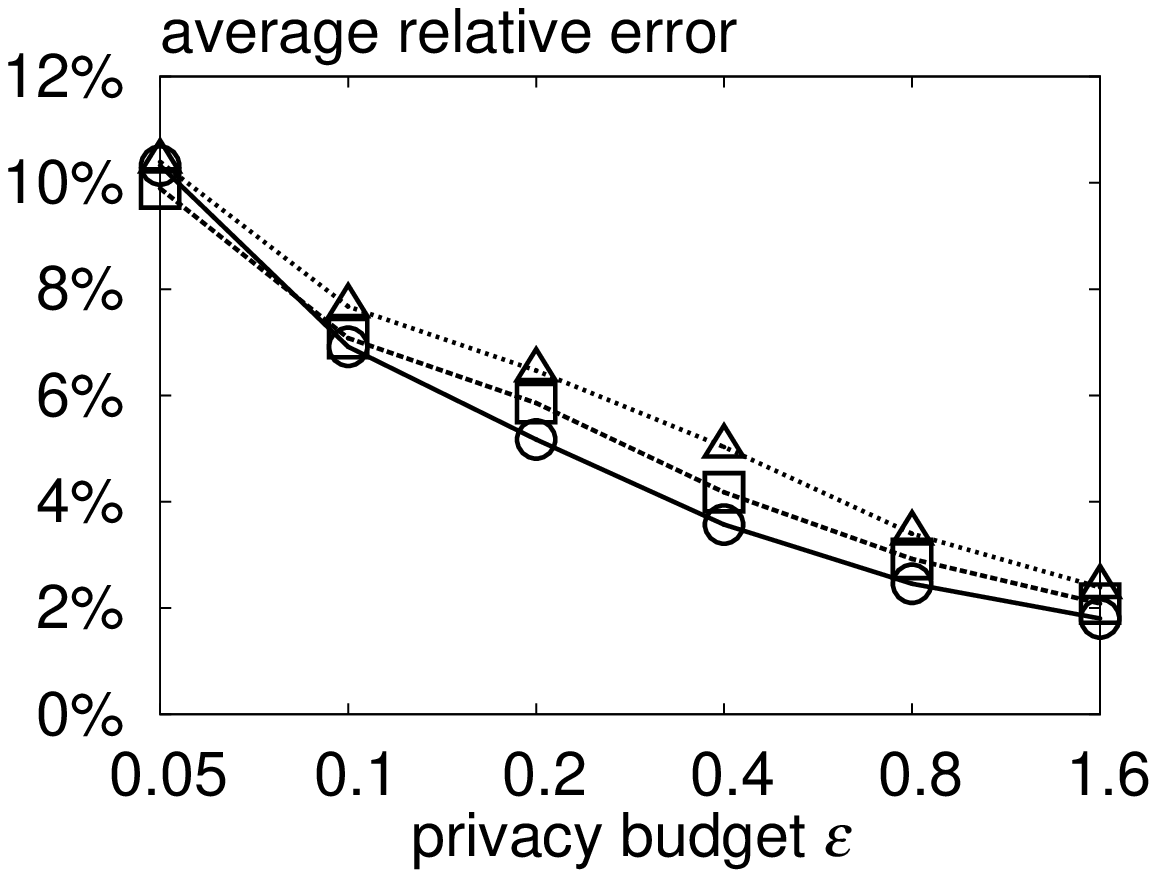}
&
\hspace{-2mm}\includegraphics[height=31.5mm]{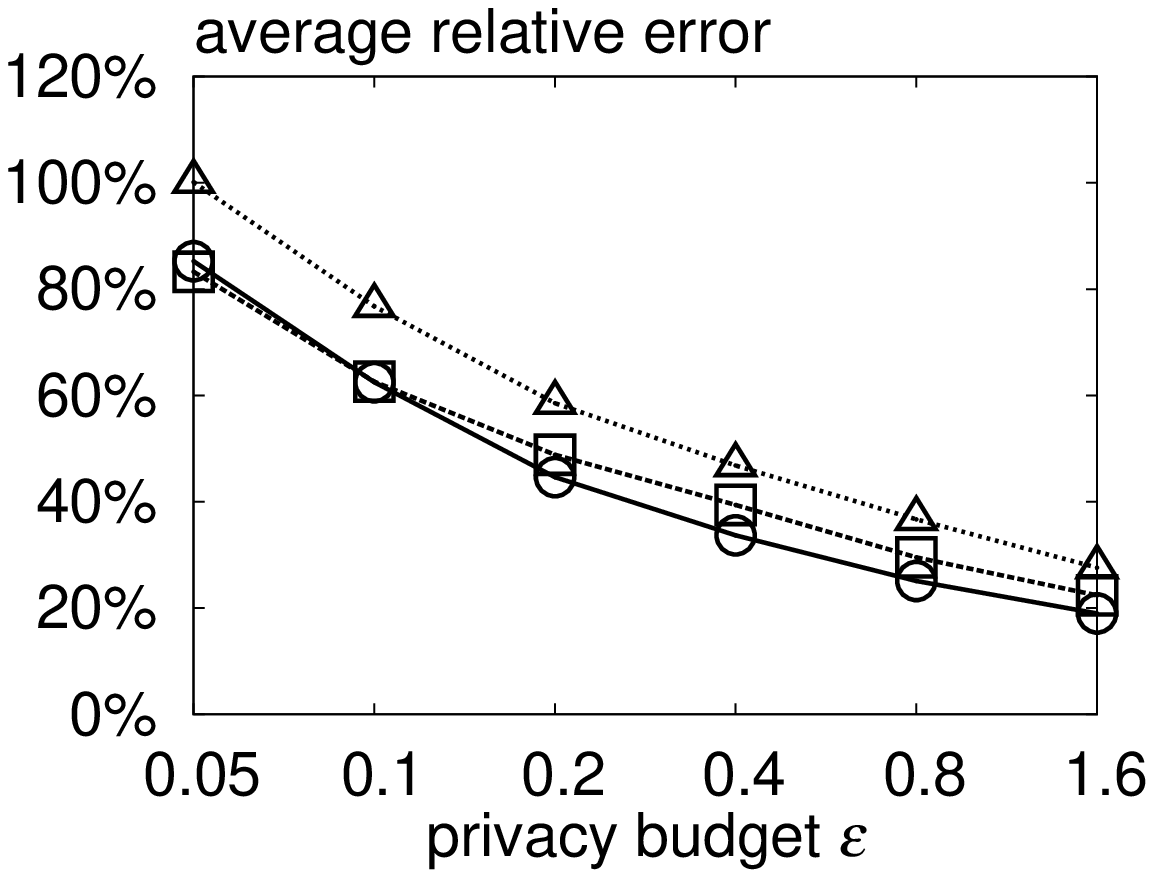}\\[0.8mm]

\hspace{-3mm} (b) {\sf road} - medium queries.
&
\hspace{-2mm} (e) {\sf Gowalla} - medium queries.
&
\hspace{-2mm}(h) {\sf NYC} - medium queries.
&
\hspace{-2mm}(k) {\sf Beijing} - medium queries. \\[2mm]

\hspace{-3mm}\includegraphics[height=31.5mm]{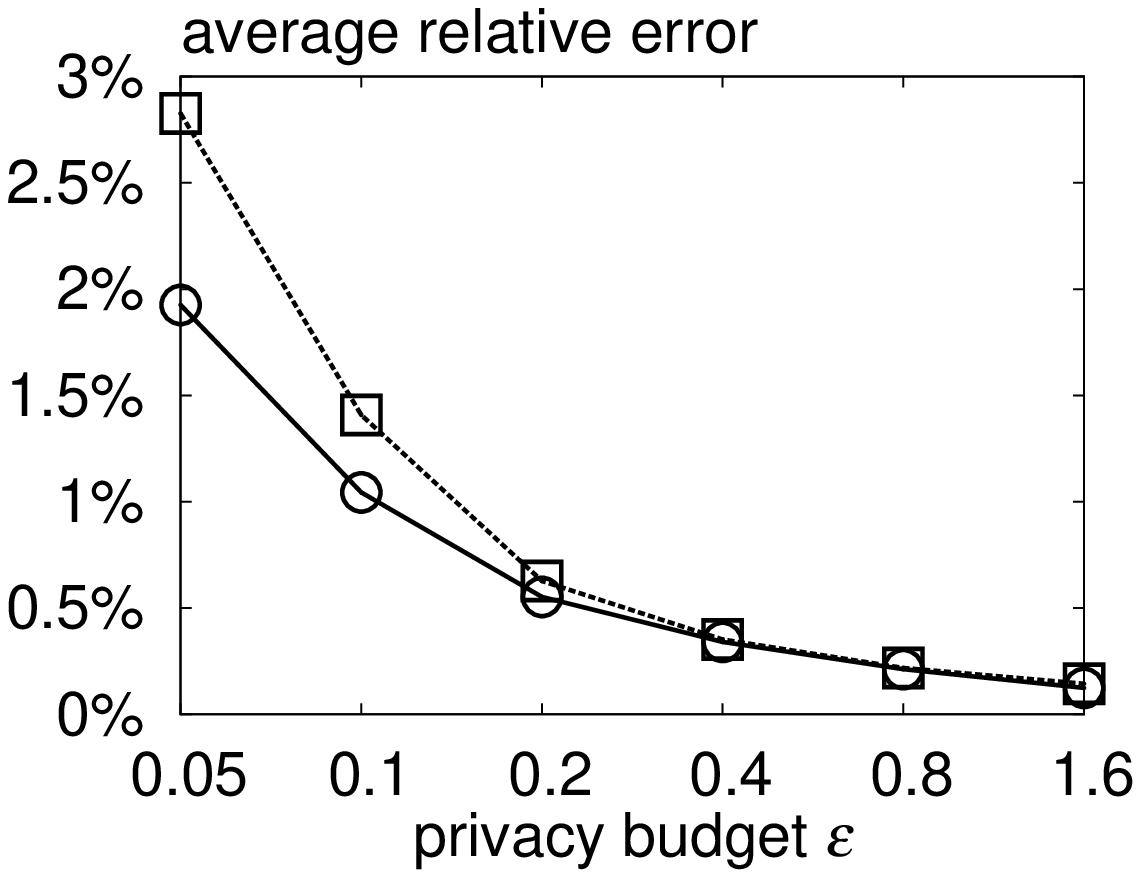}
&
\hspace{-2mm}\includegraphics[height=31.5mm]{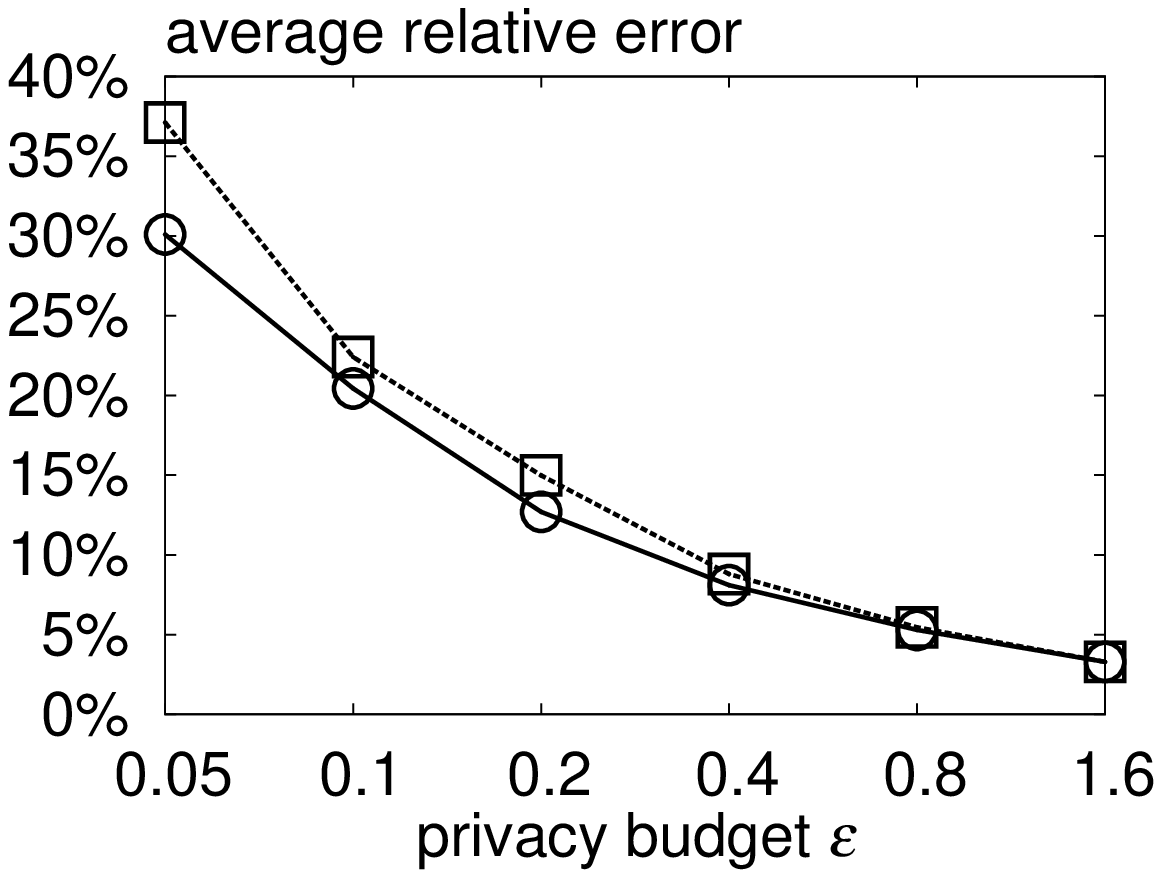}
&
\hspace{-2mm}\includegraphics[height=31.5mm]{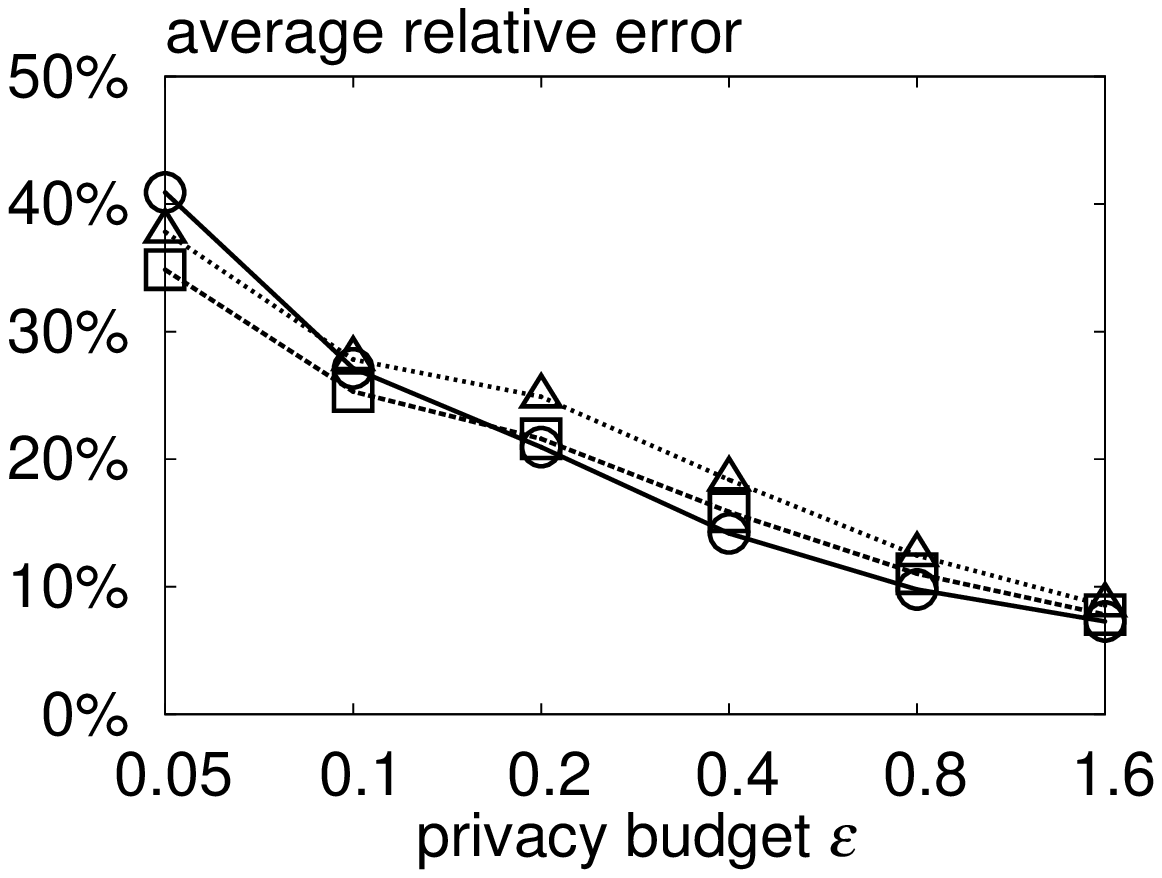}
&
\hspace{-2mm}\includegraphics[height=31.5mm]{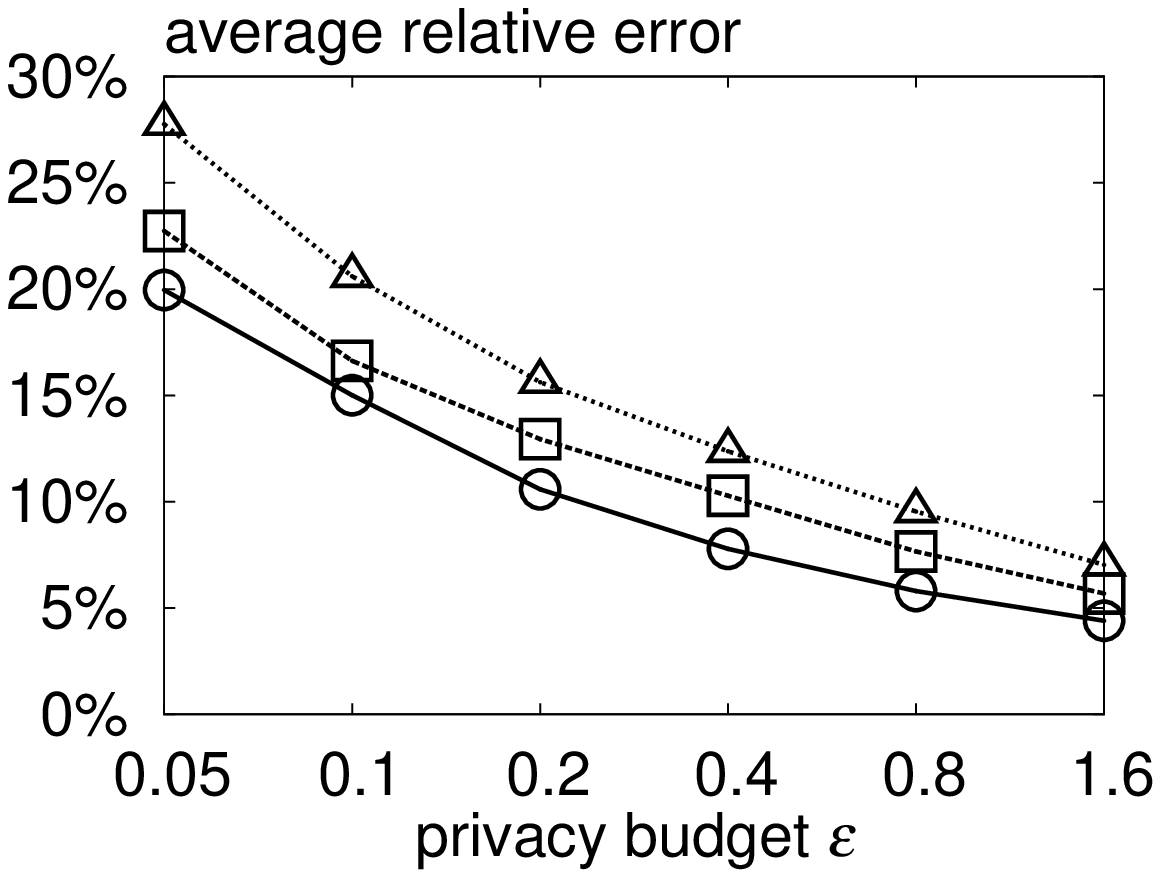} \\[0.8mm]

\hspace{-3mm} (c) {\sf road} - large queries.
&
\hspace{-2mm} (f) {\sf Gowalla} - large queries.
&
\hspace{-2mm}(i) {\sf NYC} - large queries.
&
\hspace{-2mm}(l) {\sf Beijing} - large queries.
\end{tabular}
\end{small}
\vspace{-2mm}
\caption{Impact of fanout on PrivTree.}
\label{fig:fanout}
\vspace{-2mm}
}
\end{figure*}

In this section, we evaluate the computation efficiency of \tree, the impact of $\beta$ (i.e., tree fanout) on the accuracy of \tree, as well as the effect of the tree height $h$ on the performance of four methods tested in our paper: UG~\cite{QYL13, SCLBJ15, QYL13hierarchy}, AG~\cite{QYL13}, Hierarchy\cite{QYL13hierarchy}, and N-gram~\cite{CAC12}. (We do not consider the other methods (i.e., \tree, DAWA~\cite{LHMW14}, and Privelet$^*$~\cite{XWG11}) in our experiments, because they do not take $h$ as input.)

Table~\ref{tbl:time} shows the processing time of \tree on each dataset (averaged over $100$ runs), with $\e$ varying from $0.05$ to $1.6$. The running time of \tree on {\sf road} and {\sf msnbc} are larger than that on the other datasets, since {\sf road} and {\sf msnbc} are larger in size than the others. In addition, the computation cost of \tree increases with $\e$. To understand this, recall that when \tree decides whether or not to split a node $v$, it first subtracts a bias term $\depth(v) \cdot \delta$ from the score of $v$, and then injects noise into the biased score, after which it splits $v$ if the noisy score is larger than the threshold $\theta$. As $\delta$ is inversely proportional to $\e$ (see Corollary~\ref{coro:alg:privacy}), the bias term increases when $\e$ decreases, in which case the noisy score of $v$ is less likely to be larger than $\theta$. Therefore, when $\e$ is small, \tree has lower probabilities to split nodes, which leads to a small running time.

\begin{table}[t]
\centering
\edit{
\caption{Running time of PrivTree (seconds).}\label{tbl:time}
\vspace{2mm}
\begin{tabular}{@{}l@{$\:$}|l@{$\:\:\:$}l@{$\;\;\;$}l@{$\;\;\;$}l@{$\;\;\;$}l@{$\;\;\;$}l}
\hline
{\bf Dataset} & $\e \!=\! 0.05$ & $\e \!=\! 0.1$ & $\e \!=\! 0.2$ & $\e \!=\! 0.4$ & $\e \!=\! 0.8$ & $\e \!=\! 1.6$ \\
\hline \hline
{\sf road} & $0.97$ & $1.15$ & $1.35$ & $1.61$ & $1.93$ & $2.52$ \\
\hline
{\sf Gowalla} & $0.044$ & $0.055$ & $0.073$ & $0.093$ & $0.12$ & $0.17$ \\
\hline
{\sf NYC} & $0.032$ & $0.040$ & $0.051$ & $0.072$ & $0.10$ & $0.15$ \\
\hline
{\sf Beijing} & $0.0085$ & $0.012$ & $0.013$ & $0.019$ & $0.030$ & $0.047$ \\
\hline
{\sf mooc} & $0.22$ & $0.26$ & $0.30$ & $0.35$ & $0.41$ & $0.46$ \\
\hline
{\sf msnbc} & $1.73$ & $2.03$ & $2.21$ & $2.50$ & $2.72$ & $3.05$ \\
\hline
\end{tabular}
}
\end{table}

Previously, in Section~\ref{sec:exp-spatial}, we evaluate the query accuracy of \tree on spatial data with its fanout $\beta$ set to $2^d$, where $d$ is the dataset dimensionality. In that case, whenever \tree splits a node $v$, the sub-domain $\dom(v)$ of $v$ is divided into $2^d$ parts by bisecting all dimensions of $\dom(v)$. Figure~\ref{fig:fanout} illustrates the results of the same experiments when $\beta$ varies. In particular, when we set $\beta = 2^i$ with $i < d$, \tree would split the dimensions of each node in a round robin fashion, with $i$ dimensions being bisected each time. Observe that, in general, the query error of \tree slightly increases when $\beta$ decreases. This is mainly due to the bias term $\depth(v) \cdot \delta$ that \tree subtracts from the score $c(v)$ of each node $v$, when it decides whether $v$ should be split. Specifically, a decreased $\beta$ increases the height of \tree's decomposition tree, in which case the nodes towards the leaf level of the tree would be given a larger bias term. In turn, the increased bias term renders it more difficult for \tree to correctly decide whether a node should be split, thus degrading the quality of \tree's output.

Nevertheless, on a few settings on {\sf NYC} and {\sf Beijing}, $\beta = 2^{d/2}$ entails smaller errors than $\beta = 2^d$. The reason is that, when $\beta$ is large, any incorrect decisions made by \tree in node splitting would have a more pronounced negative effect, e.g., a quadtree node with a small count would be divided into a larger number of child nodes, each of which would have an even smaller count that is likely to be overwhelmed by the noise subsequently added. The increased number of noise-dominated nodes would then lead to less accurate query answers, which explains why $\beta = 2^{d/2}$ sometimes outperforms $\beta = 2^d$. That said, the overall result in Figure~\ref{fig:fanout} indicates that $\beta = 2^d$ is still a preferable choice for \tree.
}

Our third set of experiments concerns UG~\cite{QYL13, SCLBJ15, QYL13hierarchy}, a method that partitions the data domain into $m^d$ grid cells of equal size (where $d$ is the dataset dimensionality and $m$ is a parameter), and releases a noisy count for each cell. Conceptually, UG can be regarded as a hierarchical decomposition method that (i) uses a decomposition tree $T$ with a fanout $2^d$ and a height $h = \log_2 m$, and (ii) releases noisy counts only for the leaves of $T$. \cite{SCLBJ15} suggests setting $m^d = (n\e/10)^{2d/(d+2)}$, where $n$ is the dataset cardinality. To evaluate this choice of the number of cells, we multiply it by a factor $r$, and evaluate the query accuracy of UG when it uses a grid of roughly $r \cdot m^d$ cells. In particular, we set the number of bins per dimension to $\lceil r^{1/d} \cdot m \rceil$.

Figure~\ref{fig:h:UG} shows the relative query errors of UG on the four spatial datasets, when $r$ and $\e$ vary. Evidently, there is no single value of $r$ that consistently outperforms the other values. However, in terms of overall performance, $r = 1$ is one of the best choices, which indicates that the setting of $m$ recommended in \cite{SCLBJ15} is near-optimal. Interestingly, the error of UG does not always change monotonically with $\e$, especially on {\sf NYC}. To explain, observe that $m = (n\e/10)^{2/(d+2)}$ varies with $\e$, i.e., the grid granularity of UG changes with $\e$. If the change in grid granularity happens to align the grid cells with the dense regions of the data domain $\Omega$ (i.e., there are not too many cells that cover both dense and sparse regions), then the published cell counts would provide a relatively good approximation of the underlying data. However, if the grid cells happen to be misaligned with the dense regions, then the resulting query error would be large. Therefore, the error of UG may fluctuate when $\e$ increases.

\begin{figure*}[!t]
\centering
\begin{small}
\begin{tabular}{c@{\hspace{-0.5mm}}:c@{\hspace{-0.5mm}}:c@{\hspace{-0.5mm}}:c}
\multicolumn{4}{c}{ \vspace{-0mm}\hspace{-5mm} \includegraphics[height=3.6mm]{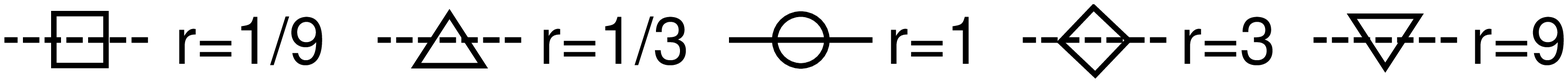}} \\

\hspace{-3mm}\includegraphics[height=31.5mm]{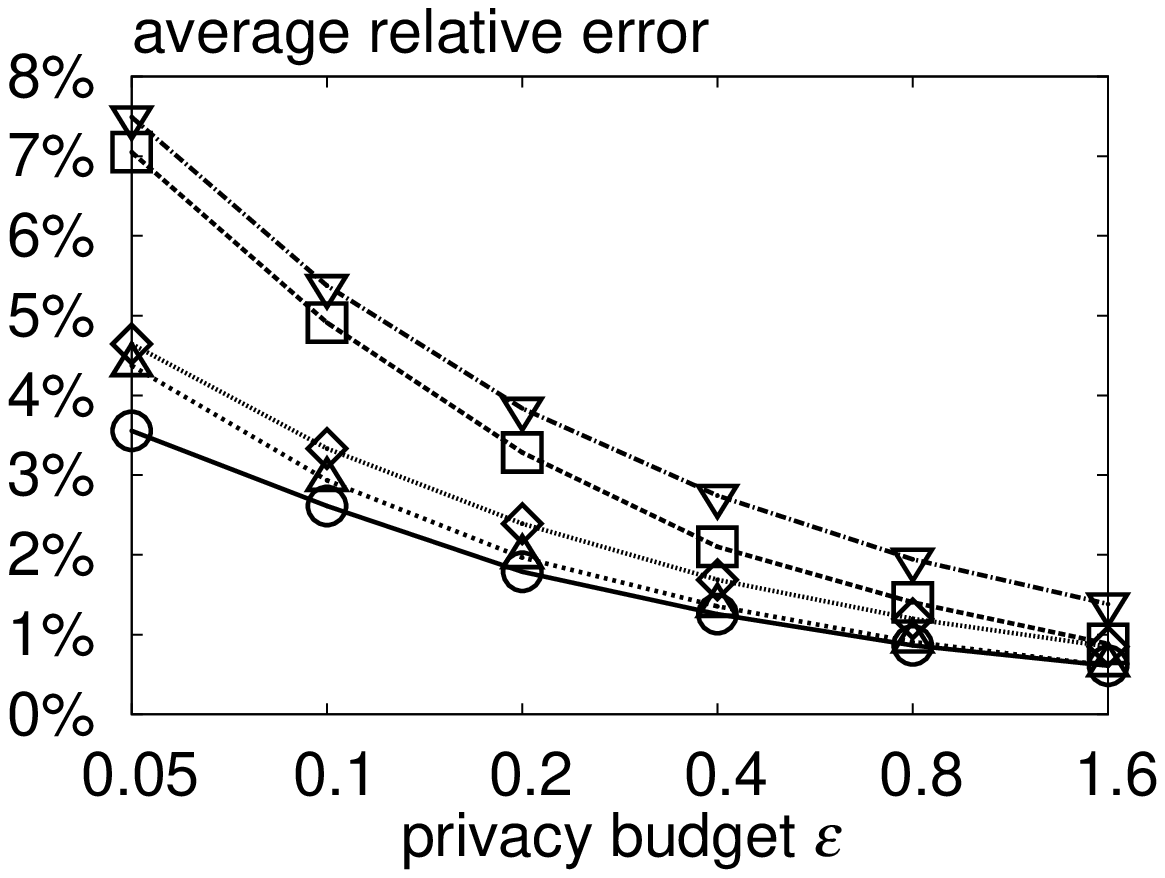}
&
\hspace{-1mm}\includegraphics[height=31.5mm]{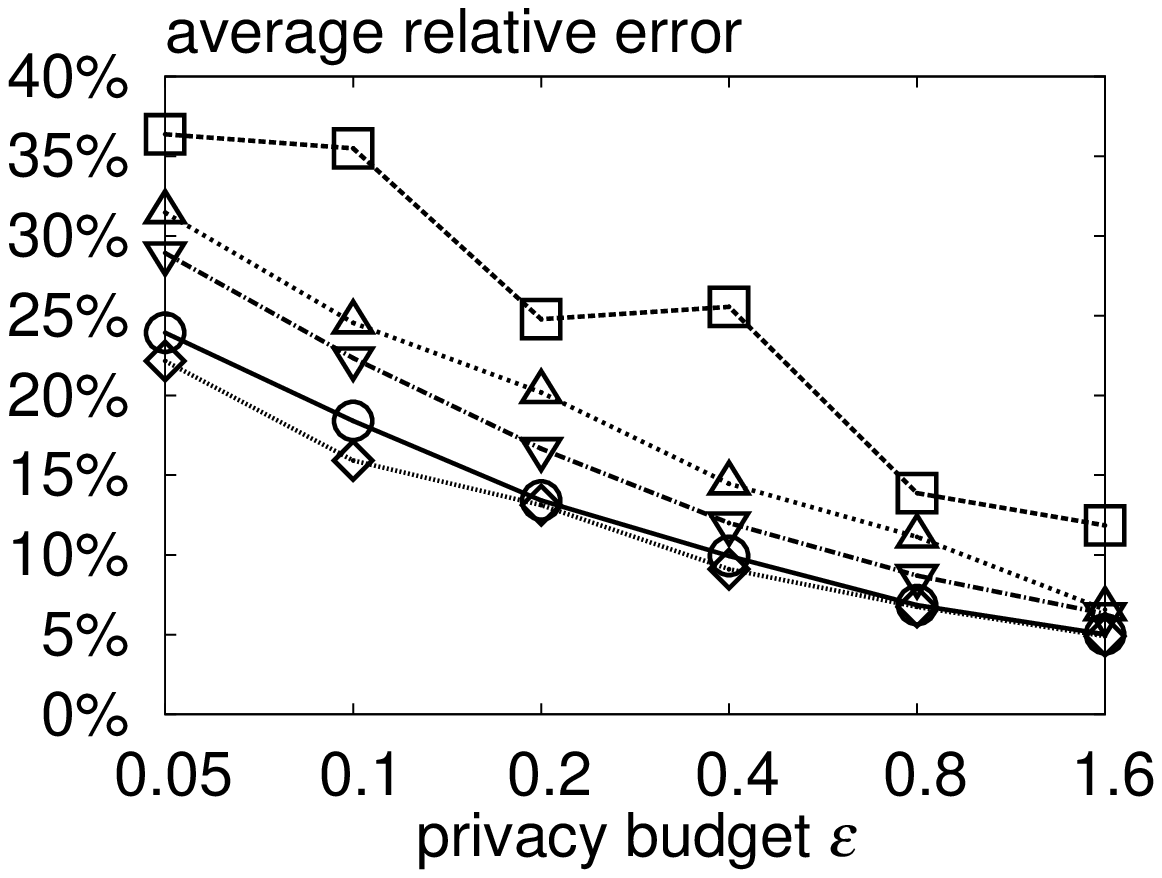}
&
\hspace{-1mm}\includegraphics[height=31.5mm]{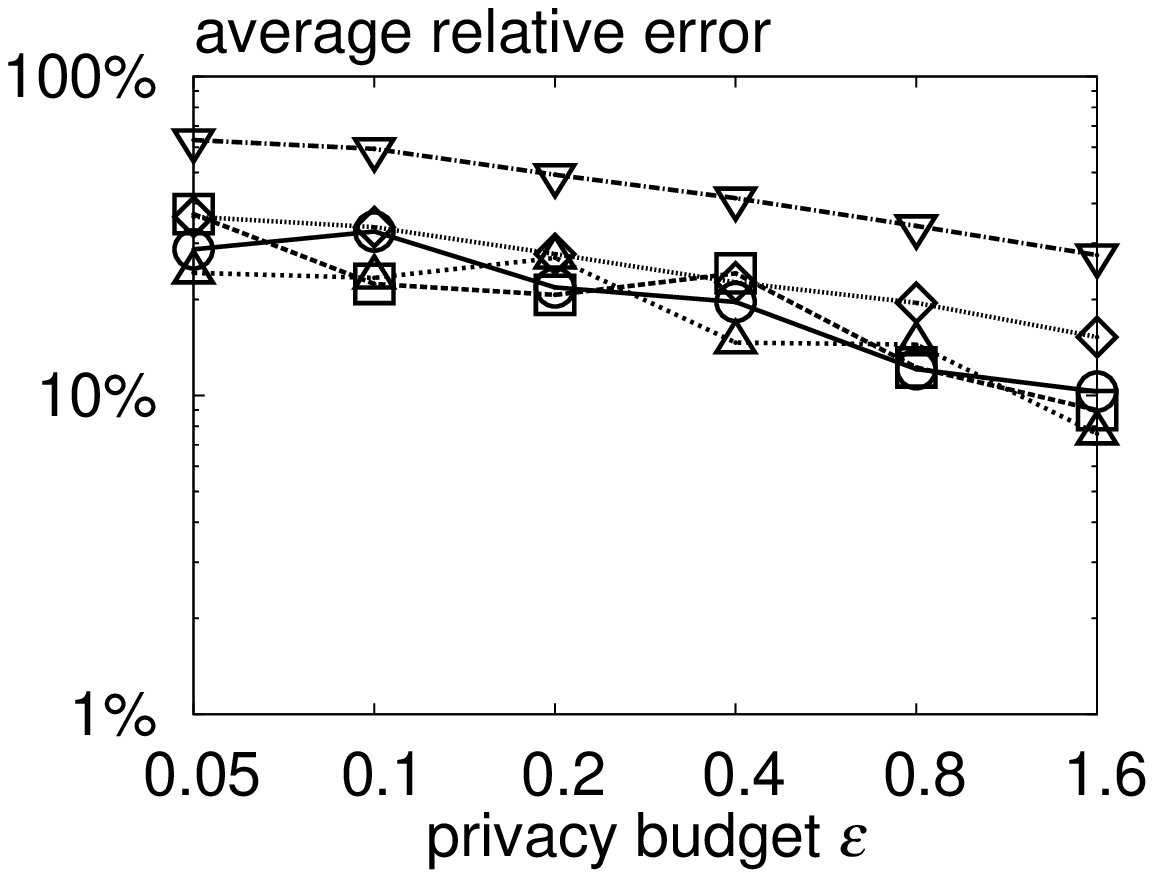}
&
\hspace{-1mm}\includegraphics[height=31.5mm]{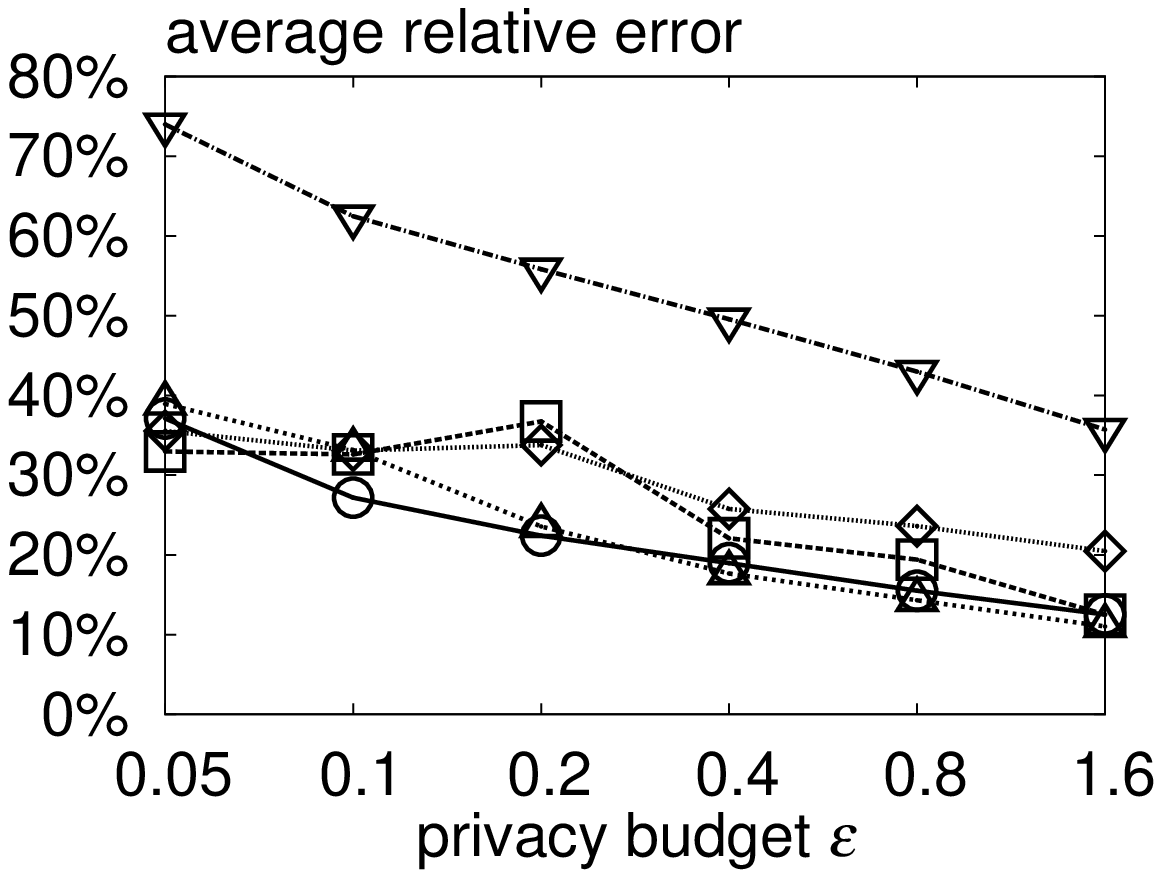} \\[0.8mm]

\hspace{-3mm} (a) {\sf road} - small queries.
&
\hspace{-2mm} (d) {\sf Gowalla} - small queries.
&
\hspace{-2mm} (g) {\sf NYC} - small queries.
&
\hspace{-2mm} (j) {\sf Beijing} - small queries. \\[2mm]

\hspace{-3mm}\includegraphics[height=31.5mm]{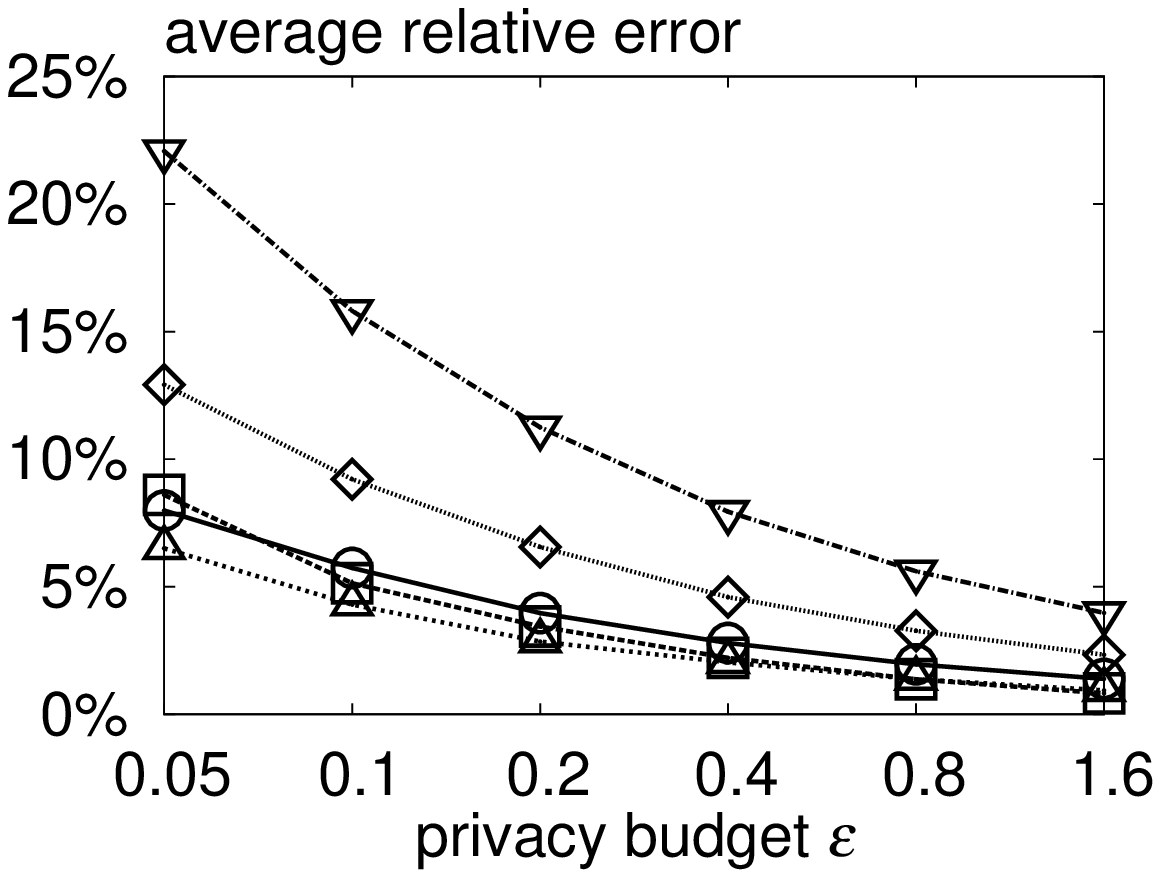}
&
\hspace{-2mm}\includegraphics[height=31.5mm]{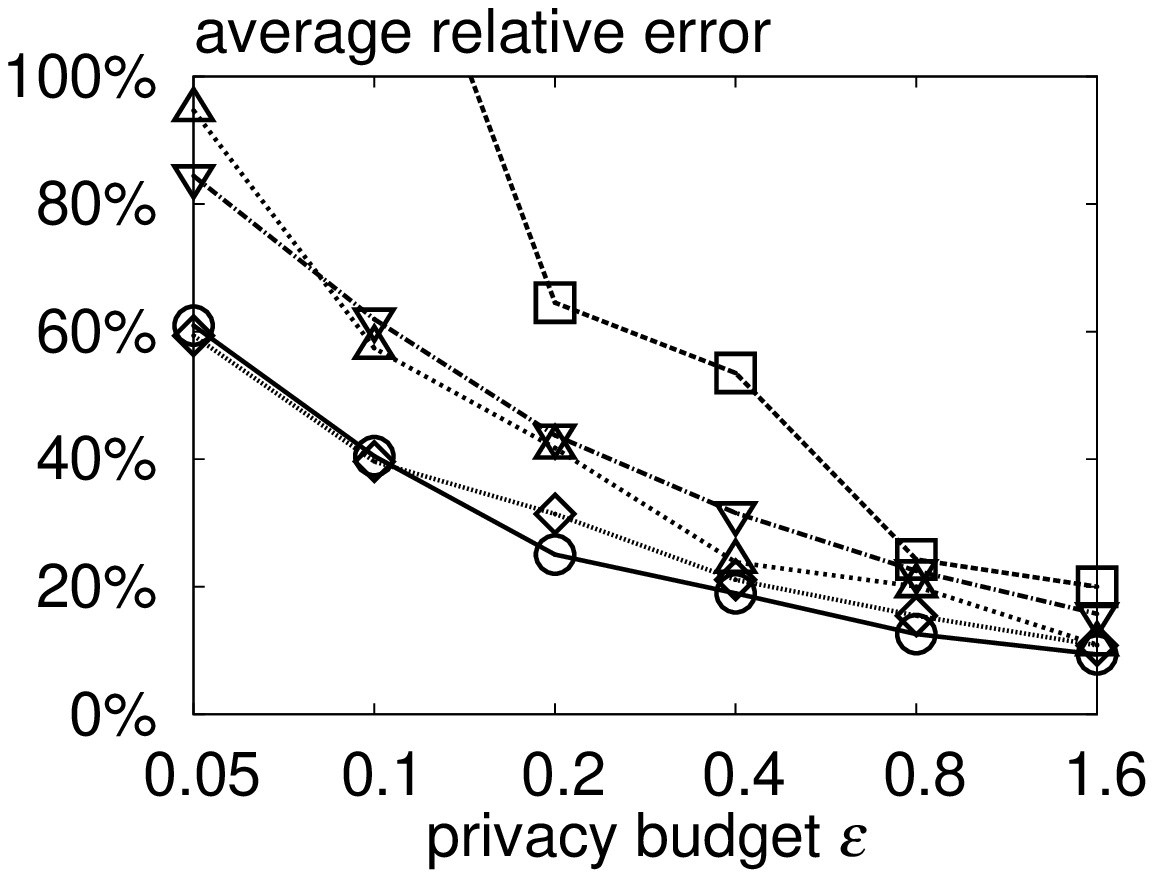}
&
\hspace{-2mm}\includegraphics[height=31.5mm]{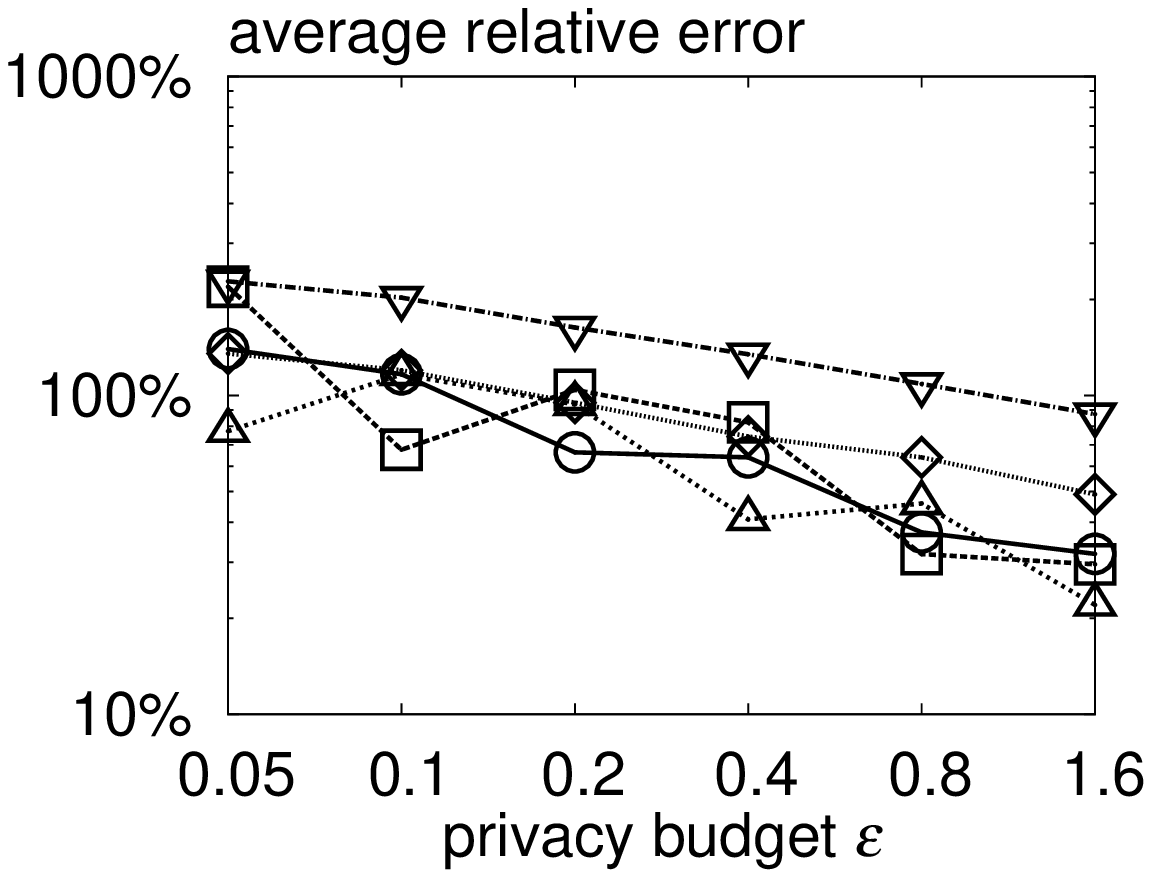}
&
\hspace{-2mm}\includegraphics[height=31.5mm]{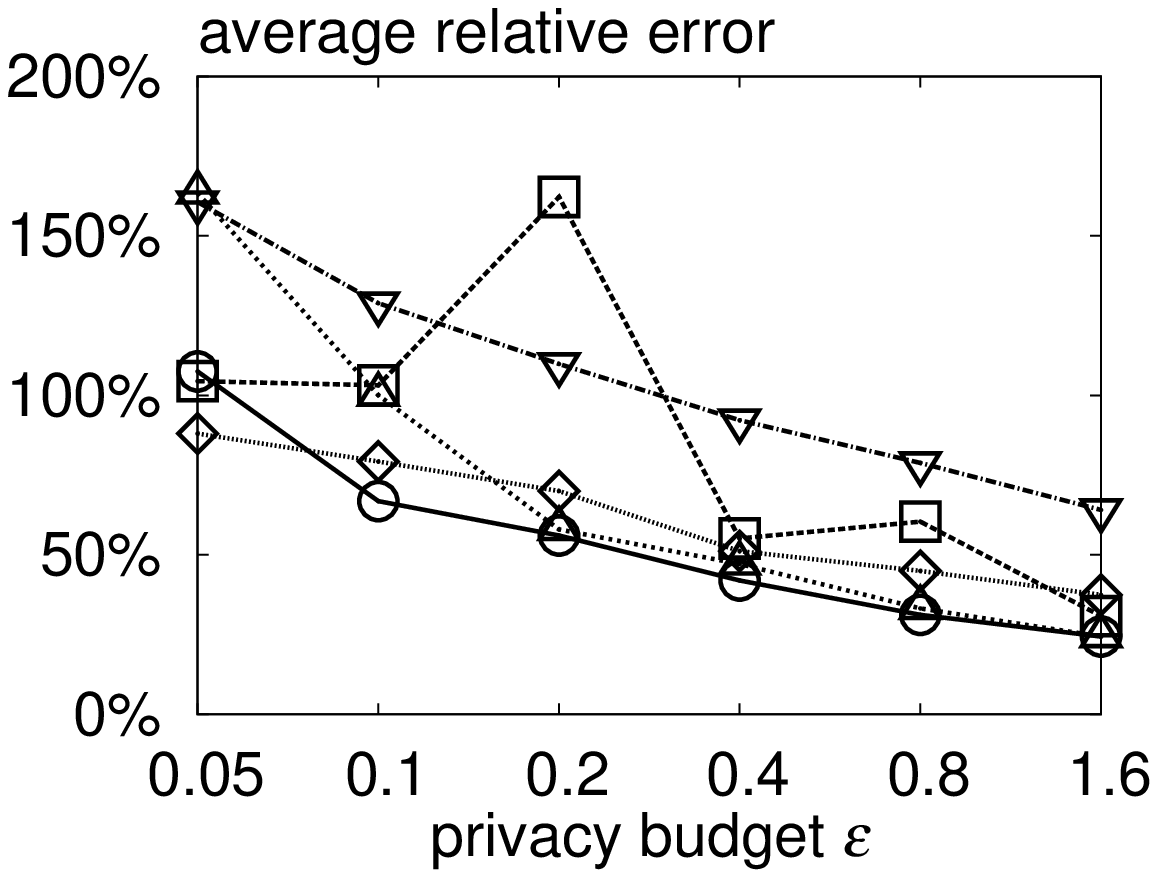}\\[0.8mm]

\hspace{-3mm} (b) {\sf road} - medium queries.
&
\hspace{-2mm} (e) {\sf Gowalla} - medium queries.
&
\hspace{-2mm}(h) {\sf NYC} - medium queries.
&
\hspace{-2mm}(k) {\sf Beijing} - medium queries. \\[2mm]

\hspace{-3mm}\includegraphics[height=31.5mm]{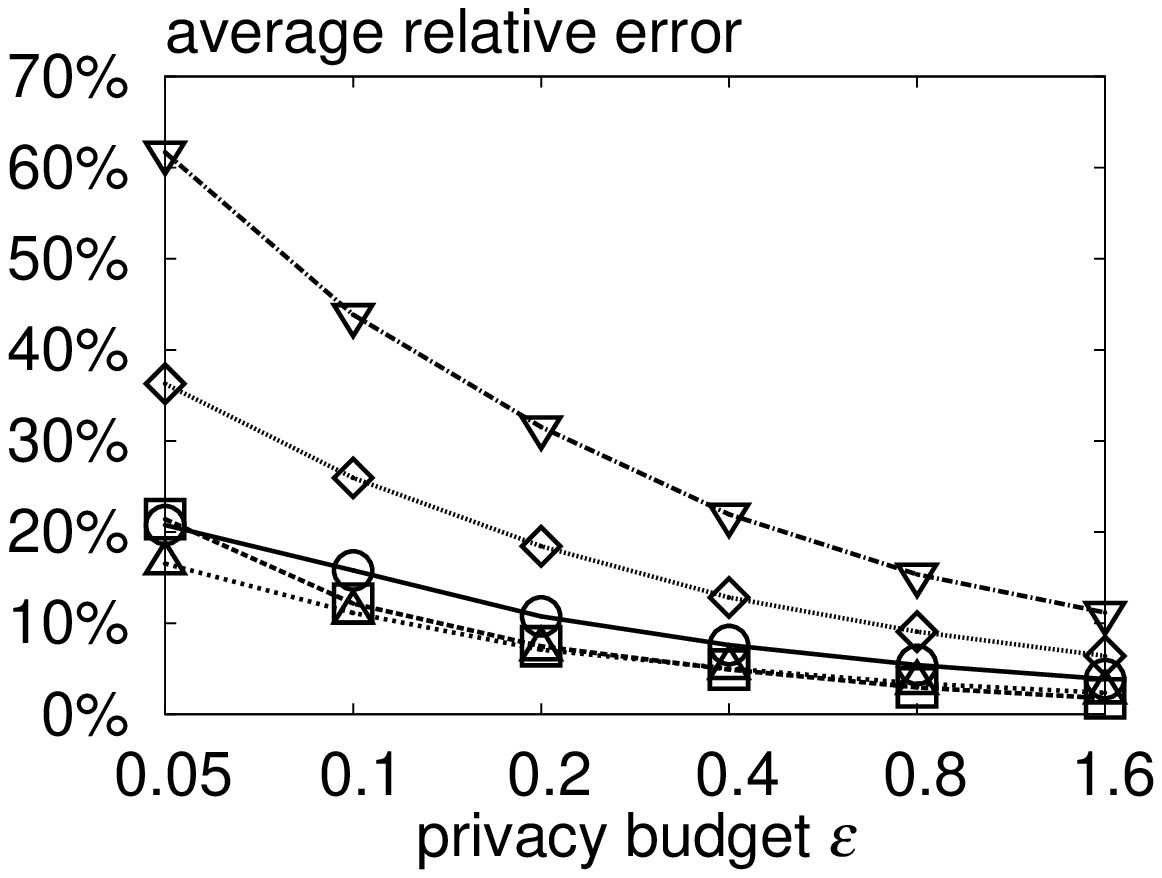}
&
\hspace{-2mm}\includegraphics[height=31.5mm]{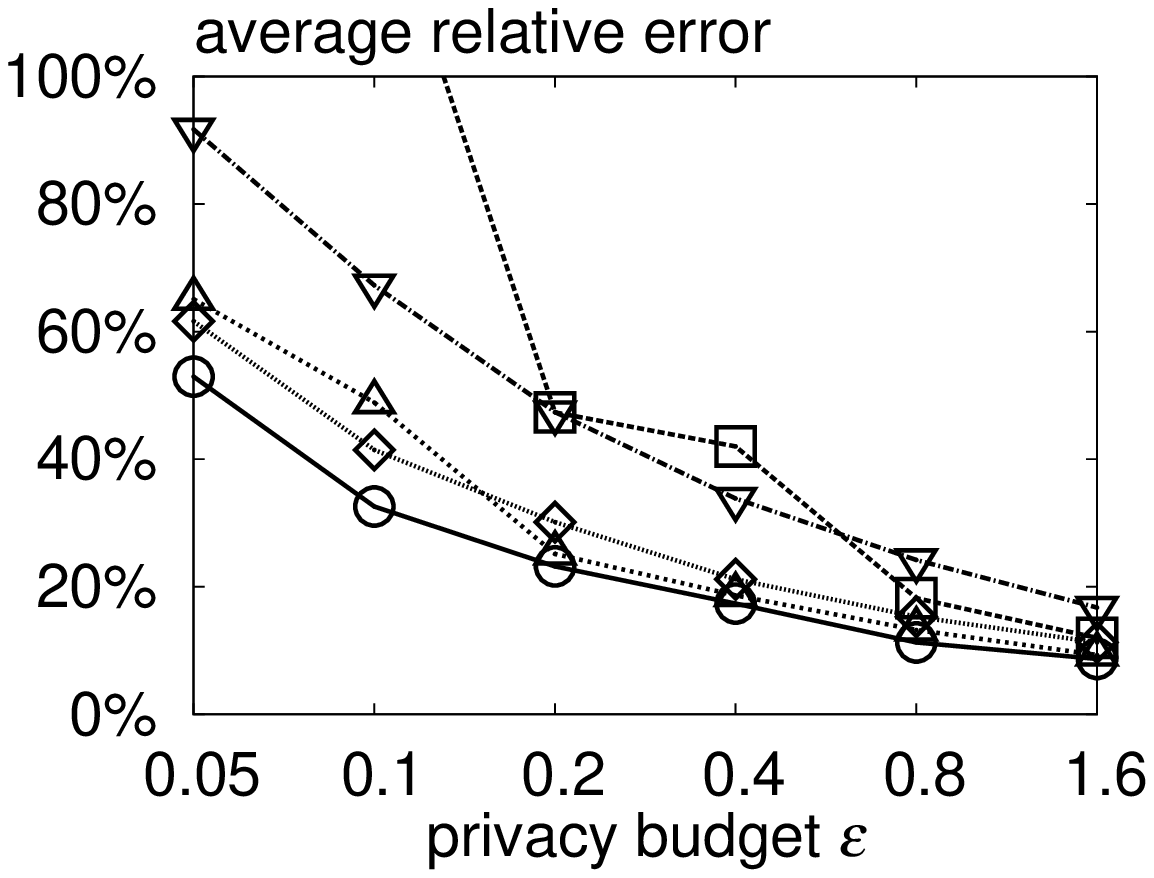}
&
\hspace{-2mm}\includegraphics[height=31.5mm]{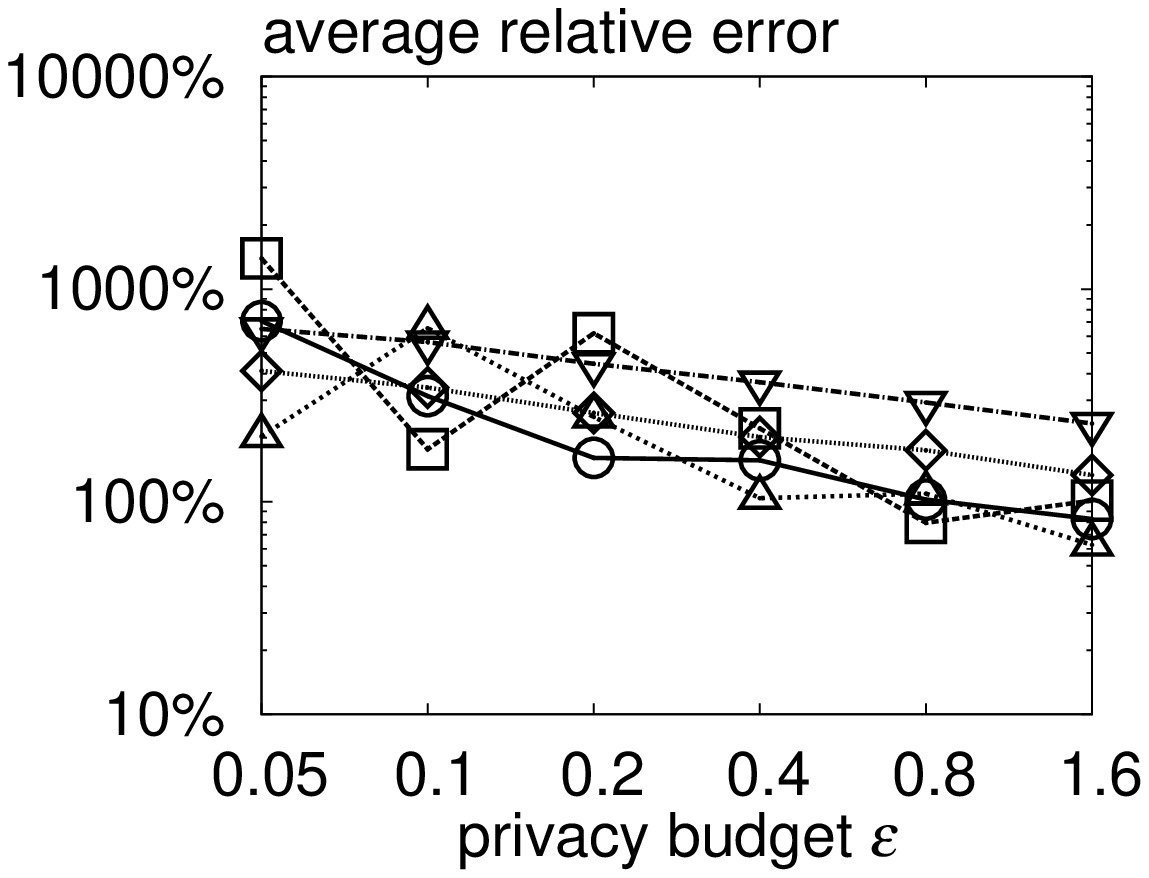}
&
\hspace{-2mm}\includegraphics[height=31.5mm]{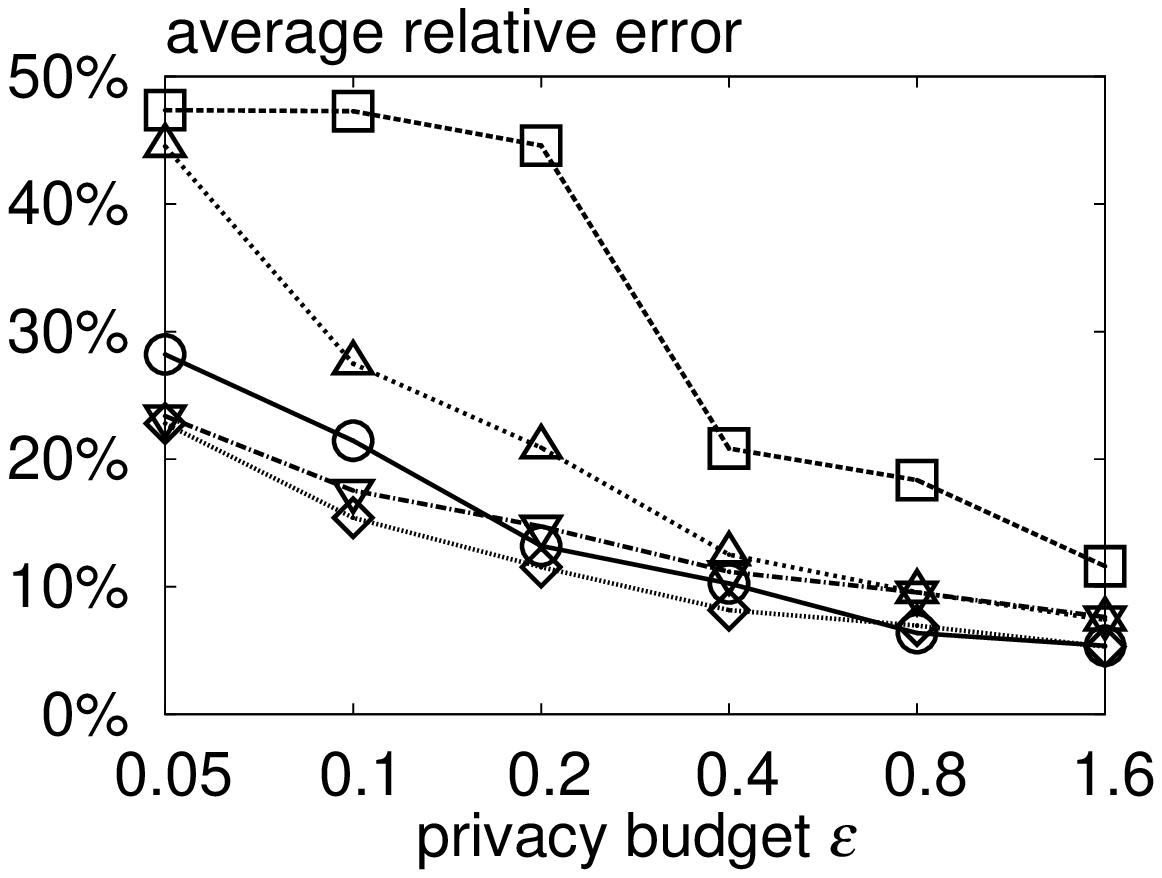} \\[0.8mm]

\hspace{-3mm} (c) {\sf road} - large queries.
&
\hspace{-2mm} (f) {\sf Gowalla} - large queries.
&
\hspace{-2mm}(i) {\sf NYC} - large queries.
&
\hspace{-2mm}(l) {\sf Beijing} - large queries.
\end{tabular}
\end{small}
\vspace{-2mm}
\caption{Impact of the tree height on UG.}
\label{fig:h:UG}
\end{figure*}

\begin{figure*}[t]
\centering
\begin{small}
\begin{tabular}{ccc}
\multicolumn{3}{c}{ \hspace{0mm} \includegraphics[height=3.6mm]{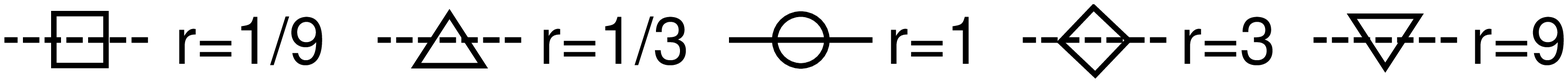}} \\

\hspace{-3mm}\includegraphics[height=31.5mm]{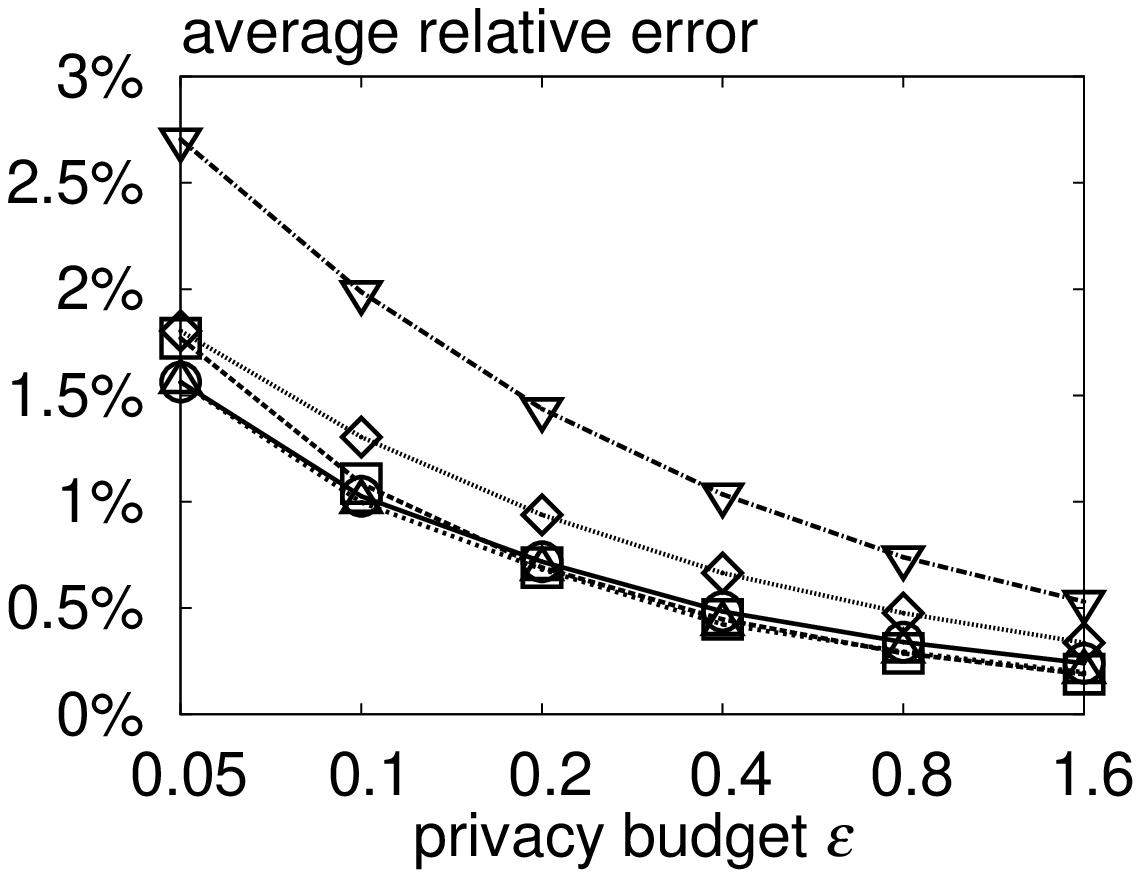}
&
\hspace{-2mm}\includegraphics[height=31.5mm]{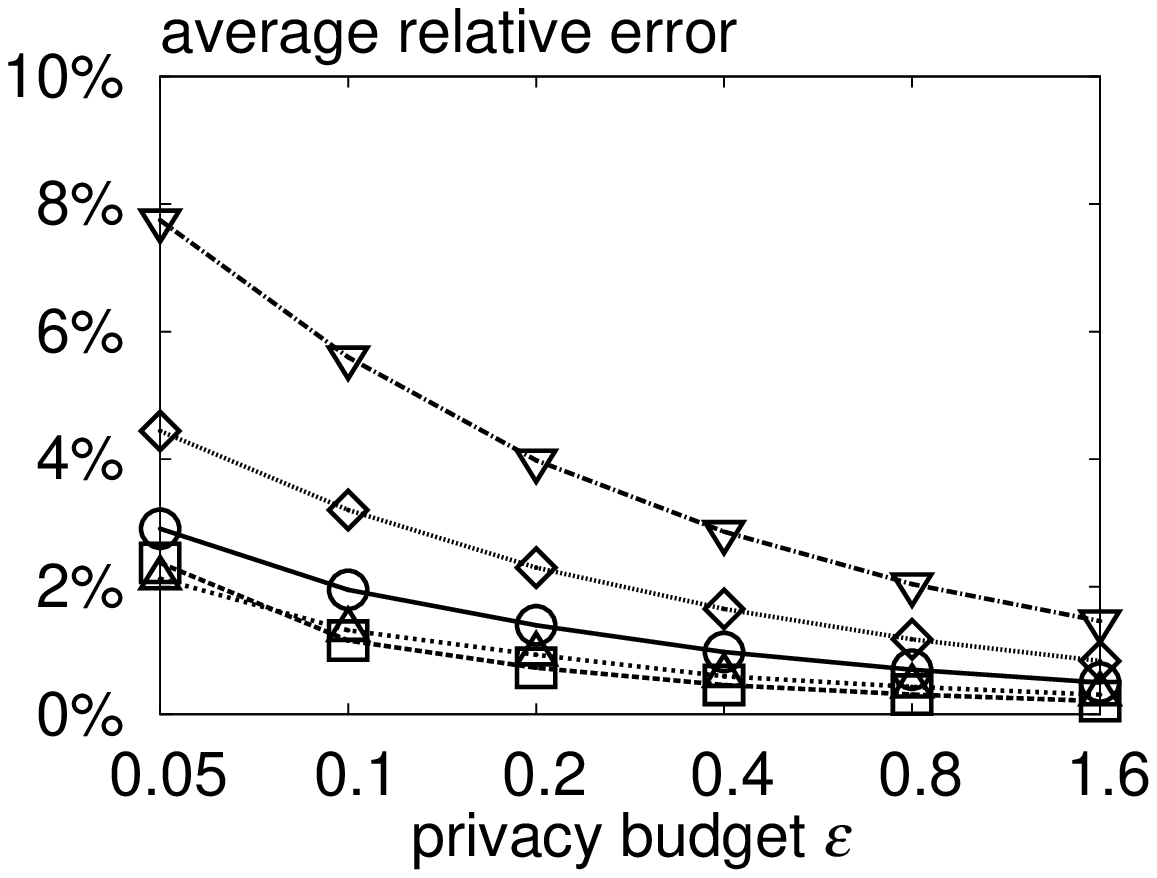}
&
\hspace{-2mm}\includegraphics[height=31.5mm]{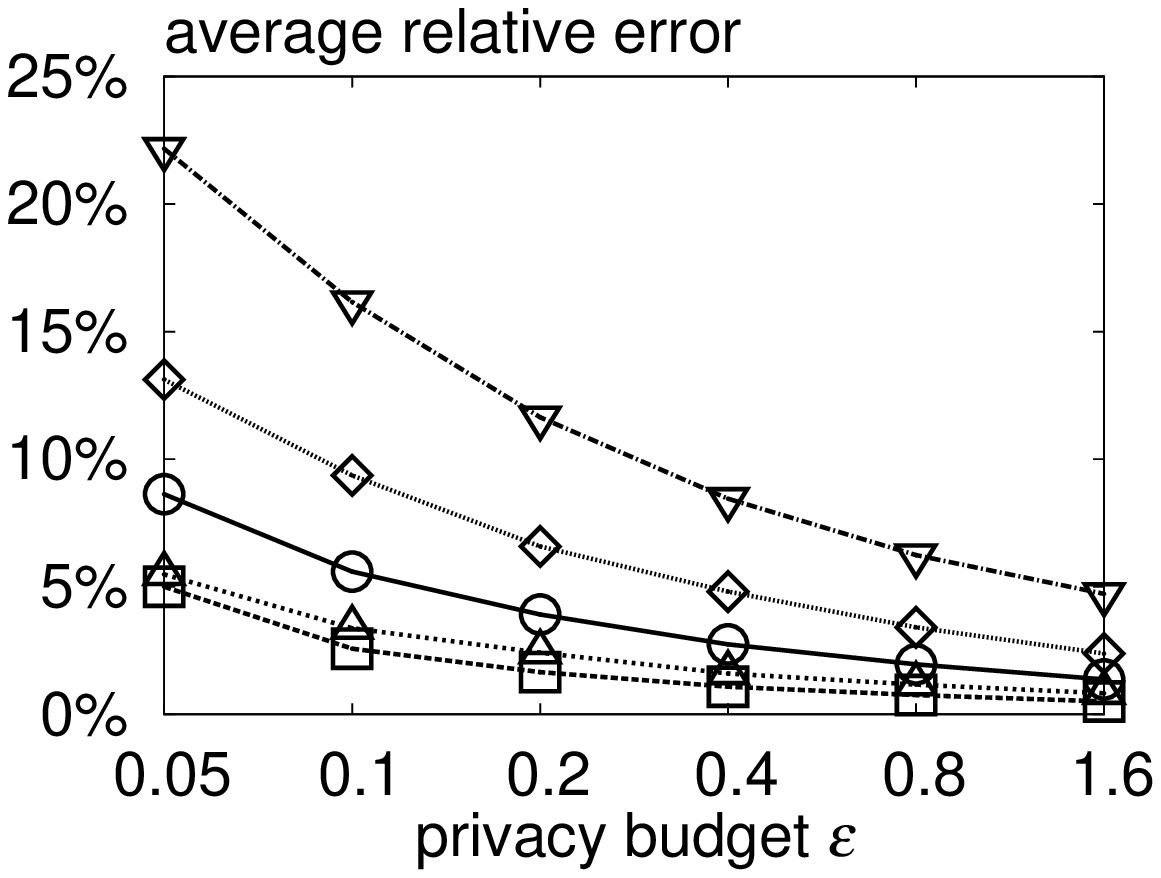}\\

\hspace{-3mm}(a) {\sf road} - small queries.
&
\hspace{-2mm}(b) {\sf road} - medium queries.
&
\hspace{-2mm}(c) {\sf road} - large queries. \\[2mm]

\hspace{-3mm}\includegraphics[height=31.5mm]{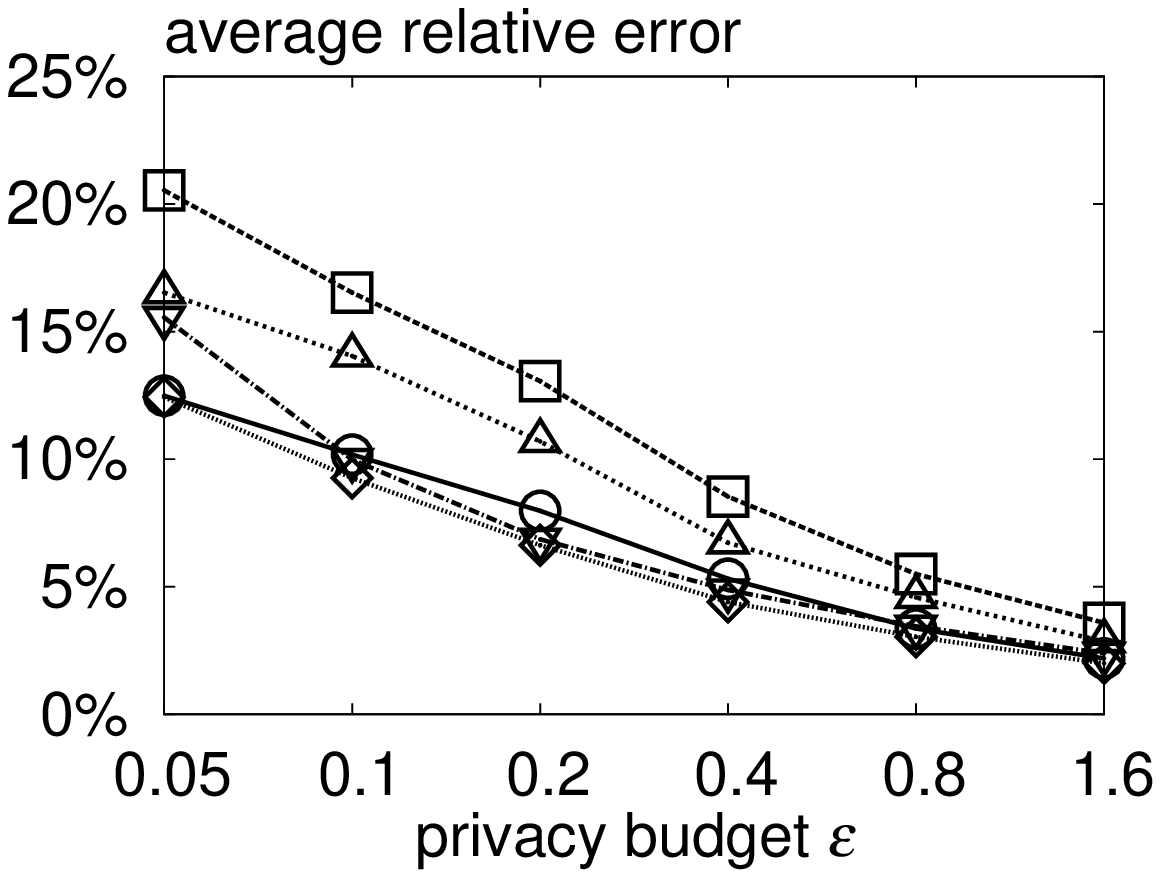}
&
\hspace{-2mm}\includegraphics[height=31.5mm]{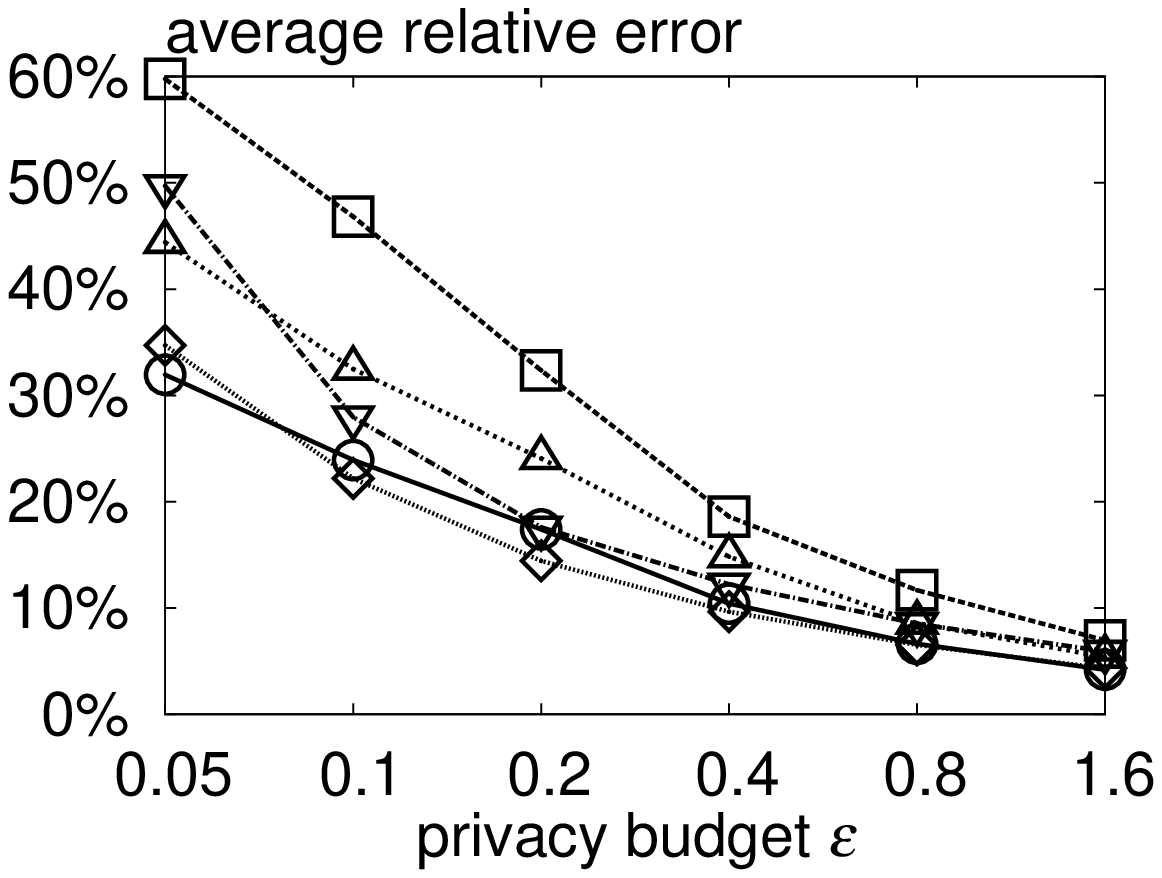}
&
\hspace{-2mm}\includegraphics[height=31.5mm]{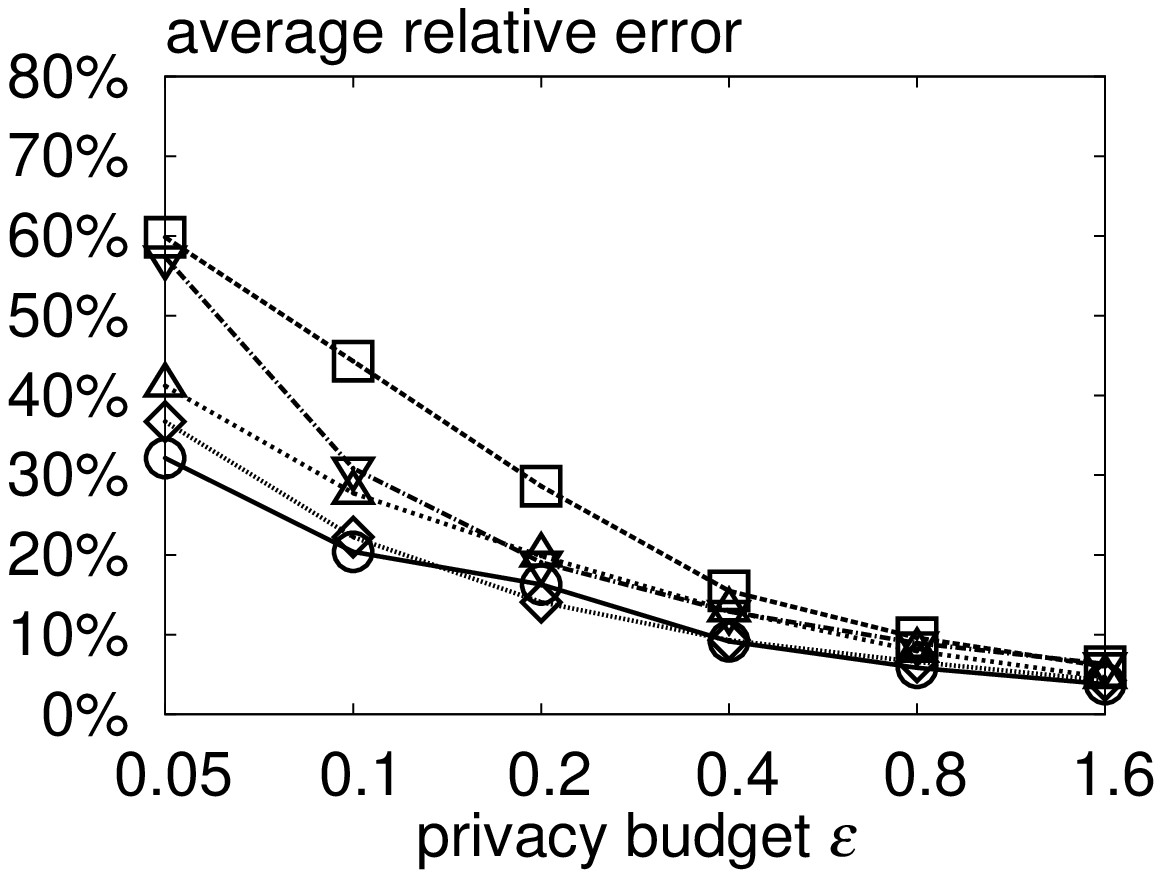}\\

\hspace{-3mm}(d) {\sf Gowalla} - small queries.
&
\hspace{-2mm}(e) {\sf Gowalla} - medium queries.
&
\hspace{-2mm}(f) {\sf Gowalla} - large queries.

\end{tabular}
\end{small}
\vspace{-1mm}
\caption{Impact of the tree height on AG.}
\label{fig:h:AG}
\end{figure*}

Our next set of experiments focuses on AG~\cite{QYL13}, an improved version of UG that (i) is specifically designed for two-dimensional data and (ii) uses two uniform grids with heuristically selected granularities. Roughly speaking, AG can be regarded as a hierarchical decomposition method that only publishes noisy counts for the nodes at the leaf level and one of the intermediate levels of the decomposition tree. Following the experimental setting for UG, we vary the grid granularities of AG, by simultaneously scaling the number of cells in each of its grids by a factor of $r$. Figure~\ref{fig:h:AG} shows the results on {\sf road} and {\sf Gowalla}. (We omit {\sf NYC} and {\sf Beijing} since AG's heuristics are inapplicable on four-dimensional data.) Observe that $r = 1$ leads to the best overall results for AG.

In the fifth set of experiments, we evaluate Hierarchy~\cite{QYL13hierarchy}, which (i) uses a decomposition tree of heuristically decided tree height $h$ and fanout $\beta$ and (ii) publishes a noisy count for every non-root node. For two-dimensional data, the heuristics in \cite{QYL13hierarchy} suggest setting $\beta = 64$ (i.e., splitting a node results in $8 \times 8$ child nodes) and $h = 3$ (i.e., the decomposition tree contains a root, an intermediate level, and a leaf level). For four-dimensional data, the heuristics in \cite{QYL13hierarchy} lead to $\beta = 1296$ and $h = 4$, which result in a prohibitively large number of leaf nodes. Therefore, we only apply Hierarchy on our two-dimensional datasets. Figure~\ref{fig:h:hierarchy} illustrates the query error of Hierarchy on {\sf road} and {\sf Gowalla}, when $h$ varies from $3$ to $8$. (We do not consider $h = 2$, since Hierarchy degrades to UG in that case.) Observe that $h = 3$ leads to the smallest query error in most settings, except for the small query set on {\sf road} when $\e \ge 0.4$. Therefore, the tree height suggested in \cite{QYL13hierarchy} is generally the best choice for Hierarchy.

\begin{figure*}[t]
\centering
\begin{small}
\begin{tabular}{ccc}
\multicolumn{3}{c}{ \hspace{0mm} \includegraphics[height=3.6mm]{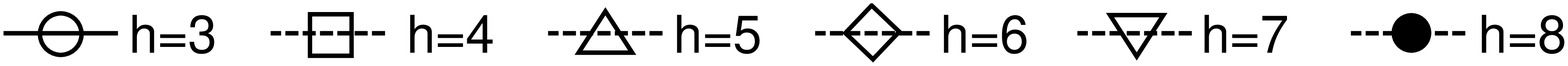}} \\

\hspace{-3mm}\includegraphics[height=31.5mm]{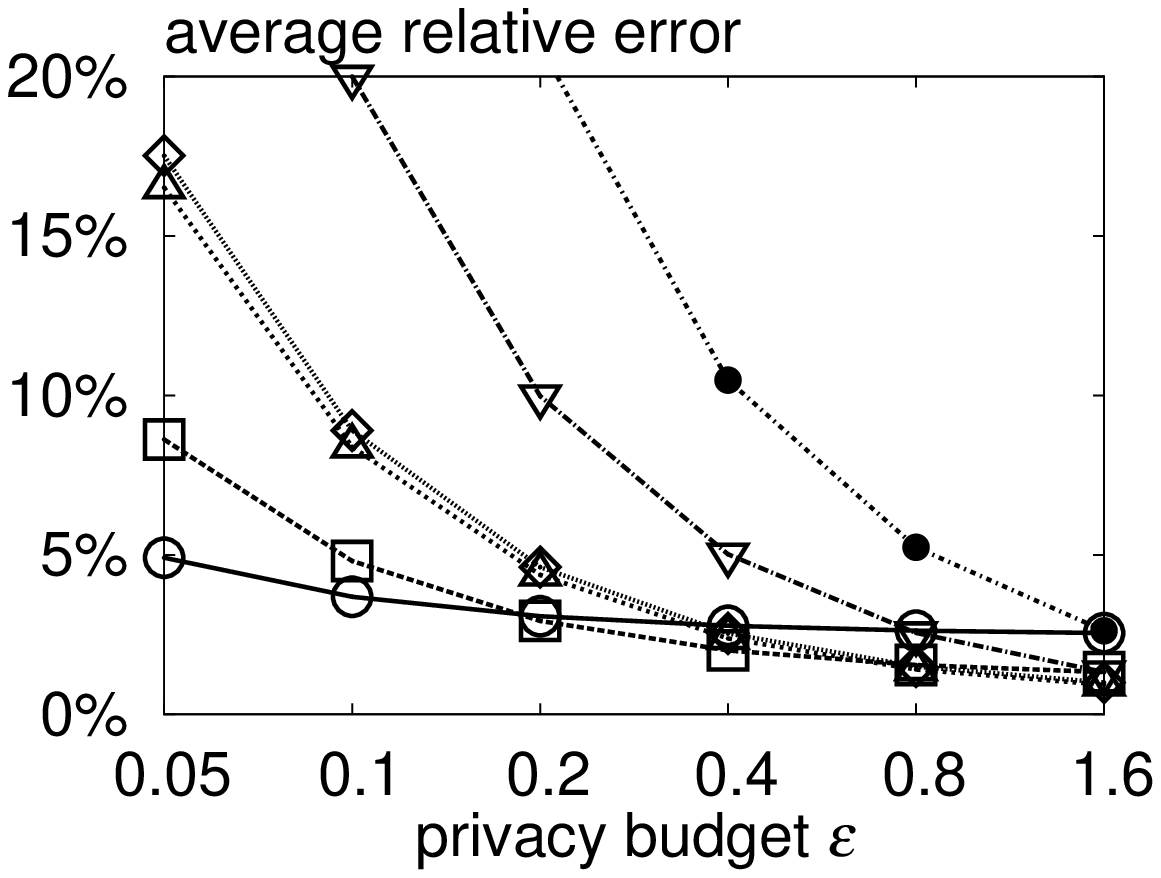}
&
\hspace{-2mm}\includegraphics[height=31.5mm]{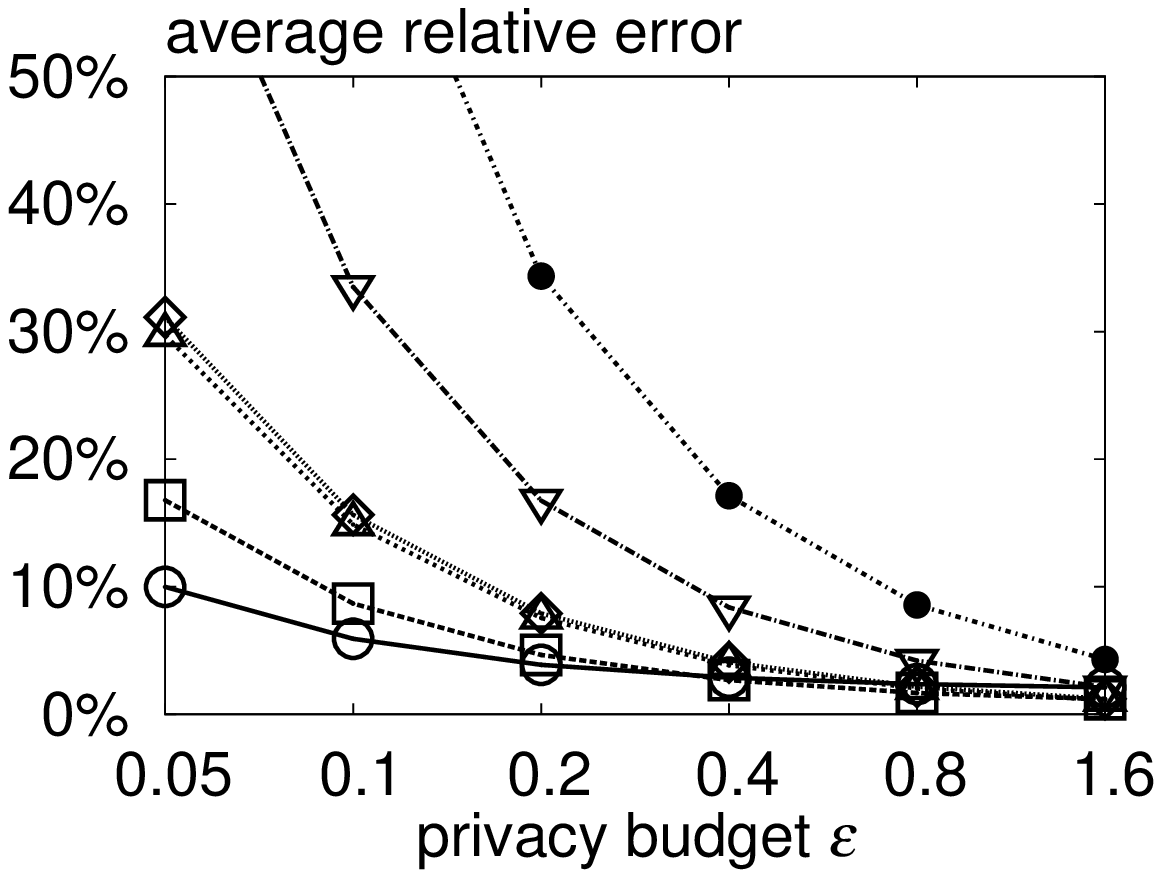}
&
\hspace{-2mm}\includegraphics[height=31.5mm]{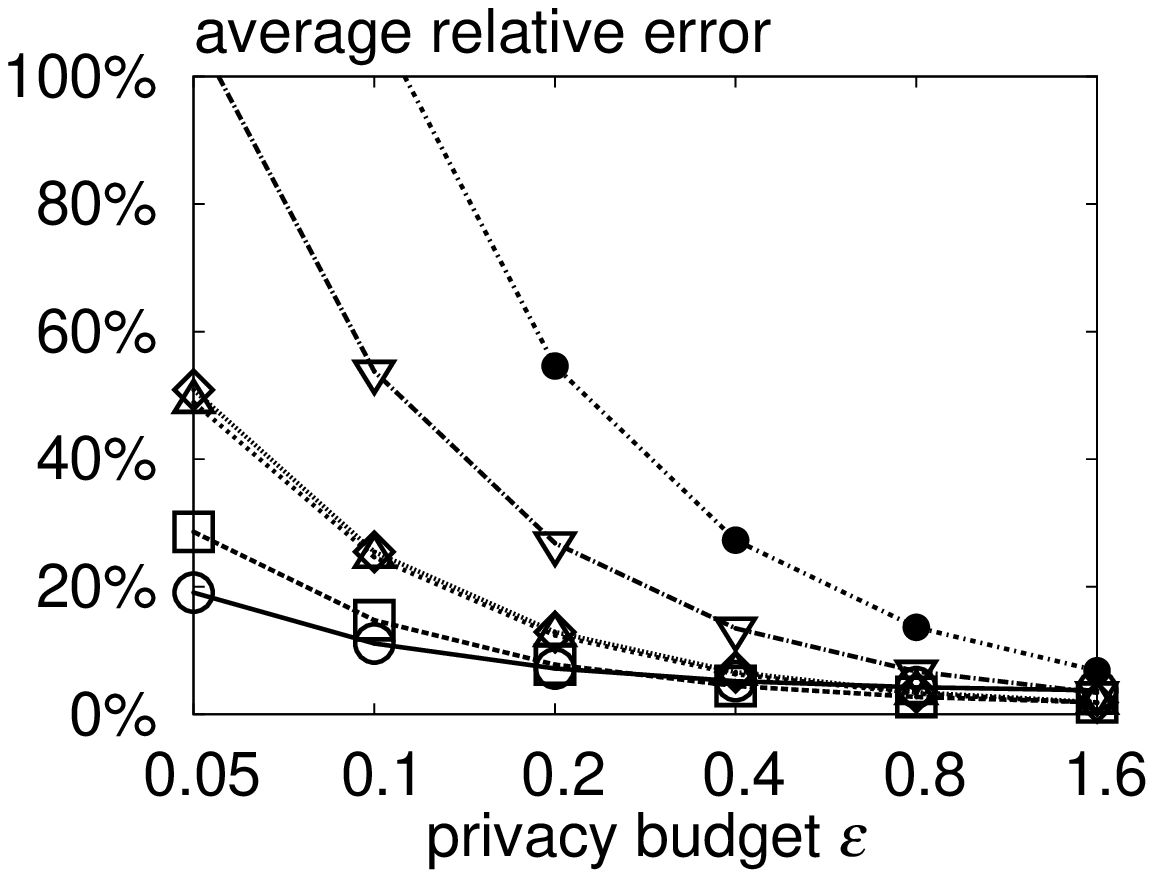}\\

\hspace{-3mm}(a) {\sf road} - small queries.
&
\hspace{-2mm}(b) {\sf road} - medium queries.
&
\hspace{-2mm}(c) {\sf road} - large queries. \\[2mm]

\hspace{-3mm}\includegraphics[height=31.5mm]{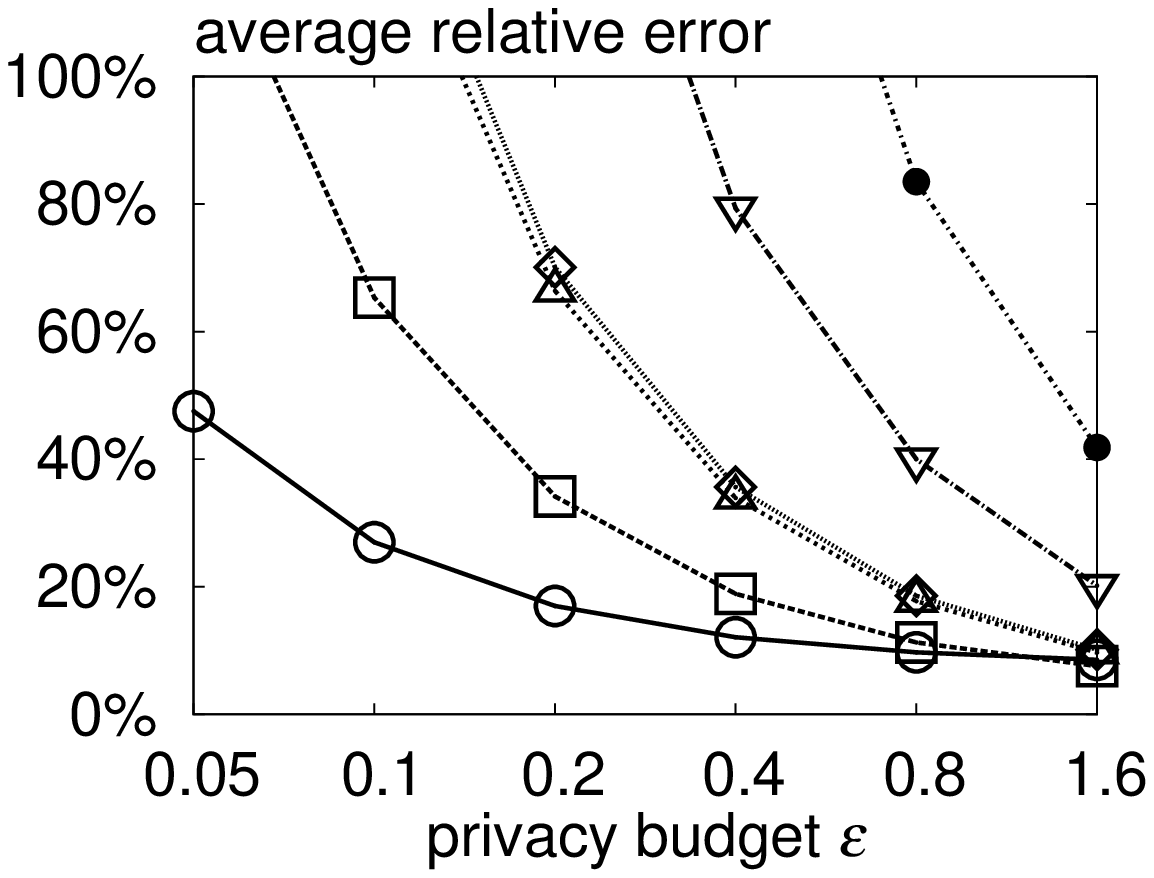}
&
\hspace{-2mm}\includegraphics[height=31.5mm]{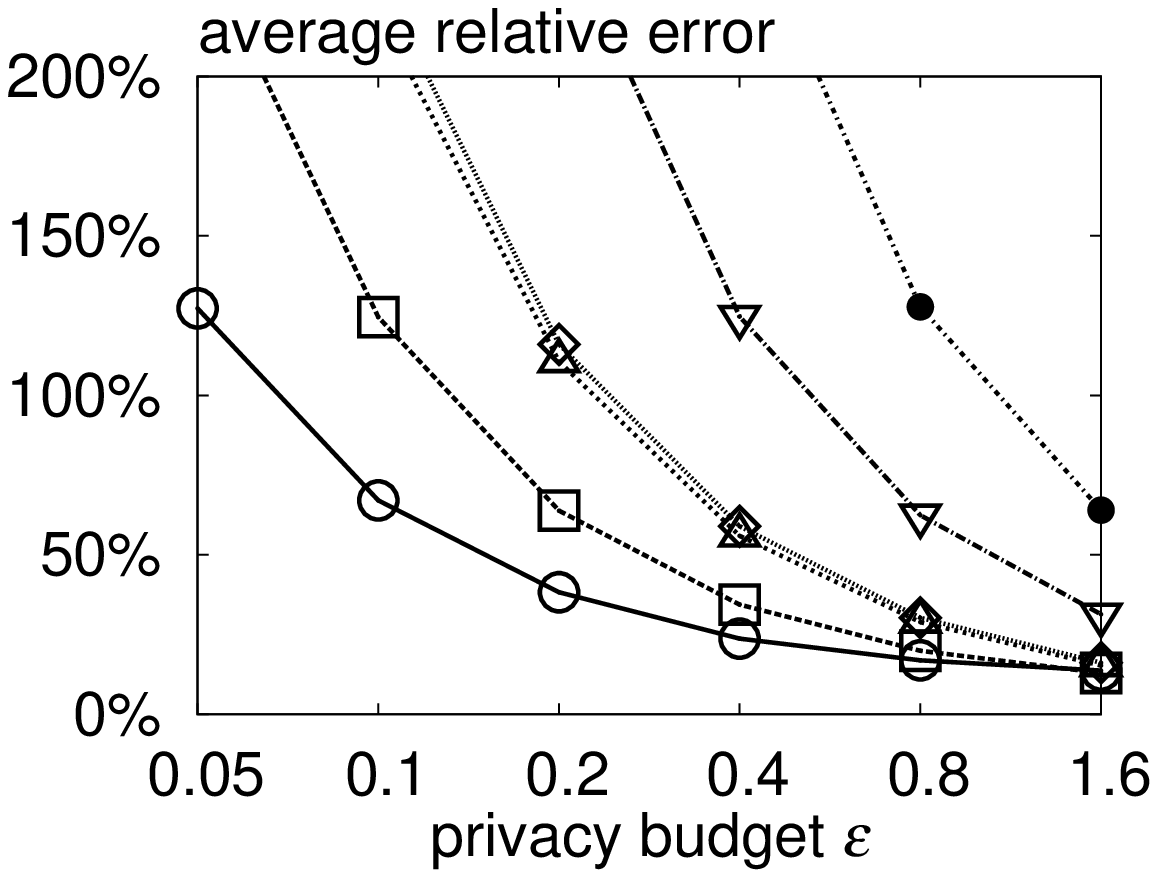}
&
\hspace{-2mm}\includegraphics[height=31.5mm]{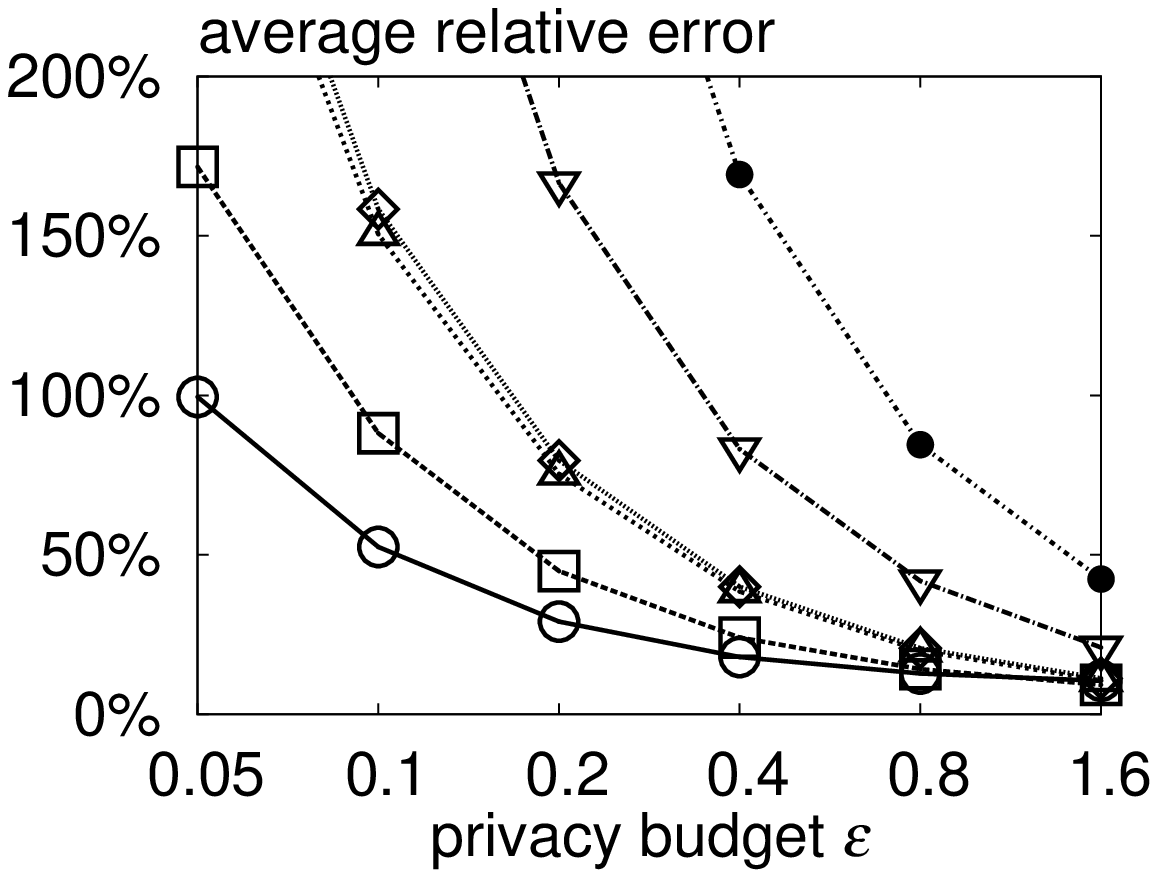}\\

\hspace{-3mm}(d) {\sf Gowalla} - small queries.
&
\hspace{-2mm}(e) {\sf Gowalla} - medium queries.
&
\hspace{-2mm}(f) {\sf Gowalla} - large queries.

\end{tabular}
\end{small}
\vspace{-1mm}
\caption{Impact of the tree height $h$ on Hierarchy.}
\label{fig:h:hierarchy}
\end{figure*}

Our last set of experiments examines N-gram~\cite{CAC12}, a hierarchical decomposition method for modeling sequence data. \cite{CAC12} suggests using a decomposition tree of height $h = 5$. Figure~\ref{fig:h:ngram} illustrates the accuracy of N-gram in answering top-$k$ frequent string queries, when $h$ varies from $3$ to $7$. Notice that although $h = 5$ does not yield the highest query accuracy in all cases, it does provide one of the best overall results (with $h = 4$ being a close competitor). This justifies our choice of using $h = 5$ in Section~\ref{sec:exp} of our paper.

\begin{figure*}[!t]
\centering
\begin{small}
\begin{tabular}{ccc}
\multicolumn{3}{c}{ \hspace{0mm} \includegraphics[height=4mm]{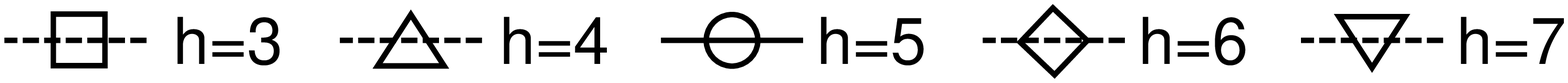}} \\

\hspace{-3mm}\includegraphics[height=31.5mm]{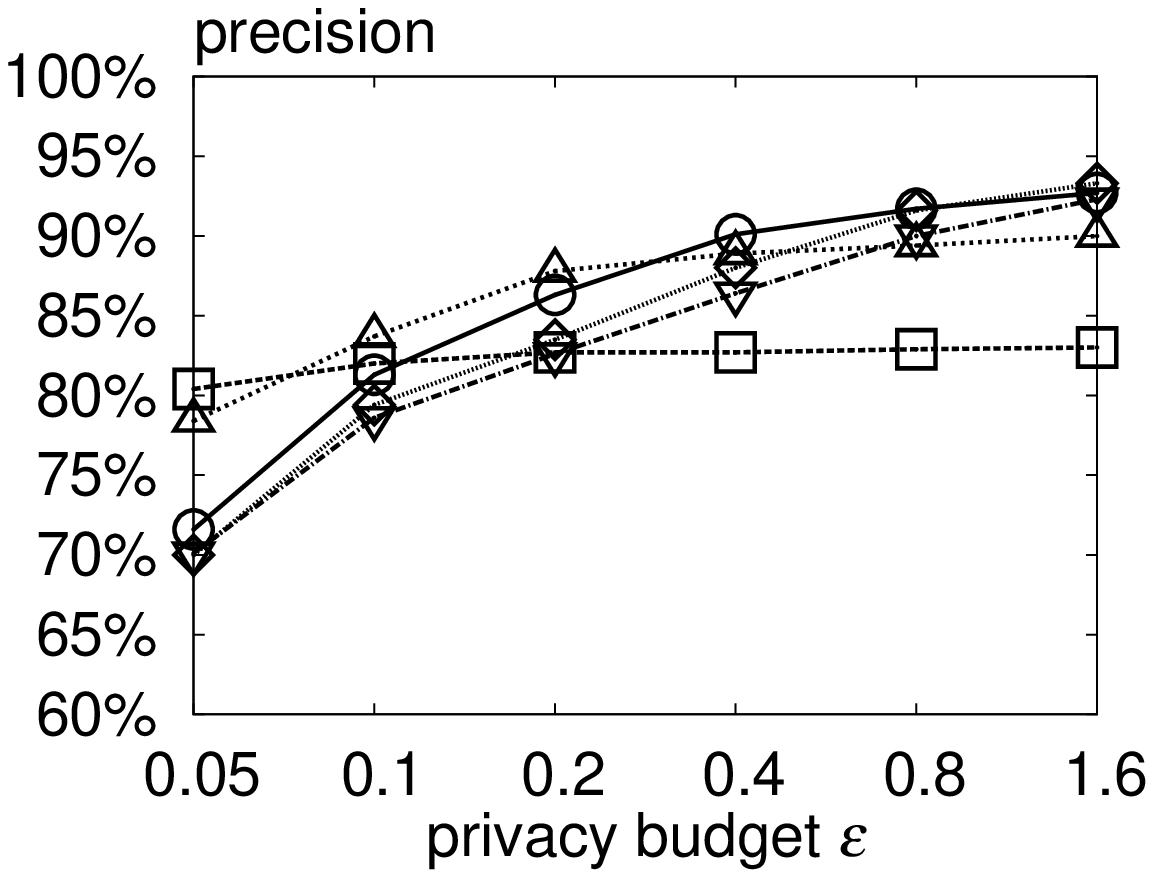}
&
\hspace{-2mm}\includegraphics[height=31.5mm]{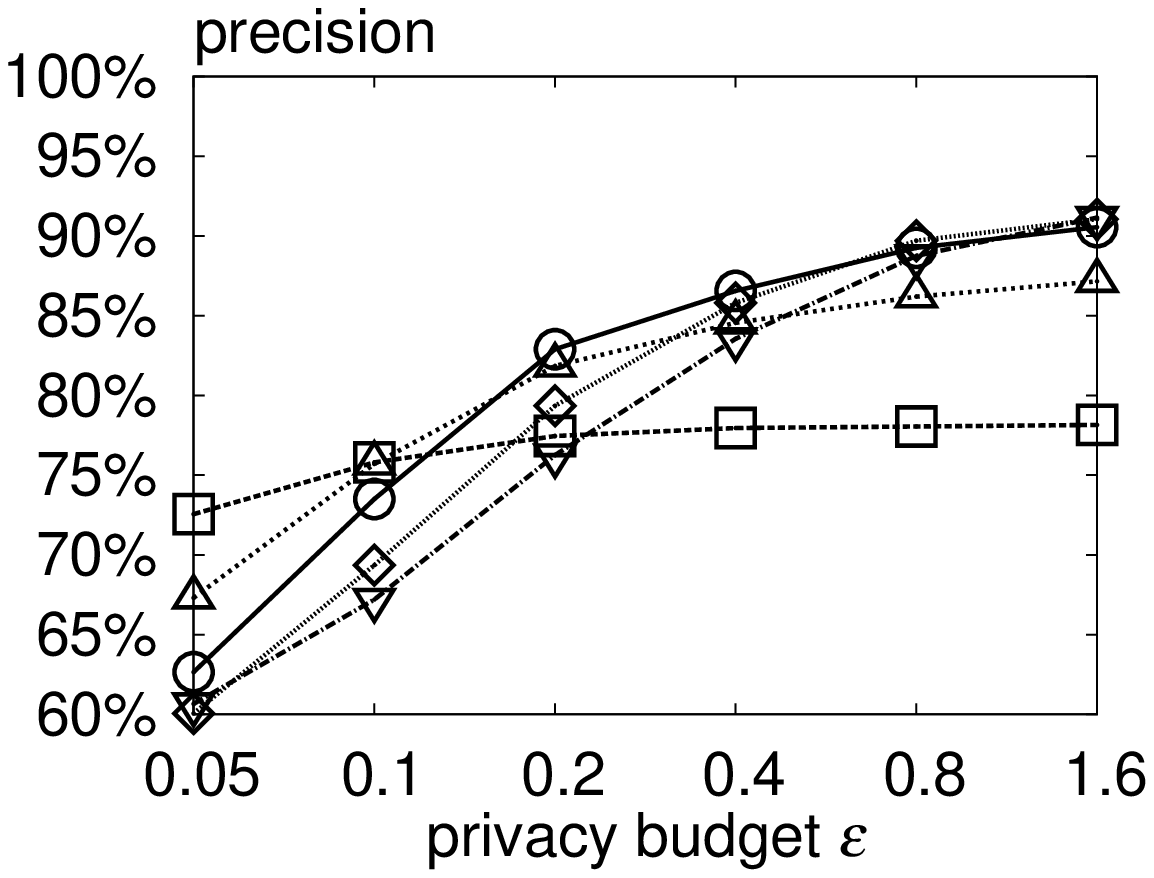}
&
\hspace{-2mm}\includegraphics[height=31.5mm]{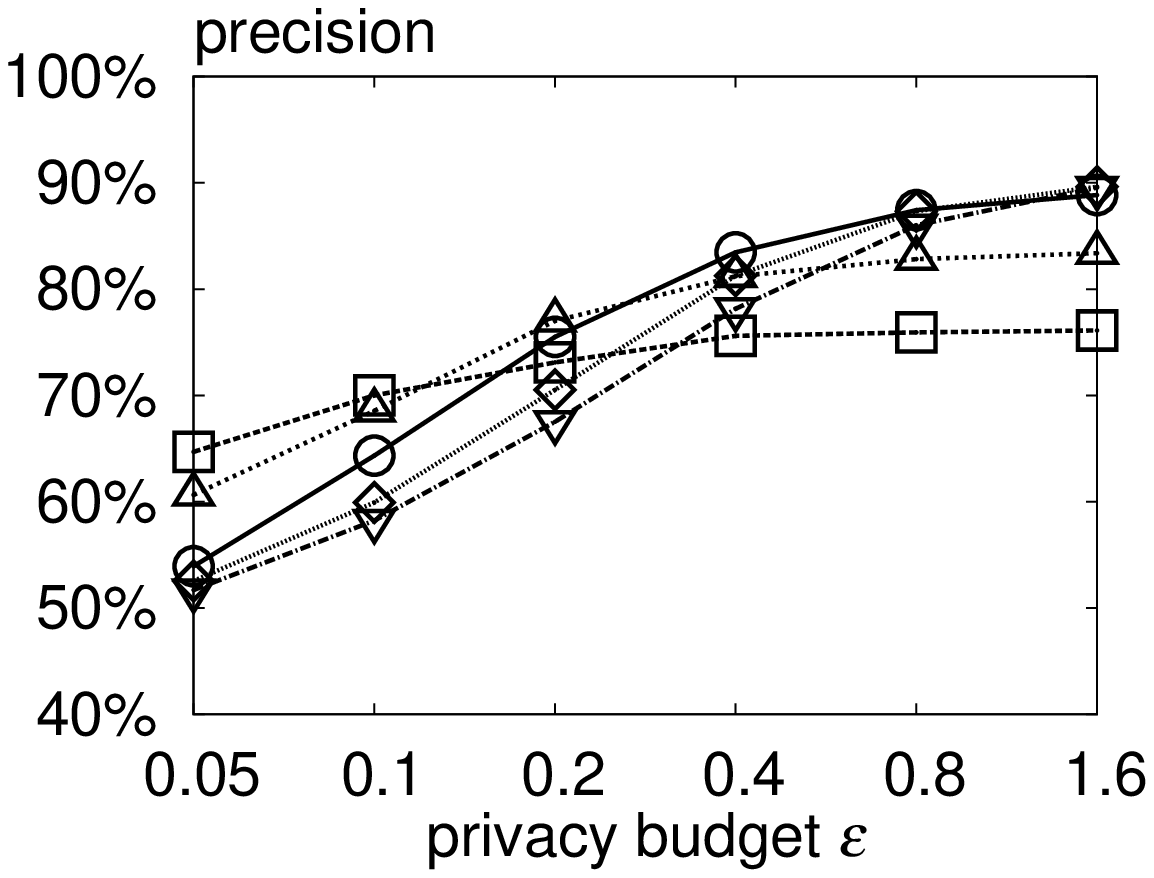}\\

\hspace{-3mm}(a) {\sf mooc} - top$50$.
&
\hspace{-2mm}(b) {\sf mooc} - top$100$.
&
\hspace{-2mm}(c) {\sf mooc} - top$200$.\\[2mm]

\hspace{-3mm}\includegraphics[height=31.5mm]{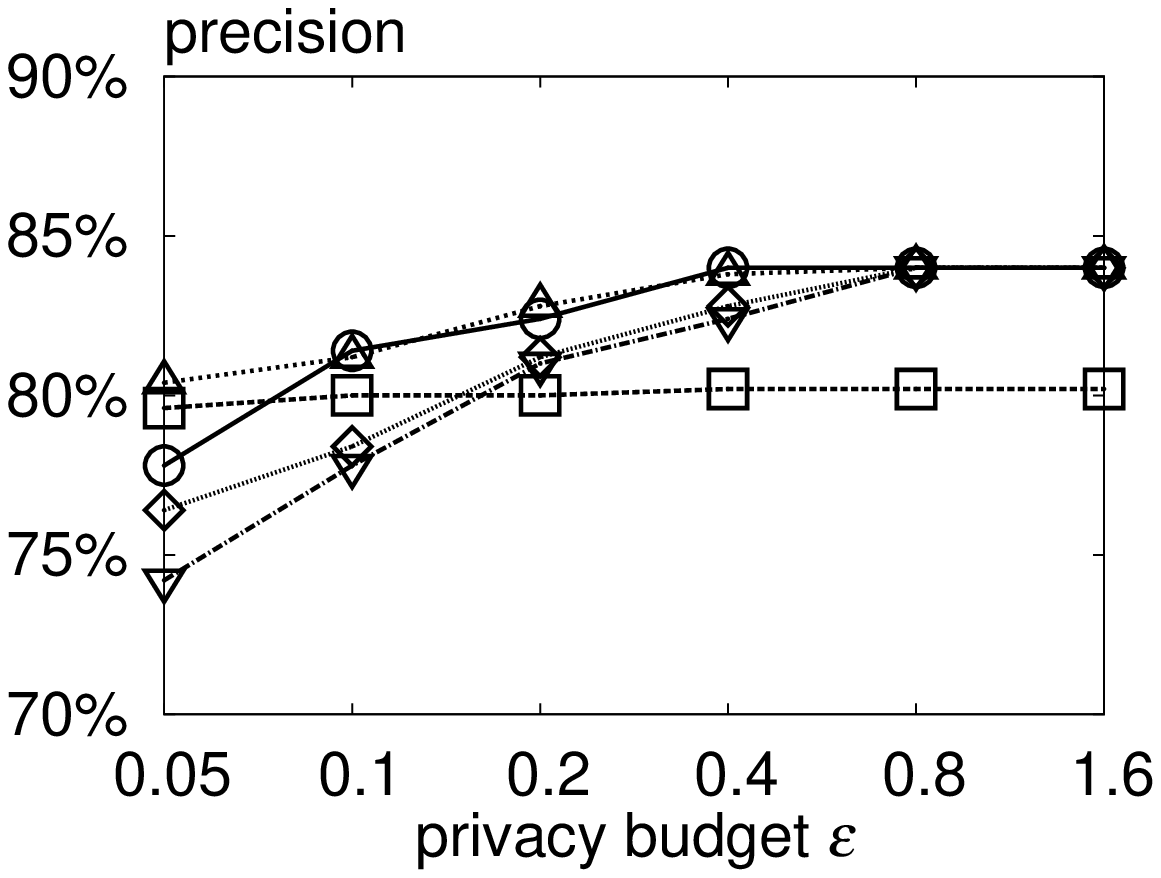}
&
\hspace{-2mm}\includegraphics[height=31.5mm]{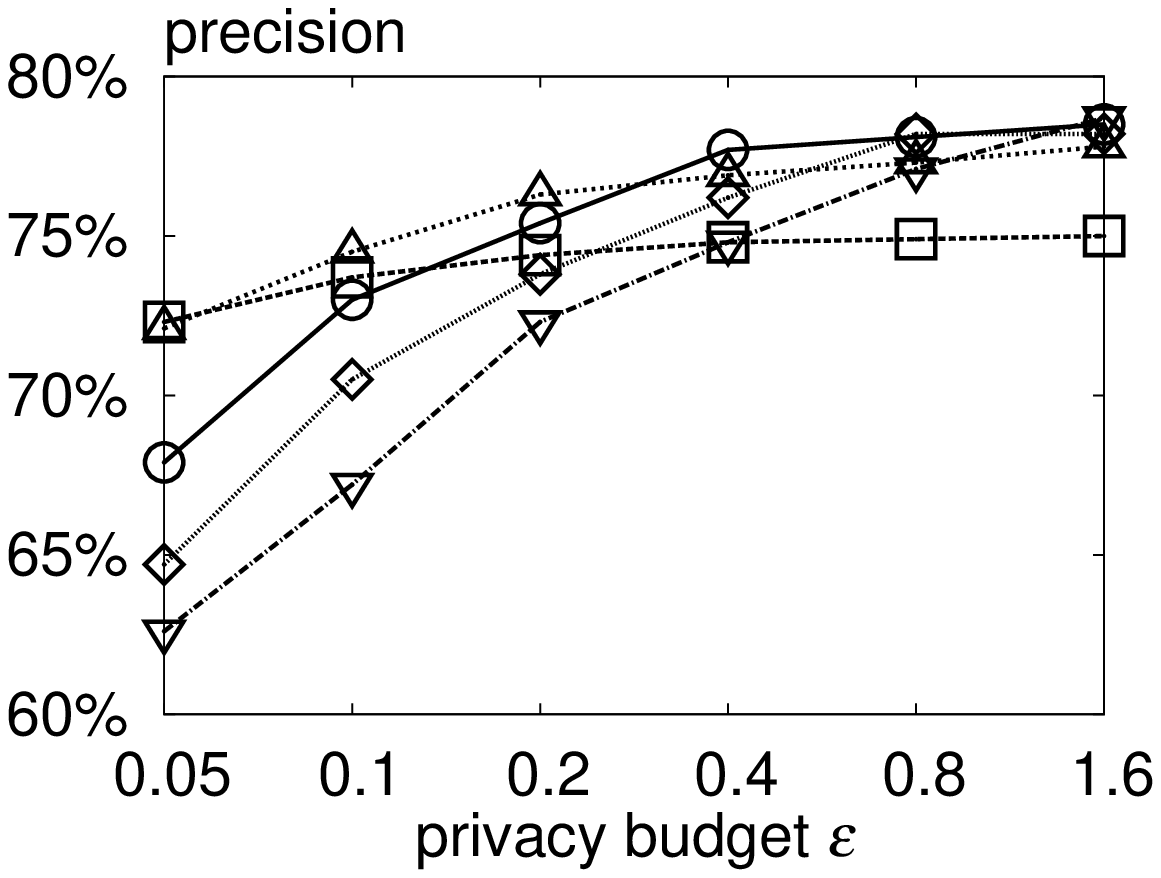}
&
\hspace{-2mm}\includegraphics[height=31.5mm]{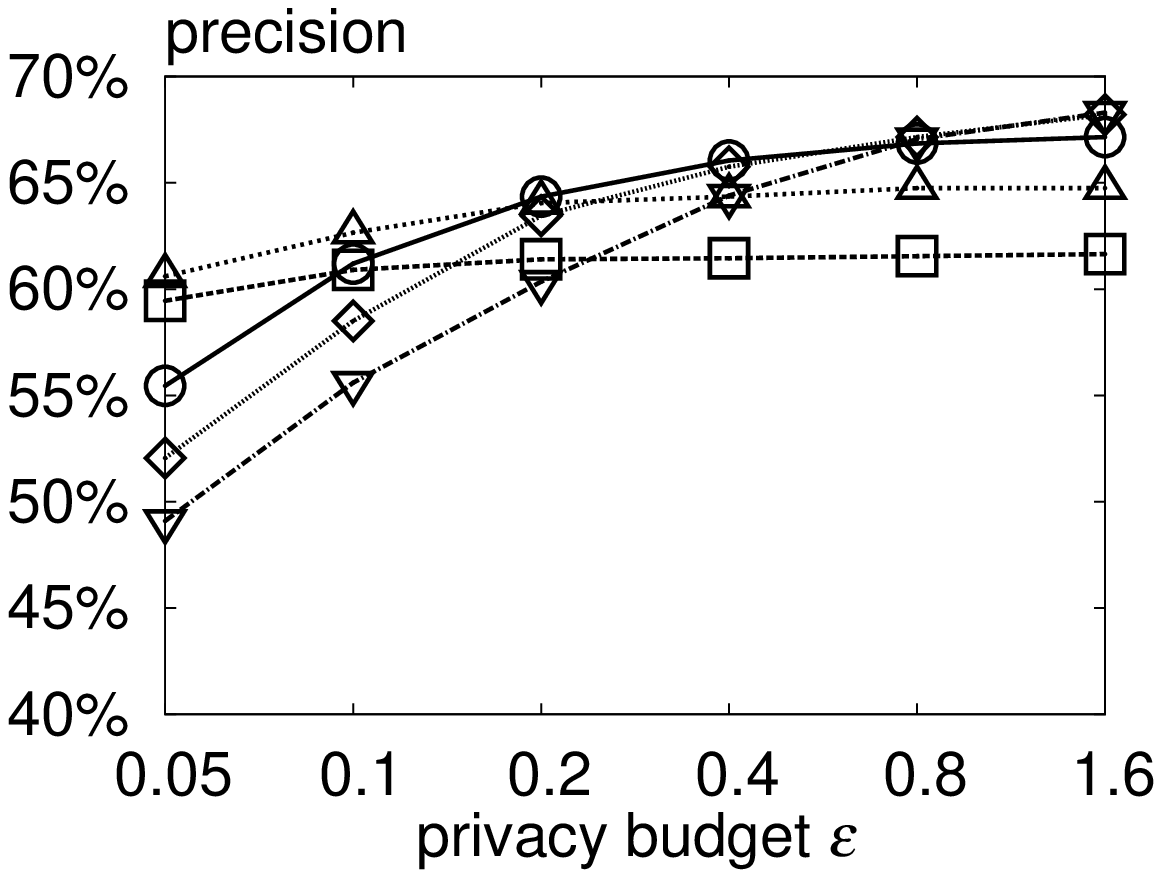}\\

\hspace{-3mm}(d) {\sf msnbc} - top$50$.
&
\hspace{-2mm}(e) {\sf msnbc} - top$100$.
&
\hspace{-2mm}(f) {\sf msnbc} - top$200$.
\end{tabular}
\end{small}
\vspace{-1mm}
\caption{Impact of the tree height $h$ on N-gram.}
\label{fig:h:ngram}
\end{figure*}

\balance

\end{sloppy}
\end{document}